\documentclass[aps,
preprint,
superscriptaddress,
nofootinbib,eqsecnum,floats]{revtex4}
\usepackage{graphicx,color}
\usepackage{float}
\usepackage{amsmath}
\usepackage{amssymb}
\usepackage{pifont}
\usepackage{graphicx}
\usepackage{caption}
\usepackage{fancybox}
\usepackage{bm}
\usepackage{color,colordvi}
\usepackage{epsfig}
\usepackage{natbib}
\usepackage{mciteplus}
\usepackage{mathcomp}
\usepackage{cancel}  
\usepackage{ulem}    
\usepackage[colorlinks=true,urlcolor=blue]{hyperref}
\usepackage{footmisc}
\usepackage{footnotebackref}
\usepackage{multirow}
\usepackage[subnum]{cases}
\usepackage[utf8]{inputenc}
\usepackage{titletoc}
\titlecontents*{subsubsection}
  [1.5em]
  {\small}
  {}
  {}
  {\ \thecontentspage}
  [,\ ]
  []
\DeclareMathOperator{\sgn}{sgn}
\hypersetup{                
backref = true,             
pagebackref = true,                 
hyperindex = true,          
colorlinks = true,          
breaklinks = true,          
urlcolor = blue,            
linkcolor = blue,           
bookmarks = true,           
bookmarksopen = true,       
citecolor=red,
}
\begin{document}
\begin{flushright}
\end{flushright}
\newcommand  {\ba} {\begin{eqnarray}}
\newcommand  {\ea} {\end{eqnarray}}
\def\cM{{\cal M}}
\def\cO{{\cal O}}
\def\cK{{\cal K}}
\def\cS{{\cal S}}
\newcommand{\mh}{m_{h^0}}
\newcommand{\mw}{m_W}
\newcommand{\mz}{m_Z}
\newcommand{\mt}{m_t}
\newcommand{\mb}{m_b}
\newcommand{\be}{\beta}\newcommand{\al}{\alpha}
\newcommand{\lam}{\lambda}
\newcommand{\no}{\nonumber}
\def\ga{\mathrel{\raise.3ex\hbox{$>$\kern-.75em\lower1ex\hbox{$\sim$}}}}
\def\la{\mathrel{\raise.3ex\hbox{$<$\kern-.75em\lower1ex\hbox{$\sim$}}}}
\newcommand{\gilbert}[1]{\textcolor{black}{ \bf{#1}}}
\newcommand{\gilbertzero}[1]{\textcolor{black}{ #1}}
\newcommand{\gilbertwhite}[1]{\textcolor{white}{ #1}}
\newcommand{\cmark}{\ding{51}}%
\newcommand{\Cmark}{\ding{52}}%
\newcommand{\xmark}{\ding{55}}%

\title{Vacuum Stability Conditions for Higgs Potentials with $SU(2)_L$ Triplets }


 \author{Gilbert Moultaka\footnote{corresponding author}}
\affiliation{Laboratoire Charles Coulomb (L2C), University of Montpellier, CNRS, Montpellier, France.}
\author{Michel C. Peyran\`ere}
\affiliation{LUPM, Univ Montpellier, CNRS, Montpellier,France}
\date{\today}

\begin{abstract}
Tree-level dynamical stability of scalar field potentials in renormalizable theories can 
in principle be expressed in terms of positivity conditions on quartic polynomial structures. 
However, these conditions cannot always be cast in a fully analytical resolved form, involving only
the couplings and being valid for all field directions. In this paper we consider such 
forms in three physically motivated models involving $SU(2)$ triplet scalar fields:
 the Type-II seesaw model, the Georgi-Machacek model, and a generalized two-triplet model. 
A detailed analysis of the latter model allows to establish the full  
set of necessary and sufficient boundedness from below conditions. These can serve as a guide, together with unitarity
and vacuum structure constraints, for consistent phenomenological (tree-level) studies. They also provide a seed for improved loop-level conditions, and encompass in particular the leading ones for the more specific Georgi-Machacek case.
 Incidentally, we present complete proofs of various properties and also derive general positivity conditions on 
quartic polynomials that are equivalent but much simpler than the 
ones used in the literature.

\end{abstract}

\maketitle
\setcounter{tocdepth}{1}
\tableofcontents

\section{Introduction}
Since the experimental discovery of a Standard Model (SM)-like 
Higgs particle at the LHC \cite{Aad:2012tfa,Chatrchyan:2012ufa} 
and the lack
so far of any direct evidence for physics beyond the standard 
model (BSM)\footnote{possible indirect "evidence" 
notwithstanding \cite{Pich:2019pzg}}, 
one might ask whether the properties of the 
discovered $125$~GeV scalar 
particle being so much close to the SM predictions (see e.g. \cite{ExpHiggsLHC})
leaves any room for BSM physics to reside below the TeV or 
at the nearby few TeV scale. If new physics is present in the electroweak symmetry
breaking sector it should either be very heavy (almost decoupled) or light but
having very weak mixing with the SM-Higgs. For the latter case, extensions
of the scalar sector of the SM by complex or real $SU(2)_L$ triplets, or further extensions
comprising Left-Right symmetric gauge groups, or possibly higher representation multiplets, are appealing possibilities.  
A typical example is the Type-II seesaw model for neutrino masses \cite{Konetschny:1977bn, Cheng:1980qt, Lazarides:1980nt, 
Schechter:1980gr,Mohapatra:1979ia, Mohapatra:1980yp}, 
for which an essentially SM-like physical Higgs state is unavoidable, a consequence of the very small mixing between the doublet and 
triplet neutral components being set off by the tiny 
(Majorana) neutrino mass scale  as compared to the electroweak scale.
Another example is the Georgi-Machacek model  \cite{Georgi:1985nv, Chanowitz:1985ug} with one complex and one real
triplet such that a tree-level custodial symmetry is preserved in the scalar sector through a global  $SU(2)_R$.

These scenarios have triggered various activities both on the phenomenological level, (including left-right symmetric or not, supersymmetric or not, scenarios) see e.g. among the recent works 
\cite{Gluza:2020qrt,Padhan:2019jlc,Primulando:2019evb,Fuks:2019clu,Ghosh:2017pxl,Ouazghour:2018mld,Ait-Ouazghour:2020slc,Dev:2013ff,
Dev:2017ouk,Dev:2018kpa,Dev:2019hev,Frank:2020mqh,Huitu:2020qxm} (and references therein), and in experimental searches at the LHC
for neutral, charged, and in particular doubly-charged scalar states that are specific to such class of models decaying either to same-sign
leptons or $W$ boson pairs \cite{Aaboud:2017qph,ATLAS:2020ius}, \cite{Chatrchyan:2012ya,CMS:2017pet}.
As for any extension of the SM, and in the absence of a unifying ultraviolet completion, these models have an increased number of 
free parameters and thus a large freedom in particular for the physical spectrum of the scalar sector. Theoretical conditions
such as the stability of the potential, a consistent electroweak vacuum, unitarity bounds, etc., are thus welcome as a guide
together with the experimental exclusion limits to narrow down future search strategies. 

The present paper focuses on the potential stability issue for three models: the Type-II seesaw model, the Georgi-Machacek model, and a generalized two-triplet model.  The aim is to address as thoroughly as possible the theoretical 
determination of necessary and sufficient (NAS)
conditions on the scalar couplings that ensure a physically sound bounded from below (BFB) potential. The NAS BFB conditions have already
been considered in the corresponding literature. Inspired by the approach of \cite{ElKaffas:2006nt}  initially proposed for the general 
two-Higgs doublet potential, the strategy consists in a change of parameterization of the field space reducing it to a minimal
set of variables corresponding to positive-valued ratios of field magnitudes and to field orientations varying in compact domains. 
It is then 
found that in contrast with the general two-Higgs doublet case, the general doublet-triplet potential leads to a simplification that allows 
a fully analytical solution.  
 A complete answer was given first in \cite{Arhrib:2011uy} and  \cite{Bonilla:2015eha} for
the Type-II seesaw model. Following the same approach the NAS BFB conditions were provided for the Georgi-Machacek model  
in \cite{Hartling:2014zca}. We will nevertheless reexamine the issue for these two models, supplementing with complete proofs,  
for reasons that will become clear in the course of the study. Encouraged by the success of the approach, we extend it in the present
paper to a generalized two-triplet model, that we will dub pre-custodial, for which we provide novel results by deriving the full NAS BFB 
conditions. Some stability constraints have already been given for this model in \cite{Blasi:2017xmc} and \cite{Krauss:2017xpj}corresponding however to specific directions in the field space, thus to a subclass
of necessary conditions. 
 This pre-custodial model can be of phenomenological interest by itself, but can also serve as a guideline for the effective
potential beyond tree-level in the Georgi-Machacek model. 

The main issue of the analysis will be to cast the conditions in a form as close as possible to a fully {\sl resolved} one. 
By `fully {\sl resolved}' we mean {\sl an analytical expression that depends solely on
 the couplings with no reference to orientations or magnitudes in field space.} A fully resolved form, when possible, is an ideal result
both technically, since no scan over the field configurations is needed, and physically, as consistency constraints are expressed directly 
 in terms of the (physical) couplings. This was the case for the conditions derived in \cite{Arhrib:2011uy},  \cite{Bonilla:2015eha}
 while in \cite{Hartling:2014zca} the conditions were resolved with respect to only one parameter, thus remaining in a partially unresolved form albeit with a residual field dependence 
reduced to a compact domain. As we will see, similar configurations arise in the pre-custodial model where the resolving occurs at different stages with respect to different parameters. 
A hindrance in the way of reaching fully resolved conditions emerges whenever dealing with a quartic polynomial that cannot be reduced to 
a biquadratic one. This fact motivated us
to investigate further a rather mathematical question, the positivity of general quartic polynomials, for which we determine 
NAS conditions that are simpler than the ones found in the literature.

A word of caution is in order here: The NAS BFB conditions we are considering are obtained by requiring the tree-level potential not 
be unbounded from below in any direction in the field space. It is only in that sense that they are necessary and sufficient. Obviously
they might be only necessary in a wider physical sense when taking into account the structure of the vacua. Moreover, going beyond tree-level would modify these conditions. As alluded to above and will be briefly discussed towards the end of the paper,
the tree-level conditions can, however, encapsulate in some cases the leading loop corrections.

Several methods to treat the stability of the potential have been conceived in the literature, e.g. specifically for 
multi-Higgs-doublets models \cite{Maniatis:2006fs,Maniatis:2014oza,Maniatis:2015gma} including elegant geometric approaches 
\cite{Ivanov:2006yq}, or more general methods relying on copositivity \cite{Kannike:2012pe,Chakrabortty:2013mha,Kannike:2016fmd} or on other powerful mathematical techniques \cite{Ivanov:2018jmz} (and references therein). As attractive as it may seem, the ability of the latter 
systematic methods to treat in principle any model through ready-to-use packages \cite{BFBpackage}, can yet in practice run into technical difficulties when dealing with extended scalar sectors as noted in \cite{Ivanov:2018jmz}. Also to the best of our knowledge a model with one triplet has
been treated using copositivity \cite{Chakrabortty:2013mha} but for which only specific
directions in field space where considered in agreement with \cite{Arhrib:2011uy}, while \cite{Babu:2016gpg} obtained with this method
the all-directions conditons for the Type-II seesaw model in a form different from that of \cite{Arhrib:2011uy}. It should however be stressed that the copositivity method cannot always be applied to potentials with extended Higgs multiplets when all $4$-dimensional 
operators allowed by the gauge symmetries and renormalizability are considered.
This was the case for the doublet extensions studied in \cite{Maniatis:2006fs,Maniatis:2014oza,Maniatis:2015gma}, and will
be the case here for the extensions with two Higgs triplets.\footnote{More precisely, the potential cannot always be cast in a bilinear form involving positive-definite independent vector components {\sl and} 
an optimal space dimension to make the method advantageous; more on this at the end of Section \ref{sec:disc}. 
Models with increased symmetries can be more tractable, see e.g. \cite{Deshpande:1977rw}, 
\cite{Chauhan:2019fji}.} Thus, the more pedestrian and somewhat mathematically lowbrow approach we adopt in this paper remains in our opinion an efficient way of tackling the stability problem specifically for the
 models under consideration with two triplets.

The paper is organized as follows. In Section \ref{sec:typeII} we revisit the derivation of the NAS-BFB conditions for
the Type-II seesaw model finding equivalence with the conditions of \cite{Bonilla:2015eha} that corrected \cite{Arhrib:2011uy}, but 
stress that the conditions of \cite{Arhrib:2011uy} do remain valid {\sl necessary and sufficient} when one of the couplings is negative. 
Adding one real $SU(2)$
triplet, the approach is extended to the general pre-custodial model in Section \ref{sec:2triplets}, including the Georgi-Machacek model as 
a special case. This section contains the bulk of the new results. We recall some useful ingredients of the two models potentials 
in Sections \ref{sec:p-c} and \ref{sec:GM}.
 In Section \ref{sec:p-c-BFB}  we first identify  
six field dependent variables that provide a reduced parameterization of the field space suitable 
for the BFB study, four of which, dubbed {\sl $\alpha$-parameters}, vary in compact domains. We then investigate the NAS-BFB conditions 
following a procedure where the resolving with respect to these six field-dependent variables is performed step-by-step. 
Section \ref{sec:glob-corr} deals with the analytical determination of the 
domains of variation of the {\sl $\alpha$-parameters} as well as all 2,3,4-dimensional analytical correlations between them. In Section \ref{sec:resolved-p-c-BFB} we derive the main results identifying the fully and partially resolved branches of the NAS-BFB conditions.
The special case of the Georgi-Machacek model is reconsidered in Section \ref{sec:GM-BFB} where we relate the reduced parameters to those of the pre-custodial model and provide a proof of their domain of variation that was only conjectured in the literature. Section \ref{sec:peeling} illustrates an unexpected feedback of the Georgi-Machacek model on the pre-custodial one. A wrap-up with further illustrations, comments and a user's guide, is given in Section \ref{sec:disc} and we conclude in Section \ref{sec:conclusion}.
Further material and detailed proofs, either missing in the literature for known properties, or for the new results found in this paper are 
given in appendices \ref{appendix:A} -- \ref{subsec:resolved}. Special attention is payed, in appendices \ref{appendix:generalquartic} and \ref{appendix:generalquarticplus}, to the mathematical issue of deriving simple forms for the NAS positivity conditions of quartic polynomials .



\section{The Type-II seesaw doublet-triplet Higgs potential
\label{sec:typeII}}
We first sketch the main ingredients, relying on the detailed analysis and notations
of \cite{Arhrib:2011uy} to which the reader may refer for more
details.

The potential reads
\begin{eqnarray}
V(H, \Delta) &=& -m_{{}_H}^2{H^\dagger{H}}+\frac{\lambda}{4}(H^\dagger{H})^2+M_{{}_\Delta}^2Tr(\Delta{\Delta}^{\dagger})
+[\mu(H^T{i}\sigma^2\Delta^{\dagger}H)+{\rm h.c.}] \nonumber\\
&&+\lambda_1(H^\dagger{H})Tr(\Delta{\Delta}^{\dagger})+\lambda_2(Tr\Delta{\Delta}^{\dagger})^2
+\lambda_3Tr(\Delta{\Delta}^{\dagger})^2 +\lambda_4{H^\dagger\Delta\Delta^{\dagger}H} \ . \nonumber \\
\label{eq:Vpot}
\end{eqnarray}
$H$ denotes the standard scalar field $SU(2)_L$ doublet
and $\Delta$ a colorless  $SU(2)_L$ complex triplet 
scalar field,
with charge assignments  
$H \sim (1, 2, 1)$ and  $\Delta \sim (1, 3, 2)$ under $SU(3)_c \times SU(2)_L \times U(1)_Y$,    
\begin{eqnarray}
H=\left(\begin{array}{c}
                      \phi^+ \\
                      \phi^0 \\
                    \end{array}
                  \right),~~~~~~~~\Delta &=\left(
\begin{array}{cc}
\delta^+/\sqrt{2} & \delta^{++} \\
\delta^0 & -\delta^+/\sqrt{2}\\
\end{array}
\right)  . \label{eq:HDelta}
\end{eqnarray}
We have used the $2 \times 2$ traceless matrix representation for 
the triplet and wrote the two multiplets in terms of
their complex valued scalar components and indicated a choice
of electric charges
with the conventional electric charge assignment for the doublet
and following
 $Q= I_3 + \frac{Y_{\Delta}}{2}$ with $I_3= -1,0, 1$ and $Y_{{}_\Delta}=2$ 
for the triplet. $\sigma^2$ denotes the second Pauli matrix. The 
potential $V(H, \Delta)$ is invariant
under $SU(2)_L \times U(1)_Y$ field transformations
$H \to e^{i \alpha} {\cal U}_L H$  and $\Delta \to e^{i 2 \alpha}
{\cal U}_L \Delta {\cal U}_L^\dag$ where  ${\cal U}_L$ denotes 
an arbitrary element of $SU(2)_L$
in the fundamental representation. Since we are only interested
in the issue of boundedness from below of the potential, we 
need not go further here into the details of the dynamics
of spontaneous electroweak symmetry breaking, the structure of the physical Higgs states and the generation of Majorana neutrino masses. 

\subsection{The BFB conditions \label{subsec:BFB}}

In order to cope generically with the shape of $V(H, \Delta)$ along all possible directions of the $10$-dimensional field space, we adopt a reduced parameterization for the fields that will 
turn out to be
particularly convenient to entirely solve the problem analytically. Following
\cite{Arhrib:2011uy} we define:
\begin{eqnarray}
 r &\equiv& \sqrt{H^\dagger{H} + Tr \Delta{\Delta}^{\dagger}} , \label{eq:paramr0} \\
 H^\dagger{H} &\equiv& r^2 \cos^2 \gamma ,  \\
 Tr \Delta{\Delta}^{\dagger} &\equiv& r^2 \sin^2 \gamma ,  \\
Tr (\Delta{\Delta}^{\dagger})^2/(Tr \Delta{\Delta}^{\dagger})^2 &\equiv& \zeta \label{eq:paramzeta} , \\
 (H^\dagger\Delta\Delta^{\dagger}H)/(H^\dagger{H} Tr\Delta{\Delta}^{\dagger} ) &\equiv& \xi \, . \label{eq:paramxi}
\end{eqnarray}
\noindent
Obviously, when $H$ and $\Delta$ scan all the field space, the radius $r$ scans the domain $[0, +\infty)$ and
the angle $\gamma \in [0, \frac{\pi}{2}]$. 
\noindent
With this parameterization it is straightforward to cast the quartic part of the potential, denoted hereafter by $V^{(4)}$ and given by 
the second line of Eq.~(\ref{eq:Vpot}), in the following simple form,
\begin{equation}
V^{(4)}(r, \tan \gamma, \xi, \zeta)= \frac{r^4}{ 4 (1+ \tan^2 \gamma)^{2}} ( \lambda + 
4 (\lambda_1 + \xi \lambda_4) \tan^2 \gamma  + 4 (\lambda_2 + \zeta \lambda_3) \tan^4 \gamma)
\label{eq:V4general}
\end{equation}
We stress here that the crux of the matter is the existence
of a parameterization, Eqs~(\ref{eq:paramr0} -\ref{eq:paramxi}),
which allows to scan {\sl all} the field space and in the same
time recasting the relevant part of the potential into a 
{\sl biquadratic form} in $\tan \gamma$.
It is the concomitance of these two facts that allows a tractable 
and complete analytical solution for the {\sl necessary and sufficient} boundedness from below conditions.
Indeed, the absence of linear and/or cubic powers of $\tan \gamma$
in Eq.~(\ref{eq:V4general}) is anything but generic.
(For instance, in a similar parameterization initially proposed
in \cite{ElKaffas:2006nt} to study two-Higgs-doublet models 
such terms do remain, hindering an easy fully analytical treatment.)

One can thus  consider only the range  $0 \le \tan \gamma <  +
\infty$ in accordance with the above stated range for $\gamma$. 
Boundedness from below is then equivalent to requiring $V^{(4)} > 0$
 {\sl for all} $\tan \gamma \in [0, +\infty)$ 
and {\sl all} $\xi, \zeta$ in their allowed domain. 
The $\gamma$-free necessary and sufficient conditions on the $\lambda_i$'s have already been 
given in \cite{Arhrib:2011uy}\footnote{\label{foot:Bool} We use the conventional notations
$\land, \lor$ and $\neg$
for the Boolean operators 'AND', 'OR' and 'NOT', respectively.}:
\begin{equation}
\lambda > 0 \;\;\land\;\; \lambda_2+ \zeta \lambda_3 \geq 0 \;\;
\land \;\;\lambda_1+ \xi \lambda_4 + \sqrt{\lambda(\lambda_2+\zeta \lambda_3)} > 0 \ .
\label{eq:NAS-BFB}
\end{equation}
Note that the second inequality above is non-strict. This accounts rigorously for the only possible
equality among the NAS conditions that is compatible with requiring
$V^{(4)}$ to be strictly positive.\footnote{In Section \ref{sec:p-c-BFB} we will 
elaborate further on the meaning of the condition $V^{(4)} > 0$, as well as on the fact
that the parameter $\tan \gamma$ varies independently of $\zeta$ and $\xi$.}
These inequalities are a subset of the general necessary and sufficient (NAS) positivity
conditions for a quartic polynomial (see Appendix 
\ref{appendix:generalquartic}). We stress here that 
Eq.~(\ref{eq:NAS-BFB}) answers fully the question of (tree-level)
boundedness from below {\sl in the totality
of the $\it 10$-dimensional field space}. There remains however the
dependence on $\xi$ and $\zeta$ that parameterize the relative
magnitudes of the dimension four gauge invariant operators
in Eq.~(\ref{eq:Vpot}) 
that are not controlled solely by $r$ and $\gamma$.    
 
One can, however, show that 
\begin{eqnarray}
 0 \le \xi \le 1  & {\rm and} &  \frac{1}{2} \le \zeta \le 1 . 
\label{eq:xizetaranges}
\end{eqnarray}
(See Appendices \ref{appendix:xi}, \ref{appendix:zeta} for a proof.)

In \cite{Arhrib:2011uy} the authors relied on this allowed 
range and on the
monotonic dependence on $(\xi, \zeta)$ in Eq.(\ref{eq:NAS-BFB})
to obtain equations  (4.21),(4.22) and (4.23) of \cite{Arhrib:2011uy} reproduced in Appendix \ref{appendix:almostNAS} 
for later discussions. 
The authors of \cite{Bonilla:2015eha} rightly observed that \cite{Arhrib:2011uy} had actually overlooked the
fact that $(\xi, \zeta)$ being correlated, cannot reach an arbitrary point
in the rectangle defined by Eq.(\ref{eq:xizetaranges}). Starting 
from Eq.~(\ref{eq:NAS-BFB}) and using the constraint
\begin{equation}
2 \xi^2 - 2 \xi + 1 \leq \zeta \leq 1, 
\label{eq:xizetaenv}
\end{equation}
 they showed that the set of conditions Eqs.~(\ref{eq:BFBgen1} - \ref{eq:BFBgen3}) established in \cite{Arhrib:2011uy},
  although sufficient in all field space directions, are in fact not
necessary, even though deviation from absolute necessity is typically at the few percent level. Although we totally agree
with their general observation, we will see that despite the correlation between  $\xi$ and $\zeta$ the conditions
Eqs.~(\ref{eq:BFBgen1} - \ref{eq:BFBgen3}) {\sl do remain sufficient and necessary whenever $\lambda_3 < 0$}; the modification
will come only for $\lambda_3 > 0$. We will come back to this point in more detail later on in Appendix \ref{appendix:B}.

For now, we just add that, as shown in Appendix \ref{appendix:xizetacorr}, it is possible to cast the $\xi$ and $\zeta$ parameters as follows 
\begin{eqnarray}
\xi &=& \frac12 ( 1 + c_{{}_{2H}}  \, c_{{}_{2\Delta}}),
\label{eq:xizetacorr1}\\ 
\zeta &=& \frac12 (1 + c^2_{{}_{2\Delta}}),
\label{eq:xizetacorr2}
\end{eqnarray}
with $c_{{}_{2H}}, c_{{}_{2 \Delta}}$ two {\sl independent} cosines taking any
value in their allowed domain $[-1,1]$; note also that
Eq.~(\ref{eq:xizetaenv}) comes as a direct consequence of
these equations.

Altough the authors of \cite{Bonilla:2015eha} wrote a correct form of the necessary and sufficient
BFB conditions, they only sketched a proof of their result. In Appendix \ref{appendix:B}, we provide a detailed proof through a careful study of Eq.~(\ref{eq:NAS-BFB}) leading to an alternative form of the fully resolved NAS BFB conditions. The latter reduce to:
\begin{equation}
{\mathcal B}_0 \;\; \land  \;\; \Big\{  {\mathcal B}_1 \;\; \lor \;\; {\mathcal B}_2 \Big\} \label{eq:BFB}
\end{equation}
where
\begin{eqnarray}
&&\displaystyle {\mathcal B}_0 \Leftrightarrow \Big\{ \lambda > 0 \;\; \land \;\; \lambda_2+\lambda_3 \geq 0  \;\;\land \;\; \lambda_2+\frac{\lambda_3}{2} \geq 0 
\Big \}\ , \label{eq:BFBnew1} \\
&&{\mathcal B}_1 \Leftrightarrow 
 \displaystyle  
\Big\{  \lambda_1+ \sqrt{\lambda(\lambda_2+\lambda_3)} >0 \;\; \land \;\; \lambda_1+ \lambda_4 +\sqrt{\lambda(\lambda_2+\lambda_3)} >0 \;\; \land \;\; \sqrt{\lambda} \lambda_3 \leq \sqrt{(\lambda_2 + \lambda_3) \lambda_4^2} \Big\} \ , \nonumber \\ \label{eq:BFBnew3prime} \\
&& \text{and} \nonumber  \\
&& {\mathcal B}_2 \Leftrightarrow \Big\{\sqrt{\lambda} \lambda_3 \geq \sqrt{(\lambda_2 + \lambda_3) \lambda_4^2} \;\; \land \;\;
 \displaystyle  \lambda_1 +
\frac{\lambda_4}{2} + \sqrt{\lambda (\lambda_2 + \frac{\lambda_3}{2}) \big(1 - \frac{\lambda_4^2}{2 \lambda \lambda_3} \big)} > 0   
   \Big\} \label{eq:BFBnew5prime}  \ .
 \end{eqnarray}
 Note also that Eq.~(\ref{eq:BFBnew5prime})
implies $\lambda_3 >0$ and $2 \lambda \lambda_3 - \lambda^2_4 > 0$ 
so that the ${\mathcal B}_2$ part is relevant only when these conditions are satisfied simultaneously. 

The above constraints are in fact totally equivalent to \cite{Bonilla:2015eha} although they have a slightly different form.
Indeed the equivalence is not straightforward as the two involved Boolean forms are in general not equivalent to each other.
However, they become equivalent due to the implication given by Eq.~(\ref{eq:implication}). 
The above constraints:
\begin{itemize}
\item constitute an independent check of the results of \cite{Bonilla:2015eha}.
\item are written explicitly as a union of domains one of which, ${\mathcal B}_1$,  
is a necessary consequence of constraints Eqs.~(\ref{eq:BFBgen2} - \ref{eq:BFBgen3}).
\item allow to understand why in some regimes the previous constraints  Eqs.~(\ref{eq:BFBgen2} - \ref{eq:BFBgen3})
would exclude only a very small part of the allowed parameter space. This is the case in particular in the regimes where
$\lambda_4 \ll 1$ or $\lambda^2_4 \ll 2 \lambda \lambda_3$.
\item allow to see analytically that our previous constraints Eqs.~(\ref{eq:BFBgen2} - \ref{eq:BFBgen3})
were sufficient but not necessary. Indeed Eq.~(\ref{eq:BFBgen1}) is the same as Eq.~(\ref{eq:BFBnew1}) while one can easily check that 
Eqs.~(\ref{eq:BFBgen2} - \ref{eq:BFBgen3}) always imply 
Eq.~(\ref{eq:BFBnew3prime}). 
\end{itemize}

\section{Generalization adding one extra real triplet
\label{sec:2triplets}}

Such a generalization can be of phenomenological interest by itself, but is also motivated by the structure of the
Georgi-Machacek model beyond the tree-level \cite{Gunion:1990dt}.

\subsection{The pre-custodial potential \label{sec:p-c}}
Defining
\begin{eqnarray}
A=\left(\begin{array}{cc}
                       {a^+}{/\sqrt{2}} & -a^{++}  \\
                       a^0 & -{a^+}{/\sqrt{2}} \\
                    \end{array}
                  \right),~~~~~~~~B &=
                  \displaystyle \left(
\begin{array}{cc}
{b^0}{/\sqrt{2}} & -b^{+} \\
-b^{+ *} & -{b^0}{/\sqrt{2}}\\
\end{array}
\right)  \ , \label{eq:AB}
\end{eqnarray}
with $A$ a different notation for the complex triplet $\Delta$, and $H$ as defined in Eq.~(\ref{eq:HDelta}), $B=B^\dag$ a real triplet ($b^0$ real-valued), 
we write the most general renormalizable pre-custodial potential involving $H,A$ and $B$ 
as follows,  
\begin{equation}
V_{\text{p-c}} = V_{\text{p-c}}^{(2,3)} + V_{\text{p-c}}^{(4)}
\label{eq:V-p-c}
\end{equation}
where the dimension-$2$, -$3$ operators are collected in 
\begin{eqnarray}
V_{\text{p-c}}^{(2,3)}&=& - \, m_{{}_H}^2{H^\dagger{H}}+
M_{{}_A}^2Tr(A{A}^{\dagger}) + M_{{}_B}^2Tr(B^2) \nonumber \\
&& +\, [\mu_A(H^T{i}\sigma^2 A^{\dagger}H)+{\rm h.c.}] + 
\mu_B H^\dagger B H + \mu_{AB} Tr(A A^{\dagger} B), 
 \label{eq:p-c-2-3}
\end{eqnarray}
and the dimension-$4$ operators in
\begin{eqnarray}
V_{\text{p-c}}^{(4)}&=&\frac{\lambda_H}{4} (H^\dagger H)^2 + \frac{\lambda_A^{(1)}}{4} 
(Tr \,A  A^\dagger)^2 + \frac{\lambda_A^{(2)}}{4} Tr (A  A^\dagger)^2 + 
 \frac{\lambda_B}{4!} ( Tr B^2)^2  \nonumber \\
&& + \, \lambda_{AH}^{(1)} H^\dagger H \; Tr \,A  A^\dagger
   +  \lambda_{AH}^{(2)} H^\dagger A A^\dagger H  
   +  \frac{\lambda_{BH}}{2} H^\dagger H \; Tr B^2 \nonumber \\
&&   + \, \frac{\lambda_{AB}^{(1)}}{2}  Tr \,A  A^\dagger \; Tr B^2
   +    \frac{\lambda_{AB}^{(2)}}{2}  Tr \,A B \; Tr \,A^\dagger B
   \nonumber \\
&&   +  \frac{i}{2} \, \lambda_{ABH} (H^\top \sigma^2 A^\dagger B H -
    H^\dagger  B A \, \sigma^2 H^*) . \label{eq:p-c-4}
\end{eqnarray}
$V_{\text{p-c}}$  is invariant
under $SU(2)_L \times U(1)_Y$ field transformations
\begin{eqnarray}
H &\to& e^{i \alpha} {\cal U}_L H , \nonumber \\
A &\to& e^{i 2 \alpha}{\cal U}_L A {\cal U}_L^\dag , \label{eq:p-c-transfos}\\ 
B &\to& {\cal U}_L B {\cal U}_L^\dag, \nonumber
\end{eqnarray}  
where  ${\cal U}_L$ denotes an arbitrary element of $SU(2)_L$
in the fundamental representation. This potential was written
in \cite{Gunion:1990dt} and later on in \cite{Blasi:2017xmc} with which we agree up to different normalizations
and notations\footnote{\label{foot:2} with the field correspondence as given by Eq.~(\ref{eq:PhiX}) and couplings correspondence: $\lambda_H=4 \lambda$, $\lambda_A^{(i=1,2)}=16 \rho_{i}$, $\lambda_B =4!\times 4 \rho_3$, 
 $\lambda_{AB}^{(i=1,2)}=8 \rho_{i+3}$, $\lambda_{AH}^{(i=1,2)}=2 \sigma_{i}$, $\lambda_{BH} = 4 \sigma_3$ and $\lambda_{ABH}= 4 \sigma_4$. 
 Note that our normalization factors for the various couplings are chosen such that they cancel out for at least one vertex
 originating from each operator when symmetry factors are taken into account in the Feynman rules.}. 
All other dimension-$3$,-$4$ gauge invariant operators are either vanishing or can be expressed
in terms of the ones listed above.
(For completeness we give a proof of this in Appendix~\ref{appendix:pre-c}.)


\subsection{The Georgi-Machacek potential \label{sec:GM}} 

This model \cite{Georgi:1985nv, Chanowitz:1985ug}, a special 
setup  of the model presented in the previous subsection, allows
to extend the validity of the  SM tree-level (approximate) 
custodial symmetry in the presence of $SU(2)_L$ triplet scalar 
fields. In particular the potential reads
\begin{equation}
V_{\text{G-M}}= V_{\text{G-M}}^{(2,3)} + V_{\text{G-M}}^{(4)},
\end{equation}
\begin{eqnarray}
V_{\text{G-M}}^{(2,3)}&=&\frac{\mu_2^2}{2} Tr \Phi^\dagger \Phi +\frac{\mu_3^2}{2} Tr X^\dagger X -\left(M_1 Tr\Phi^\dagger \tau^a \Phi \tau^b + M_2 TrX^\dagger t^a X t^b \right)(U X U^\dagger)_{a b} ,  \label{eq:G-M-2-3}\\
V_{\text{G-M}}^{(4)}&=&\hat\lambda_1 (Tr\Phi^\dagger \Phi)^2 + 
\hat\lambda_2 Tr (\Phi^\dagger \Phi) Tr(X^\dagger X) 
 +\hat\lambda_3 Tr(X^\dagger X X^\dagger X) \nonumber \\
&& + \hat\lambda_4 (Tr X^\dagger X)^2 -
  \hat\lambda_5 Tr(\Phi^\dagger \tau^a \Phi \tau^b)  
  Tr(X^\dagger t^a X t^b), \label{eq:G-M-4}
\end{eqnarray}
where we followed the notations of \cite{Hartling:2014zca}.\footnote{In Eqs.~(\ref{eq:G-M-2-3}, \ref{eq:G-M-4}) $\tau^a=\sigma^a/2$ with $\sigma^a$ the Pauli matrices are the usual $SU(2)$ generators in the fundamental representation, $t^a$ the generators in the triplet (adjoint) representation, with $a=1,2,3$, and $U$ some rotation matrix about which we skip here the details 
(see \cite{Aoki:2007ah} and \cite{Hartling:2014zca}) as Eq.~(\ref{eq:G-M-2-3}) will
not be relevant to our study.} We hat the $\lambda$'s to distinguish them from those of Sec.~\ref{sec:typeII}, and define
the scalar bi-doublet and bi-triplet as
\begin{eqnarray}
\Phi \equiv \left(\begin{array}{cc}
                       \phi^{0 *} & \phi^+  \\
                       - \phi^{+ *} & \phi^0 \\
                    \end{array}
                  \right) =  
                  \left(\begin{array}{cc}
                       i \sigma^2 H^{*}, & H  \\
                       \end{array}
                  \right) , \\
X \equiv \left(
\begin{array}{ccc}
\chi^{0 *} & \xi^+ & \chi^{++}\\
-\chi^{+ *} & \xi^0& \chi^+\\
\chi^{++ *} & -\xi^{+ *}& \chi^0\\
\end{array}
\right)  = \sqrt{2} \left(
\begin{array}{ccc}
a^{0 *} & b^+ & a^{++}\\
-a^{+ *} & b^0& a^+\\
a^{++ *} & -b^{+ *}& a^0\\
\end{array}
\right ) , \label{eq:PhiX}
\end{eqnarray}
so that the normalization of the VEVs 
are the same as in \cite{Hartling:2014zca}.  
Note also the sign difference
in $a^{++}$ and $b^+$ between  Eq.~(\ref{eq:AB}) and Eq.~(\ref{eq:PhiX}). The potential $V_{\text{p-c}}$ is then mapped onto 
$V_{\text{G-M}}$ through the following correspondence
among the couplings 
\begin{eqnarray}
&& \hat\lambda_1 = \frac{1}{16} \lambda_H, \
\hat\lambda_2 = \frac{1}{8} \lambda_{BH}, 
 \ 
\hat\lambda_3 = -\frac{1}{64} \lambda_{A}^{(2)}, \nonumber \\ 
&& \hat\lambda_4 = \frac{1}{32} \lambda_{AB}^{(1)}, \
\hat\lambda_5 = -\frac{1}{4 \sqrt{2}} \lambda_{ABH},
\label{eq:correspond}
\end{eqnarray}
provided, however, the following correlations hold for the pre-custodial 
potential couplings: 
\begin{equation}
\begin{aligned} 
& \lambda_A^{(1)} = 2 \lambda_{AB}^{(1)} +  3 \lambda_{AB}^{(2)}, \  \lambda_{A}^{(2)} = - 2 \lambda_{AB}^{(2)}, \
\lambda_{ABH} =  \sqrt{2} \lambda_{AH}^{(2)},  \\
& \lambda_B = 3 (\lambda_{AB}^{(1)} +   \lambda_{AB}^{(2)}), \ \lambda_{BH} =  \lambda_{AH}^{(1)} + \frac12 \lambda_{AH}^{(2)} \ .
\label{eq:correlate}
\end{aligned}
 \end{equation}
 The potential $V_{\text{G-M}}$ enjoys an increased symmetry as compared to that of $V_{\text{p-c}}$, Eq.~(\ref{eq:p-c-transfos}),
 with an invariance under an extra global $SU(2)$,
 \begin{eqnarray}
 \Phi \to {\cal U}^{(2)}_L \Phi {\cal U}^{(2)}_R, \\
 X \to {\cal U}^{(3)}_L X {\cal U}^{(3)}_R,
 \end{eqnarray}
 where ${\cal U}^{(n)}_{L,R}$ denotes $n$-dimensional representation
 of $SU(2)_{L,R}$. 
The correlations given by Eq.~(\ref{eq:correlate}) can thus be viewed as encoding the tree-level constraints imposed by the $SU(2)_R$ global symmetry on the potential. We come back to this point in 
Sec.~\ref{sec:disc} when discussing briefly quantum effects. 
References \cite{Gunion:1990dt}, \cite{Blasi:2017xmc} considered such correlations.\footnote{We agree with \cite{Blasi:2017xmc}
except for a factor two difference on the right-hand side of the first equation of the second line of Eq.~(\ref{eq:correlate}) as
compared to the first equation of the second line of Eq.~(10) of \cite{Blasi:2017xmc}.}

 \subsection{The pre-custodial BFB conditions \label{sec:p-c-BFB}}
Being a polynomial in the fields, the tree-level potential has no singularities at finite values of the fields; it follows that 
boundedness from below means that the potential does not become infinitely negative at infinitely large field values.
This is equivalent to requiring {\sl strict} positivity of the quartic part of the
potential, Eq.~(\ref{eq:p-c-4}), for {\sl all field values in all field directions}. The latter requirement 
is {\sl sufficient} as it implies that at infinitely large field values, where 
$|V_{\text{p-c}}^{(2,3)}| \ll |V_{\text{p-c}}^{(4)}|$ in 
Eq.~(\ref{eq:V-p-c}), the potential does not become infinitely
negative. That it is also {\sl necessary} might not seem obvious since the last term in Eq.~(\ref{eq:p-c-4}) is linear in $A$ and in $B$, 
so that $V_{\text{p-c}}^{(4)}$ might be negative for some finite values of the fields without being unbounded from below. 
That this does not happen, and the above requirement is indeed necessary, can be easily seen as follows: If there existed a point in field 
space where $V_{\text{p-c}}^{(4)} \equiv V_{\text{p-c}}^{(4) \star} \leq 0$, then scaling
all the fields at that point by the same real-valued amount $s$ would have lead to $V_{\text{p-c}}^{(4)} \equiv s^4 V_{\text{p-c}}^{(4) \star} \leq 0$, implying unboundedness from below since 
$s$ can be chosen infinitely large. 
Note finally that {\sl strict} positivity is important here because a vanishing $V_{\text{p-c}}^{(4)}$ at very large field values
would generically lead to the dominance of $V_{\text{p-c}}^{(3)}$ which, barring accidental cancellations in some field directions, 
always possesses unbounded from below directions!

The BFB conditions are thus the necessary and
sufficient conditions on the nine couplings $\lambda$ of Eq.~(\ref{eq:p-c-4}) 
that ensure
\begin{eqnarray}
&&~~~~ V_{\text{p-c}}^{(4)} > 0, \, \forall  A, B, H. \label{eq:BFB-p-c}
\end{eqnarray}

Of course, loop corrections  
will modify the conditions on the couplings resulting from Eq.~(\ref{eq:BFB-p-c}),
although the effects can be partly encoded in the runnings of the couplings through a
renormalization group improvement of the potential. (We will come back briefly to this point in Section \ref{sec:disc}.)
Note also that the above definition of boundedness from below does not take into account the actual pattern of spontaneous 
symmetry breaking that would typically lead to more stringent constraints.

The condition in Eq.~(\ref{eq:BFB-p-c}) should be verified in
the full $13$-dimensional 
space of the real-valued field components of the $A,B$ and $H$ multiplets.
However, symmetries of the model (and possibly accidental symmetries
akin to $V_{\text{p-c}}^{(4)}$) will help reduce the number of
relevant degrees of freedom. 
Starting from Eq.~(\ref{eq:p-c-4}) we generalize the parameterization of Eqs.~(\ref{eq:paramr0} - \ref{eq:paramxi})
using spherical-like coordinates as follows:
\begin{eqnarray}
        H^\dagger H &\equiv& r^2 \cos^2 a, \label{eq:paramH} \\
     Tr A  A^\dagger &\equiv & r^2 \sin^2 a \cos^2 b, 
     \label{eq:paramA} \\
     Tr(B^2) &\equiv & r^2 \sin^2 a \sin^2 b, \label{eq:paramB} \\
          r^2 &=& H^\dagger H + Tr A  A^\dagger + Tr(B^2), 
          \label{eq:paramr}
\end{eqnarray}      
        where $r$ is a non negative number, and $a \in [-\pi/2, +\pi/2]$ and $b \in [-\pi, +\pi]$ two angles. It will also prove
        useful to define
        the following real-valued quantities, 
\begin{eqnarray}
&\displaystyle T \equiv \sqrt{\frac{Tr A  A^\dagger}{H^\dagger H}}=|\tan a \cos b|,  \; \; t \equiv \sqrt{\frac{Tr (B^2)}{Tr A  A^\dagger}}=|\tan b|, &  \label{eq:T} \\
&\displaystyle \alpha_A \equiv \frac{Tr A  A^\dagger A A^\dagger }{(Tr A  A^\dagger)^2}, \; \;  
\alpha_{AH} \equiv  \frac{H^\dagger A A^\dagger H}{H^\dagger H \, Tr A  A^\dagger}, \; \;
\alpha_{AB} \equiv \frac{Tr A B \, Tr A^\dagger B}{Tr A  A^\dagger \, Tr (B^2)}, & \label{eq:def-alphas1}\\
&\displaystyle \alpha_{ABH} \equiv i \frac{H^\top \sigma^2 A^\dagger B H-H^\dagger B A \sigma^2 H^*}{ H^\dagger H \sqrt{Tr A  A^\dagger \, Tr(B^2)} }. & 
\label{eq:def-alphas2}
\end{eqnarray}    
Hereafter we will refer to the latter four parameters as the {\sl $\alpha$-parameters}.
In terms of $T, t$ and the {\sl $\alpha$-parameters}, the quartic part of the potential now reads    
\begin{equation}
V_{\text{p-c}}^{(4)} = r^4 \cos^4 \! a \times (\mathfrak{a_0} + \mathfrak{b_0} T^2 + \mathfrak{c_0} T^4), \label{eq:p-c-4prime}
\end{equation}
where
\begin{equation}
\begin{aligned}
& \mathfrak{a_0} = \frac{\lambda_H}{4} , \; \mathfrak{b_0} = \lambda_{AH}^{(1)} + 
  \alpha_{AH} \lambda_{AH}^{(2)} + \frac12(\alpha_{ABH} \lambda_{ABH} t+ \lambda_{BH}  t^2) , \\
& \mathfrak{c_0} = \frac14 (\lambda_A^{(1)} + \alpha_A \lambda_A^{(2)}) + \frac12 (\lambda_{AB}^{(1)} + \alpha_{AB} \lambda_{AB}^{(2)})  t^2 + \frac{1}{24} \lambda_B t^4 \label{eq:c0} .
\end{aligned}
\end{equation}
 It becomes evident from Eqs.~(\ref{eq:p-c-4prime}--\ref{eq:c0}) that the positivity of $V_{\text{p-c}}^{(4)}$ does not depend 
 explicitly on all ten terms of the right-hand side of Eq.~(\ref{eq:p-c-4}),
 but just on the reduced set of the six combinations of gauge invariant operators defined in Eqs.~(\ref{eq:T} -- \ref{eq:def-alphas2}). 
 The sought-after NAS BFB conditions on the $\lambda$'s are thus 
 those that ensure  
\begin{eqnarray}
&&~~~~~~~~~~~~~~\mathfrak{a_0} + \mathfrak{b_0} T^2 + \mathfrak{c_0} T^4 > 0, \, \forall  T, t, \alpha_A, \alpha_{AH}, \alpha_{AB}, \alpha_{ABH}. \label{eq:NAS-BFB1}
\end{eqnarray}
It is important to note that scanning independently over all values of the thirteen real-valued components of the fields $A,B$ and $H$ 
amounts to varying $T, t$ and the {\sl $\alpha$-parameters}.
The latter, however, do not all vary independently. 
For one thing, the {\sl $\alpha$-parameters} 
vary in bounded domains: $ \alpha_A$ and $\alpha_{AH}$ are nothing but 
respectively $\zeta$ and $\xi$ defined in 
Eqs.~(\ref{eq:paramzeta}, \ref{eq:paramxi}). Hence
\begin{eqnarray}
 \alpha_A &\in& [\displaystyle \frac12,1], \label{eq:alfaArange} \\
 \alpha_{AH} &\in& [0,1],
 \end{eqnarray} 
as shown in 
appendix \ref{appendix:A}. Furthermore, one can show that
\begin{eqnarray}
\alpha_{AB} &\in&[0,1], \label{eq:alfaABrange}\\
\alpha_{ABH}&\in& [-\sqrt 2 , +\sqrt 2] ,\label{eq:alfaHABrange}
\end{eqnarray}
see Appendix \ref{appendix:alphas} for details. 
For another, the {\sl $\alpha$-parameters} are uncorrelated only {\sl locally}. But similarly to what was pointed out in \cite{Bonilla:2015eha} and discussed at length in sec.~\ref{subsec:BFB} for
the Type-II seesaw model potential, they are correlated {\sl globally} in that they cannot reach the boundaries of their respective domains 
independently of each other. The actual domain in the $4$-dimensional {\sl $\alpha$-parameters} space is certainly not the simple hyper-cube defined by Eqs.~(\ref{eq:alfaArange} --\ref{eq:alfaHABrange}). One can approach the true domain by considering the projected domains on the sub-spaces of these parameters taken two-by-two. This is not trivial to establish and will be carried out in full details in Sec.~\ref{sec:glob-corr}. 
The more difficult task of determining fully the true domain will be
discussed in Section \ref{subsec:potatoid}.
 
 In contrast, the variables $T$ and $t$ vary in $\in [0, +\infty)$ 
 {\sl independently} of each other and of the {\sl $\alpha$-parameters}.
 In essence, the {\sl $\alpha$-parameters} being ratios of 
 different gauge invariant combinations of the fields 
 can be seen as functions of cosines and sines of angles defined 
 separately in the $A$, $B$ and $H$ field spaces, where 
 $\sqrt{Tr A  A^\dagger}, \sqrt{Tr B^2}$ and $\sqrt{H^\dagger H}$ represent lengths. This hints at the obstruction to span the full
 hyper-cube as noted above. Whereas $T$ and $t$, being two 
 ratios of these three lengths, are clearly independent of each other and of the {\sl $\alpha$-parameters}. It follows that
 $T$ can be varied {\sl independently} from $\mathfrak{a_0}, \mathfrak{b_0}$ and $\mathfrak{c_0}$
 in Eq.~(\ref{eq:NAS-BFB1})  Consequently, the NAS conditions for the strict positivity  of $V_{p-c}^{(4)}$, $\forall T$, are those 
 of a biquadratic polynomial in $T$, namely conditions on the $\lambda$'s satisfying
\begin{equation}
{\mathfrak{a_0} > 0 \ \land \ \mathfrak{c_0} \geq 0 \ \land \ \mathfrak{b_0} + 2 \sqrt{\mathfrak{a_0 \mathfrak{c_0}}} > 0 },
\forall t, \alpha_A, \alpha_{AB}, \alpha_{AH}, \alpha_{ABH}. \label{eq:a0c0b0p2sqrta0c0}
\end{equation}
As noted previously, only the highest degree monomial coefficient can vanish. However, for the sake of simplicity we will
consider in the sequel only the strict inequality  $\mathfrak{c_0} > 0$.
It is convenient to  recast the above inequalities in the following 
equivalent form that disposes of the (less tractable) square root:
\begin{eqnarray}
&\mathfrak{a_0} > 0 \, \land \, \mathfrak{c_0} > 0 \  \land \ 
\{\mathfrak{b_0} > 0 \,\lor \, \{\mathfrak{b_0} < 0\, \land \, 4 \mathfrak{a_0 \mathfrak{c_0}} -\mathfrak{b_0^2} > 0 \} \}, & \\
&\forall t, \alpha_A, \alpha_{AB}, \alpha_{AH}, \alpha_{ABH}, &
\nonumber
\end{eqnarray}
which simplifies further to
\begin{eqnarray}
&\mathfrak{a_0} > 0 \, \land \, \mathfrak{c_0} > 0& \label{eq:p-c-BFB-1}\\
 &\land& \nonumber \\
&\{\mathfrak{b_0} > 0 \,\lor  \, 4 \mathfrak{a_0 c_0} -\mathfrak{b_0^2} > 0 \},& \label{eq:p-c-BFB-2} \\
&\forall t, \alpha_A, \alpha_{AB}, \alpha_{AH}, \alpha_{ABH}. &
\nonumber
\end{eqnarray}
\subsubsection{$\mathfrak{a_0} > 0 \, \land \, \mathfrak{c_0} > 0$: \label{sec:a0-AND-c0}} 
We consider first the conditions in Eqs.~(\ref{eq:p-c-BFB-1}) as they are common to the union of the two conditions
of Eqs.~(\ref{eq:p-c-BFB-2}). The coefficient $\mathfrak{c_0}$ being itself biquadratic in $t$ and the latter independent of the {\sl $\alpha$-parameters}, see Eq.~(\ref{eq:c0}), the corresponding 
NAS positivity condition is in turn of the same form as  Eqs.~(\ref{eq:p-c-BFB-1}, \ref{eq:p-c-BFB-2}). 
The two inequalities in Eq.~(\ref{eq:p-c-BFB-1}) are thus equivalent to:
\begin{eqnarray}
\lambda_H > 0 \; \land \; \lambda_B > 0 \! &\land& \! \lambda_A^{(1)} + \alpha_A \lambda_A^{(2)} > 0 \label{eq:p-c-BFB-3} 
\\
&\land&  \nonumber \\
\left\{ \lambda_{AB}^{(1)} + \alpha_{AB} \lambda_{AB}^{(2)} > 0
\ \text{(I)} \ \right.  \! &\lor& \! 
\left. \,  (\lambda_A^{(1)} + \alpha_A \lambda_A^{(2)}) \lambda_B - 
6 (\lambda_{AB}^{(1)} + \alpha_{AB} \lambda_{AB}^{(2)})^2 > 0  \ \text{(II)}  \right\} \label{eq:p-c-BFB-4} ,
\\
&&\forall  \alpha_A,  \alpha_{AB}. \nonumber
\end{eqnarray}
Note that the second inequality in Eq.~(\ref{eq:p-c-BFB-3}) and the first inequality in Eq.~(\ref{eq:p-c-BFB-4}) depend solely 
on $\alpha_A$ or on $\alpha_{AB}$. They can be easily resolved since the dependence 
on these parameters is monotonic; {\sl if} required to be valid
$\forall  \alpha_A,  \alpha_{AB}$ in the domains given by Eqs.~(\ref{eq:alfaArange}, \ref{eq:alfaABrange}), they become equivalent to requiring them simultaneously at the two edges of these domains, namely:
\begin{eqnarray}
\lambda_A^{(1)} + \frac{\lambda_A^{(2)}}{2} > 0 \; \land \; \lambda_A^{(1)} + \lambda_A^{(2)} > 0, \label{eq:p-c-BFB-3-resolved}
\end{eqnarray}
for the first, and 
\begin{eqnarray}
\lambda_{AB}^{(1)}  > 0 \; \land \; \lambda_{AB}^{(1)} +  \lambda_{AB}^{(2)} > 0, \label{eq:p-c-BFB-4-part}
\end{eqnarray}
for the second.
 Equation~(\ref{eq:p-c-BFB-4})-(II) needs more 
 care due to the nontrivial global correlation between 
 $\alpha_A$ and  $\alpha_{AB}$ (see next section and
 Fig.~\ref{fig:alfa-alfab}),
and will be kept in its present form for the time being. 
One will also have to tackle a further complication involving
the two inequalities of Eq.~(\ref{eq:p-c-BFB-4}). Indeed,
due to the `or' structure of Eq.~(\ref{eq:p-c-BFB-4}), none
of the two corresponding inequalities need to be necessarily
valid {\sl for all}  $\alpha_A,  \alpha_{AB}$ in their domains; 
it suffices that one of the two inequalities be satisfied in 
a given subset of $\alpha_A,  \alpha_{AB}$, and the other 
inequality satisfied in the complementary subset. In particular, Eq.~(\ref{eq:p-c-BFB-4-part}) is only sufficient. To reach
the NAS conditions one will have to consider all possible coverings of the domain by two 
subsets for which such a configuration holds. This issue
will be solved explicitly in Sec.~\ref{sec:resolv-a0-AND-c0}.

\subsubsection{$\mathfrak{b_0} > 0 \,\lor  \, 4 \mathfrak{a_0 c_0} -\mathfrak{b_0^2} > 0$: \label{sec:b0-AND-4a0c0-b02}}
We turn now to the two inequalities of Eq.~(\ref{eq:p-c-BFB-2}). The first is quadratic in $t$, see Eq.~(\ref{eq:c0}), 
but could in principle be treated
as a biquadratic polynomial in $\sqrt{t}$, since $t \in [0, +\infty)$. The second, $4 \mathfrak{a_0 c_0} -\mathfrak{b_0^2} > 0$, is a general quartic polynomial in this same variable. This is the first 
place where we encounter the issue of positivity conditions for a general quartic polynomial. Relying on a classic theorem about single variable polynomials that are positive on $(-\infty, +\infty)$, we derive in Appendix \ref{appendix:generalquartic} a relatively tractable form of the corresponding NAS conditions for 
 a quartic polynomial. However these conditions are not 
 directly applicable to the case at hand since the relevant variable here, $t$, is in $[0, +\infty)$. In this case the NAS conditions would obviously be less restrictive, see for instance
\cite{Powers:2000,Benoist:2017} for recent reviews.\footnote{Somewhat surprisingly, corresponding 
theorems, when the variable does not span the full $(-\infty, +\infty)$ interval, seem not to have been referenced in the mathematics
literature  before the 1970's, see \cite{Polya:1976}.}
Relying on these theorems we extend the results of Appendix \ref{appendix:generalquartic} to the domain $[0, +\infty)$ in
Appendix \ref{appendix:generalquarticplus}.
 
 However, this is not the full story.
Similarly to what we stated above in subsection \ref{sec:a0-AND-c0}  regarding Eq.~(\ref{eq:p-c-BFB-4}), the `or' structure of Eq.~(\ref{eq:p-c-BFB-2}) implies that it is
sufficient for  
the two inequalities $\mathfrak{b_0} > 0$ and   $4 \mathfrak{a_0 c_0} -\mathfrak{b_0^2} > 0$ 
to be separately valid in two complementary subsets  
of the allowed $t$ and {\sl $\alpha$-parameters} domains.
The NAS conditions will then be obtained by investigating
all possible coverings of these domains for which this happens. The upshot
is that the possibility of varying freely $t$ with respect to
the {\sl $\alpha$-parameters} is not sufficient anymore. Indeed, a
given subset of the {\sl $\alpha$-parameters} where for instance 
$\mathfrak{b_0} >0$ (or $4 \mathfrak{a_0 c_0} -\mathfrak{b_0^2} > 0$)
will be necessarily correlated with $t$.
A strategy for an explicit resolution will be given in Sec.\ref{sec:part-resolv}.
 
  
Although it will prove unavoidable to deal with positivity conditions of quartic polynomials on sub-domains of $(-\infty, +\infty)$, it will still be useful for the subsequent discussions
 to replace from the onset
$t \in [0, +\infty)$ by a variable on $(-\infty, +\infty)$
 if possible.
This is indeed the case if one considers the variable $Z$ defined
as  
\begin{equation}
Z= \alpha_{ABH} \times t \label{eq:defZ}
\end{equation}  
since $\alpha_{ABH}$ can take either signs, cf. Eq.~(\ref{eq:alfaHABrange}).  
However, 
in order to apply safely the NAS positivity conditions on a polynomial in $Z$, 
one should make sure that $Z$ is not correlated with the other parameters, $\alpha_A, \alpha_{AH}$ and $\alpha_{AB}$ appearing in 
the inequalities, even though these parameters are globally correlated with $\alpha_{ABH}$. 

\noindent
It is obviously the case for $|Z|$ since 
$t$ is uncorrelated with the other parameters and allows to scan independently of the value of $|\alpha_{ABH}|$ the full 
$[0, +\infty)$ range. 
However, the sign of $Z$ is controlled by $\alpha_{ABH}$ which {\sl is} globally correlated with $\alpha_A, \alpha_{AH}$ and 
$\alpha_{AB}$. It is thus crucial to check that the sign of $\alpha_{ABH}$ is not correlated with the latter parameters.
That this is indeed the case is easily seen by recalling that all the inequalities are required to be valid $\forall A, B, H$ 
in the field space, and noting that $\alpha_A, \alpha_{AH}$ and $\alpha_{AB}$ remain unchanged, while $\alpha_{ABH}$ flips sign, at the 
two field space points $A$ and $-A$ (or equivalently at $B$ and $-B$, or $H$ and $i H$), see Eqs.~(\ref{eq:def-alphas1}, \ref{eq:def-alphas2}). It follows that one can change freely
the sign of $\alpha_{ABH}$ for any given configuration of 
$\alpha_A, \alpha_{AH}$ and $\alpha_{AB}$. (As we will see in the next subsection, Figs.~(\ref{fig:alfah-alfabh} - \ref{fig:alfab-alfabh}), this translates into domains symmetrical around
$\alpha_{ABH}=0$.)
The variable $Z \in (-\infty, +\infty)$ is thus genuinely 
uncorrelated with the other field dependent reduced parameters. 


\subsection{Global correlations among the {\sl $\alpha$-parameters} \label{sec:glob-corr}}
In this section we first determine the allowed domains
of the {\sl $\alpha$-parameters} taken two by two, then combine
the resulting six global correlations to obtain an analytical
approximation of the full $4$D domain.  
Since the {\sl $\alpha$-parameters} are ratios of gauge invariant quantities, cf. Eqs.~(\ref{eq:def-alphas1},\ref{eq:def-alphas2}), 
it is convenient to choose a gauge that reduces the dependence on 
the set of components fields of the $A$, $B$ and $H$ multiplets.
Apart from the treatment of $\alpha_A$ versus $\alpha_{AH}$, we carry all the discussion in this section assuming
a gauge that diagonalizes the (hermitian and traceless) $B$ multiplet as defined
in Eq.~(\ref{eq:AB}), which then takes the form
\begin{eqnarray}
B &=
                  \displaystyle \left(
\begin{array}{cc}
{b} & 0 \\
0 & -{b}\\
\end{array}
\right) . \label{eq:Bdiag}
\end{eqnarray}
It follows that the dependence on $b$ cancels out
in $\alpha_{AB}$ and, up to a global sign, in $\alpha_{ABH}$.

\subsubsection{\bf \boldmath$\alpha_A$ versus $\alpha_{AH}$}
\begin{figure*}[!h]
\captionsetup{justification=raggedright,
singlelinecheck=false}
    \begin{center}
      {\includegraphics[width=0.60\textwidth, 
      height=0.3\textheight, keepaspectratio]{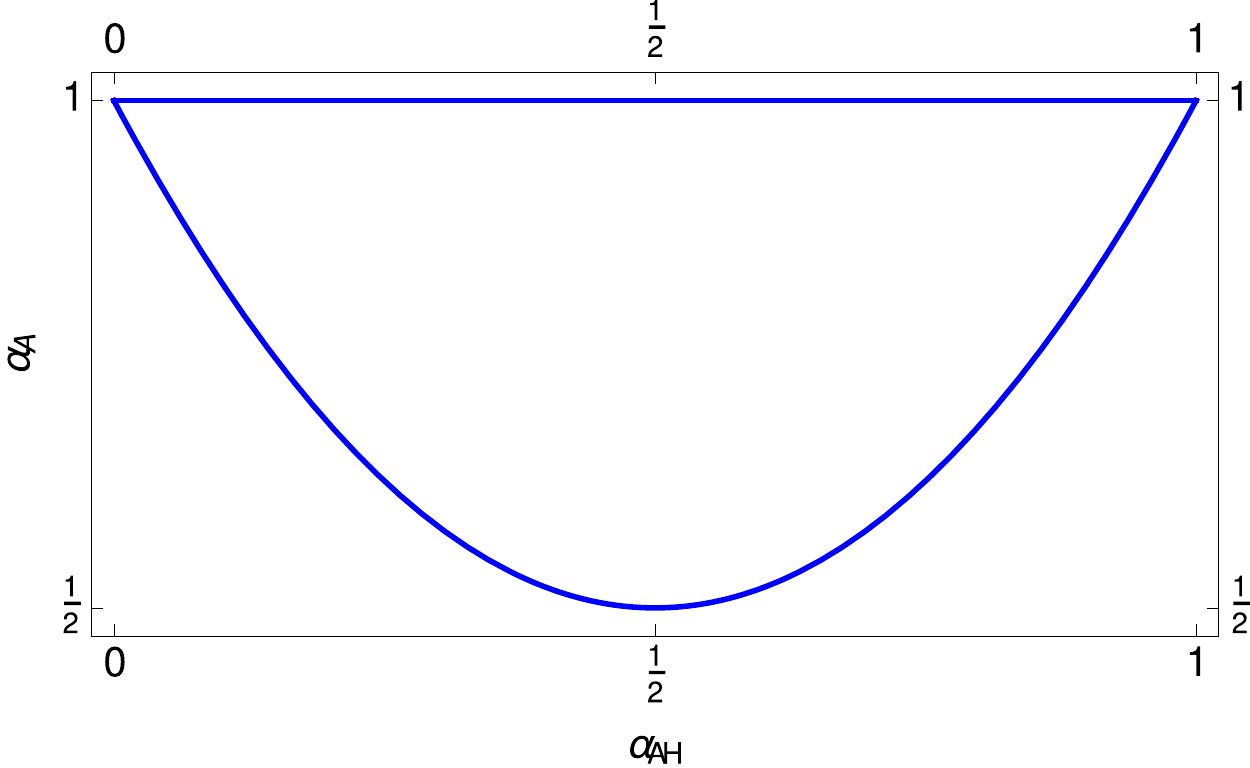}}
      \end{center}
\caption{\label{fig:alfa-alfah} \small{Projection of the 
{\sl $\alpha$-parameters} domain onto the $(\alpha_{AH}, 
\alpha_{A})$ plane.}}
\end{figure*}

These parameters are identical respectively to $\zeta$ and $\xi$ that were
defined and studied in detail in Section \ref{subsec:BFB} and Appendix \ref{appendix:xizetacorr}. We just recall here  
the corresponding domain:
\begin{eqnarray}
&(i):& 0 \leq \alpha_{AH} \leq 1 \label{eq:dom-alfa-alfah1}\\
&(ii):&  2 \, \alpha_{AH} \, ( \alpha_{AH} -1 )  +1  \leq \alpha_A \leq 1,
\label{eq:dom-alfa-alfah2}
\end{eqnarray}
illustrated in Fig.~\ref{fig:alfa-alfah}.

\subsubsection{\bf \boldmath$\alpha_A$ versus $\alpha_{AB}$ \label{subsec:alA-alAB}}
\begin{figure*}[!h]
\captionsetup{justification=raggedright,
singlelinecheck=false}
    \begin{center}
    {\includegraphics[width=0.60\textwidth, 
      height=0.3\textheight, keepaspectratio]{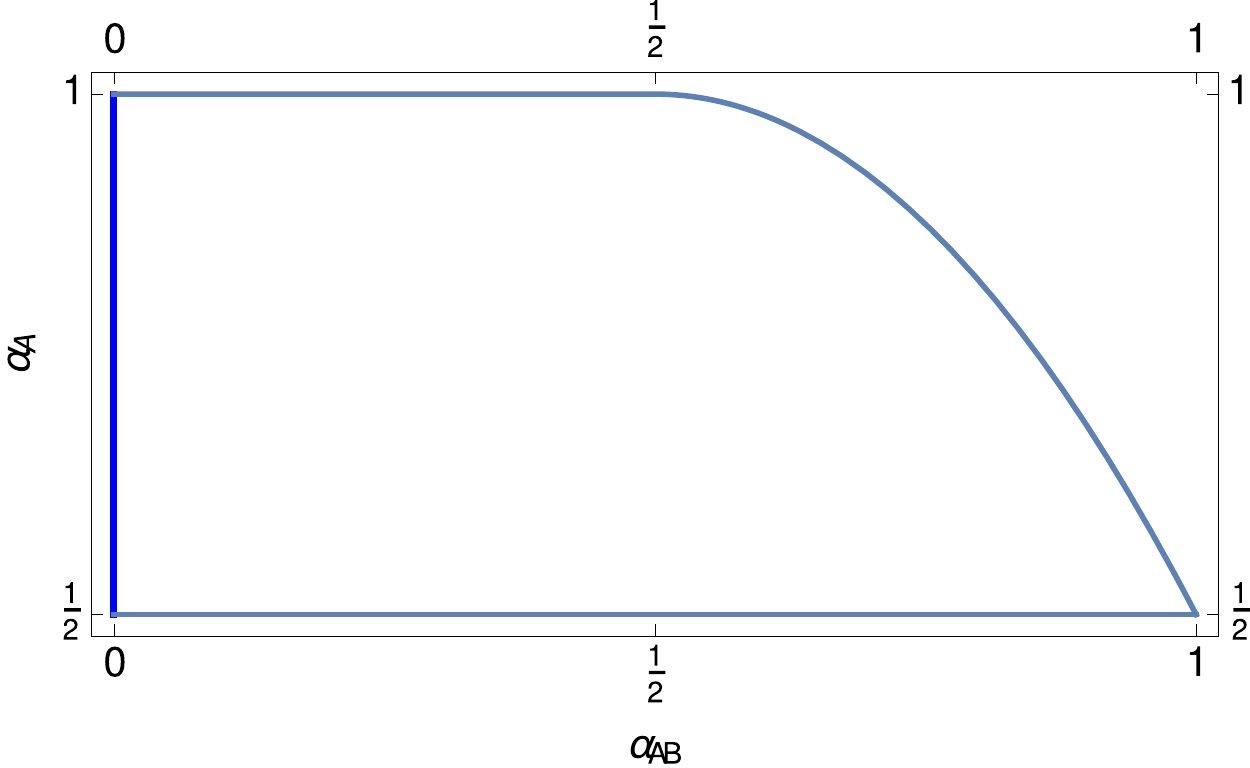}}
      
      {\includegraphics[width=0.45\textwidth, 
      height=0.3\textheight, keepaspectratio]{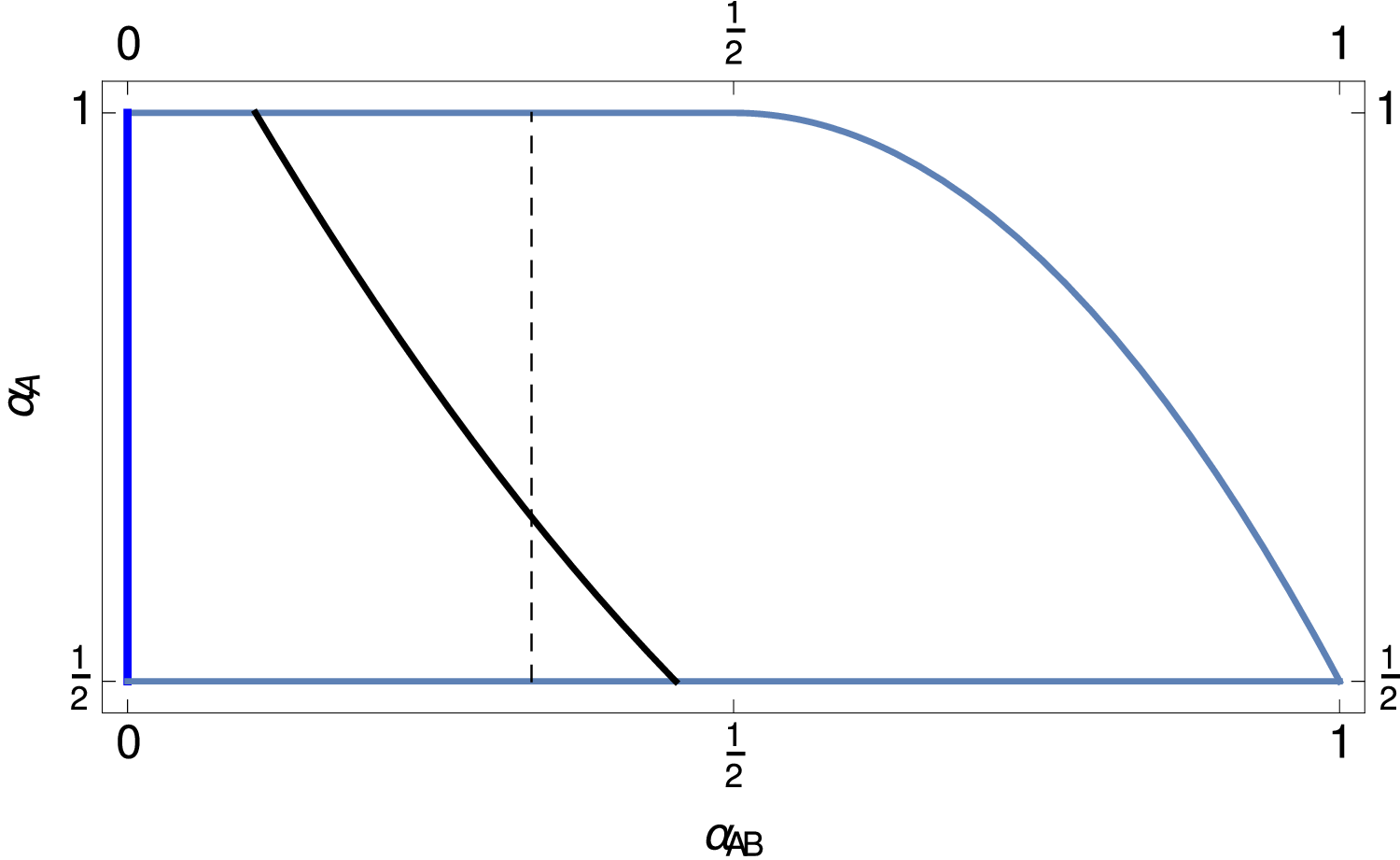}}
      {\includegraphics[width=0.45\textwidth, 
      height=0.3\textheight, keepaspectratio]{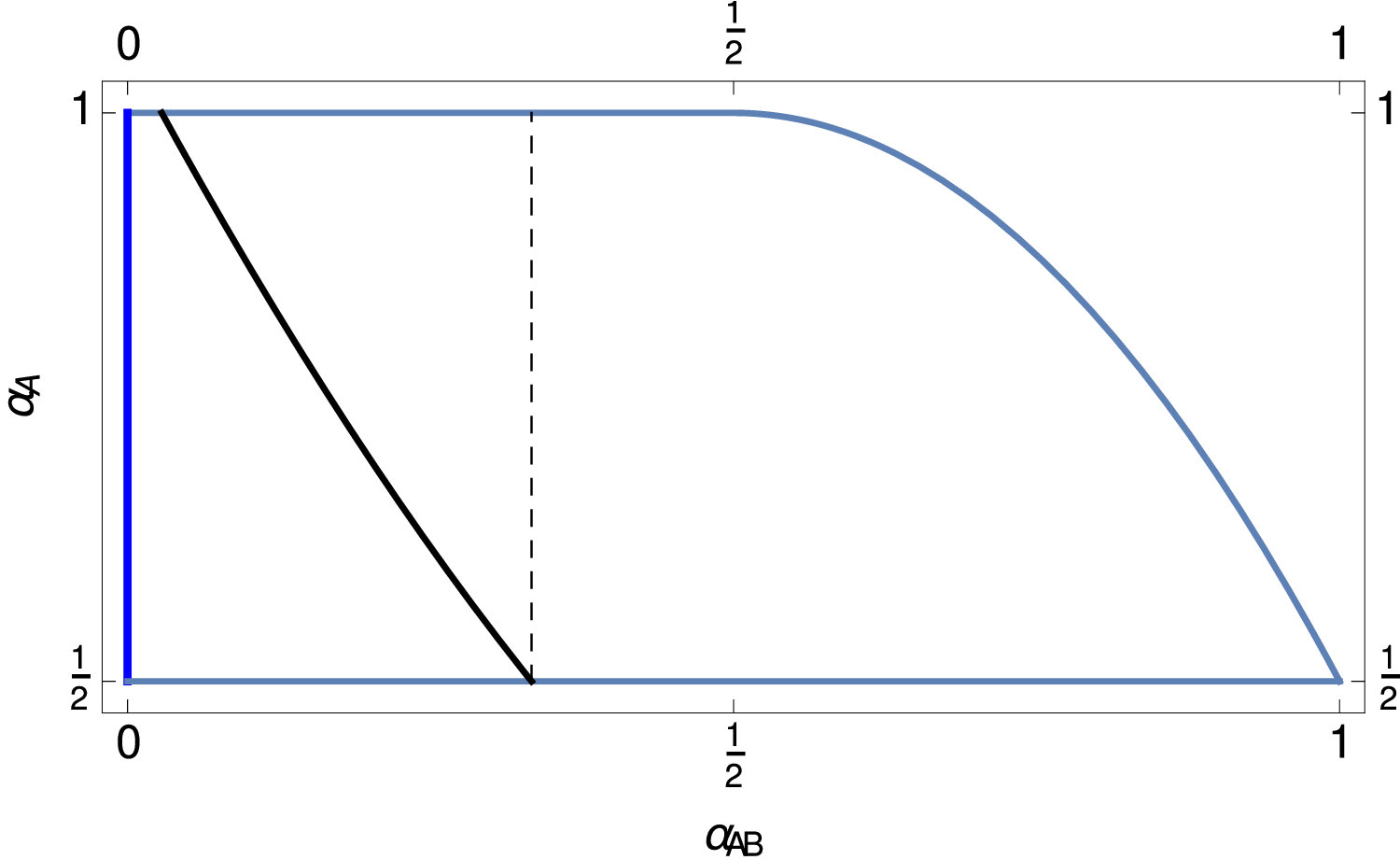}}
      {\includegraphics[width=0.45\textwidth, 
      height=0.3\textheight, keepaspectratio]{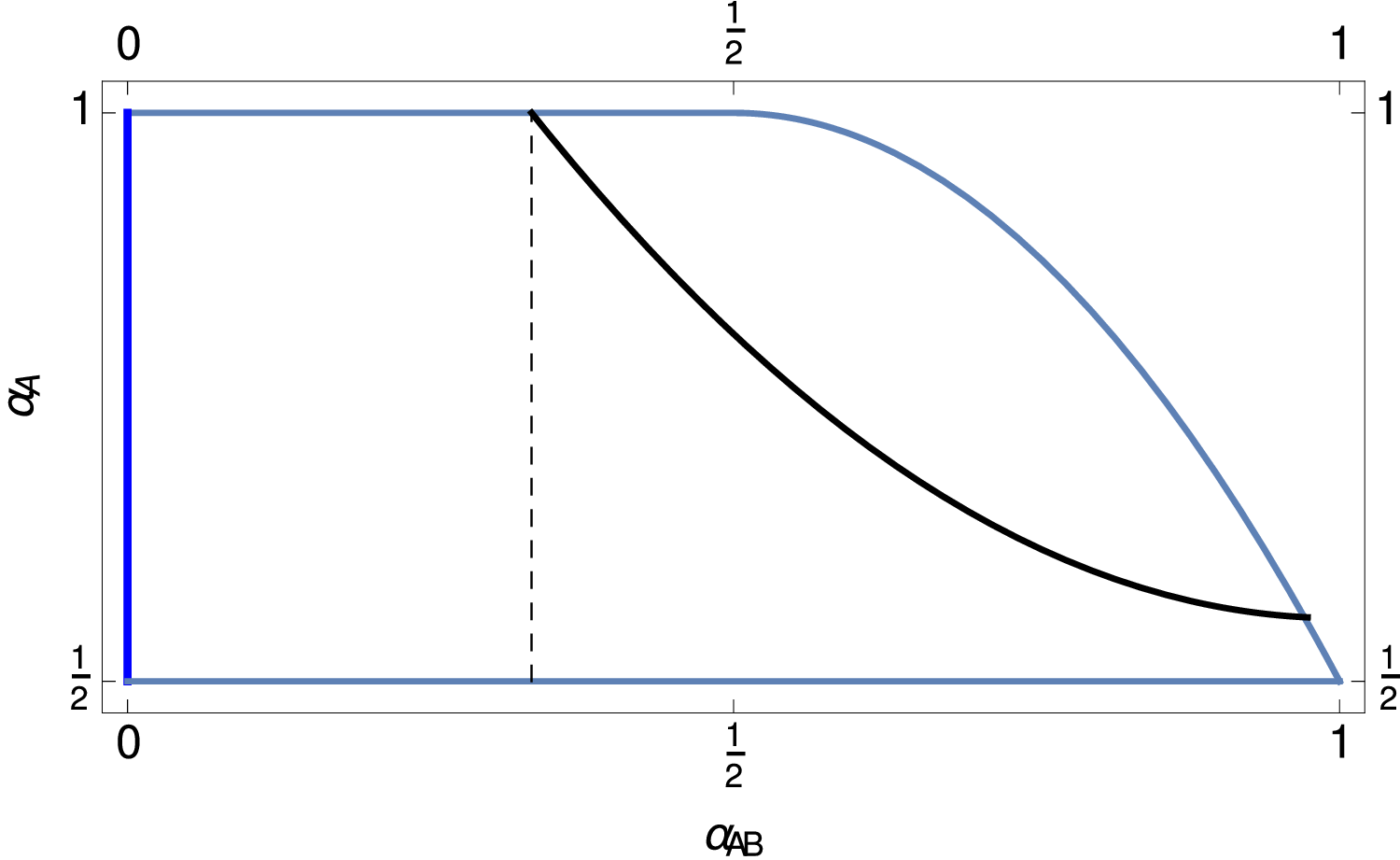}}
      {\includegraphics[width=0.45\textwidth, 
      height=0.3\textheight, keepaspectratio]{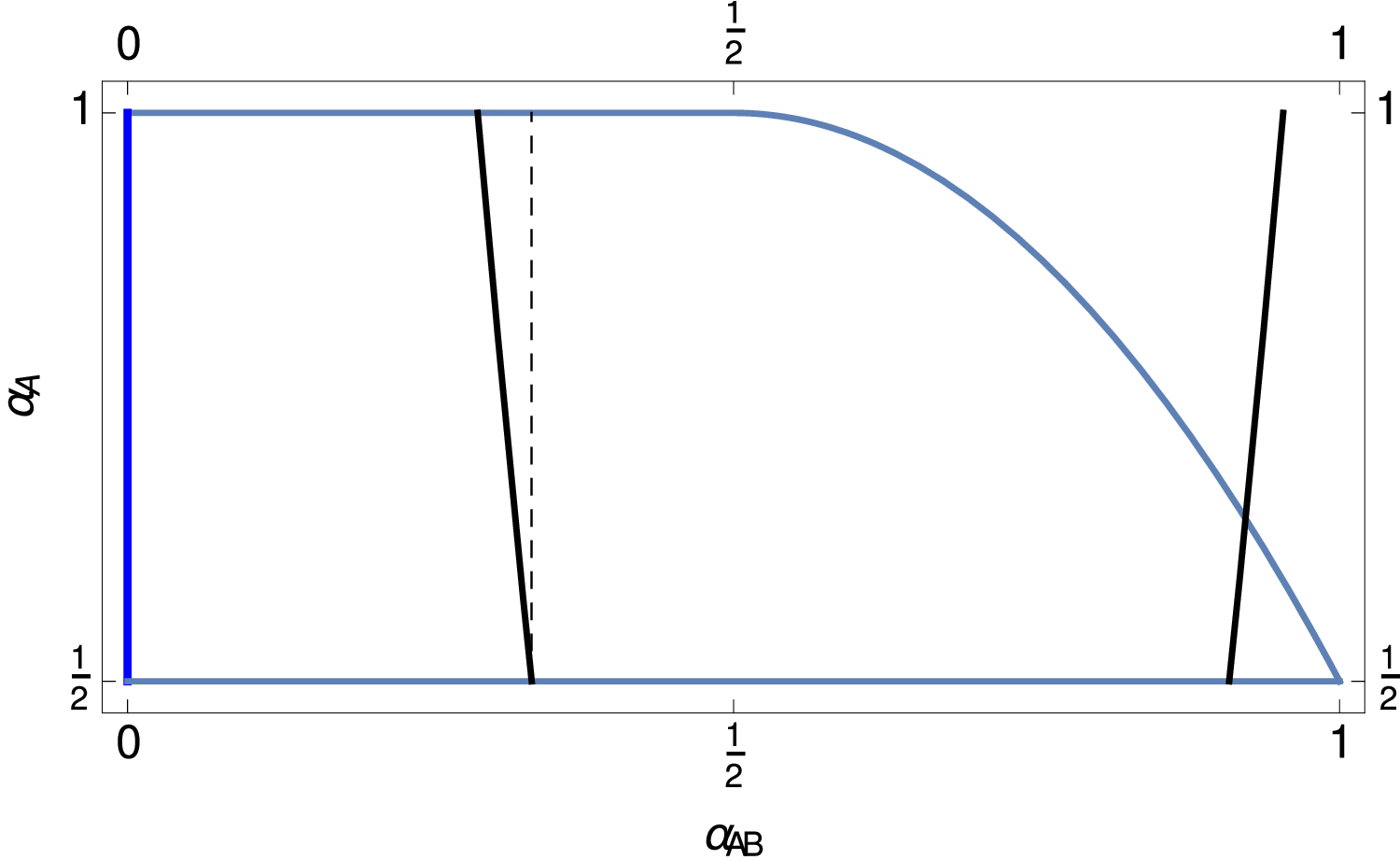}}
      \end{center}
\caption{\label{fig:alfa-alfab} \small{The upper figure blue contour indicates the projection of the {\sl $\alpha$-parameters} domain onto the $(\alpha_{AB}, \alpha_A)$ plane. The middle and lower figures are related to Section~\ref{sec:resolv-a0-AND-c0}
to which the reader is referred for more details. The dashed straight lines 
illustrate arbitrary partitions defined by Eq.(\ref{eq:p-c-BFB-4})-(I); the black solid parabolae illustrate arbitrary partitions defined~by~Eq.(\ref{eq:p-c-BFB-4})-(II).}}
\end{figure*}

With no particular gauge choice but using the fact that the parameter $\alpha_A$ is a ratio, one can recast it in terms of reduced parameters in the following form: 
\begin{eqnarray}
&&\alpha_A =\frac14 \left(2\,\cos^4 \theta + (3 + \cos 4\varphi)\,\sin^4 \theta + 
      (2 + \cos \rho\,\sin 2\varphi)\,\sin^2 2 \theta \right) \, , 
      \label{eq:alA}
\end{eqnarray}
where we defined
\begin{equation}
\begin{aligned}
&|a^0| = a \cos\varphi \sin\theta,  \\ 
&|a^+|  = a \cos\theta,  \\
&|a^{++}| = a \sin\varphi \sin\theta,  \\
& \rho = \arg(a^0) - 2 \arg(a^+) + \arg(a^{++}),\ 
\end{aligned}
\label{eq:phithetadef}
\end{equation}
with
\begin{equation}
0\leq \varphi\leq \frac{\pi}{2}, \, 
0\leq \theta \leq \frac{\pi}{2}, \, 0 \leq \rho \leq 2 \pi, \, {\rm and} \,
a= \sqrt{|a^0|^2 + |a^+|^2 + |a^{++}|^2} \, . \label{eq:angles-range}
\end{equation}
Furthermore, choosing a gauge for which Eq.~(\ref{eq:Bdiag}) is valid the $\alpha_{AB}$ parameter takes the very simple form, 
\begin{eqnarray}
&&\alpha_{AB}=\cos^2 \theta \, \label{eq:alAB} .
\end{eqnarray} 
Equations (\ref{eq:alA}, \ref{eq:alAB}) lead straightforwardly to
\begin{eqnarray}
\alpha_A = \frac14 \left( 3 + (2 - 3 \alpha_{AB}) \alpha_{AB} + ( \alpha_{AB} -1 )^2 \cos 4 \varphi\right) + 
     ( 1- \alpha_{AB} ) \alpha_{AB} \cos \rho \sin 2 \varphi .
     \label{eq:alfa-alfab}
\end{eqnarray}
To determine the boundary of the allowed domain one can for instance study the variation of $\alpha_A$ in Eq.~(\ref{eq:alfa-alfab}) as a quadratic function of $x\equiv\sin 2 \varphi$ in the domain 
$0 \leq x \leq 1$ to identify the set of maximal and minimal
possible values of $\alpha_A$ for a given $\alpha_{AB}$ depending on $\cos \rho$. The maximum is reached for 
$x=\frac{\alpha_{AB}}{1 - \alpha_{AB}}  \cos \rho$ which lies
in the allowed domain only if $\cos \rho \geq 0$ and 
$\alpha_{AB} \leq \frac12$. Otherwise, the maximum is reached
at one of the boundary values $x=0$ or $x=1$.
We find that the boundary of the domain is given 
by the following four curves:  
\begin{equation}
\begin{aligned}
{\rm (I):}& ~\alpha_{AB}=0 \ {\rm and} \ \frac12 \leq \alpha_A \leq 1, \\
{\rm (II):}& ~\alpha_A=\frac12 \ {\rm and} \ 0 \leq \alpha_{AB} \leq 1, \\
{\rm (III):}& ~\alpha_A=1 \ {\rm and} \ 0 \leq \alpha_{AB} \leq \frac12, \\
{\rm (IV):}& ~\frac12 \leq \alpha_{AB} \leq 1 \ {\rm and} \
\alpha_A=\frac12 + 2 (1-\alpha_{AB}) \alpha_{AB}, 
\end{aligned}
\label{eq:dom-alfa-alfab}
\end{equation}
see also Fig.~\ref{fig:alfa-alfab}.




\subsubsection{\bf \boldmath$\alpha_{AH}$ versus $\alpha_{ABH}$ 
\label{subsec:alAH-alABH}}
Similarly to the preceding case, we recast $\alpha_{AH}$ and $\alpha_{ABH}$ in terms of
reduced parameters and in the gauge where Eq.~(\ref{eq:Bdiag}) holds:
\begin{eqnarray}
\alpha_{AH} &=&\frac12 \, (1 + \cos 2 \varphi \, \cos 2 \psi  \, \sin^2 \theta) + \, \frac{1}{2 \sqrt{2}}\,( \cos \varphi\,\cos \theta_3 +\,\sin \varphi\, \cos \theta_4 )\,\sin 2 \psi\,\sin 2 \theta,  \nonumber \\ \label{eq:alAH} \\
\alpha_{ABH} &=& \sqrt{2} \,\sgn(b) \,(\sin \varphi\,\sin^2 \psi\,\cos \theta_2 -\cos \varphi\,\cos^2 \psi\,\cos \theta_1)\,\sin \theta 
,\label{eq:alABH}
\end{eqnarray}
where $\theta$ and $\varphi$ are as previously defined and
\begin{equation}
\begin{aligned}
&\theta_1 = \arg(a^0) - 2 \arg(\phi^0),  \\
&\theta_2 = \arg(a^{++}) - 2 \arg(\phi^+),  \\
&\theta_3 = \arg(a^0) - \arg(a^+) - \arg(\phi^0) + \arg(\phi^+), 
 \\
&\theta_4 = \arg(a^+) - \arg(a^{++}) - \arg(\phi^0) + \arg(\phi^+), 
 \\
&\cos \psi = \frac{|\phi^0|}{\sqrt{|\phi^0|^2 + |\phi^+|^2}},
\end{aligned}
\label{eq:t1-t4}
\end{equation}
with $\displaystyle 0\leq \psi \leq \frac{\pi}{2}$ and 
$ 0\leq \theta_i \leq 2 \pi$. (Note that $\theta_1= \theta_2 + \theta_3 + \theta_4$ (modulo $2 \pi$).)

\begin{figure*}[!h]
\captionsetup{justification=raggedright,
singlelinecheck=false}
    \begin{center}
      {\includegraphics[width=0.60\textwidth, 
      height=0.3\textheight, keepaspectratio]{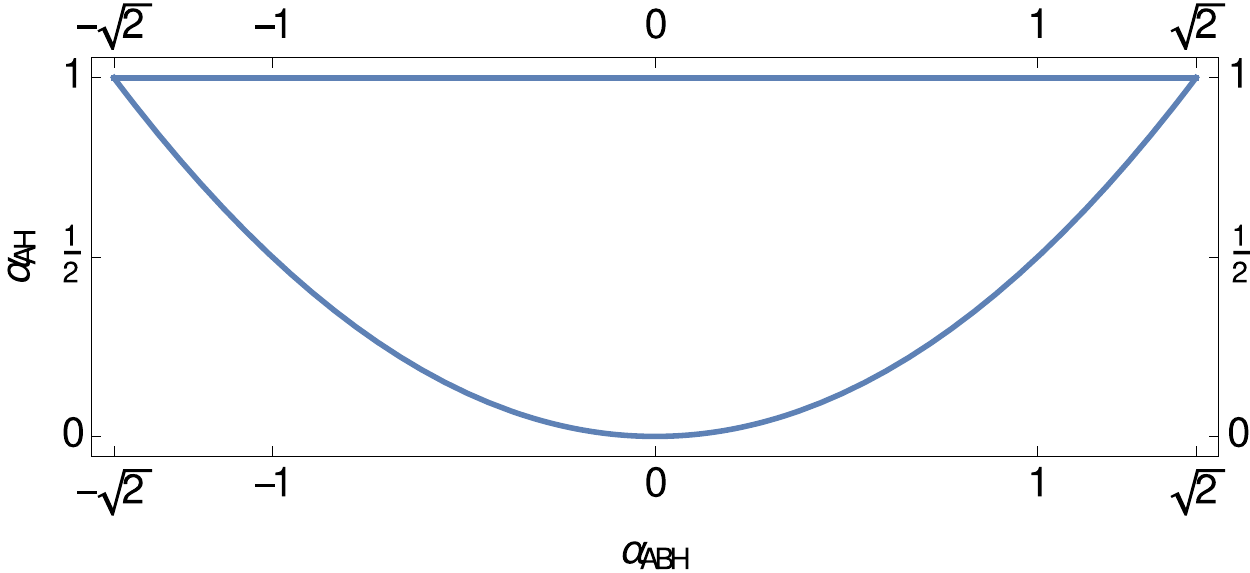}}
      \end{center}
\caption{\label{fig:alfah-alfabh} \small{Projection of the 
{\sl $\alpha$-parameters} domain onto the $(\alpha_{ABH}, 
\alpha_{AH})$ plane.}}
\end{figure*}

A numerical parametric scan over the various angles allows to 
guess the boundary of the $\alpha_{AH}$ versus $\alpha_{ABH}$ 
domain. 
The result turns out to be very simple given by the two curves:
\begin{eqnarray}
&{\rm (V):}&\alpha_{AH}=1, \ \forall  \alpha_{ABH}\in [-\sqrt 2 , +\sqrt 2], \label{eq:V}\\
&{\rm (VI):}&\alpha_{AH}=\frac12 \alpha_{ABH}^2 \label{eq:VI}\ ,
\end{eqnarray}
illustrated in Fig.~\ref{fig:alfah-alfabh}.
The proof for the upper boundary (\ref{eq:V}) is simple: It suffices to exhibit particular configurations of the various 
angles for which $\alpha_{AH}$ saturates its upper bound while
$\alpha_{ABH}$ scans all its allowed domain. An example is
$\displaystyle \varphi=\psi=\theta=\frac{\pi}{2}$, keeping all the $\theta_i$'s free. This gives $\alpha_{AH}=1$ and $\alpha_{ABH}=
\sqrt{2} \cos \theta_2$, which proves the above statement.
The lower boundary (\ref{eq:VI}) is much more difficult
to establish analytically. The proof is somewhat involved 
and will be relegated to Appendix \ref{appendix:alAH-alAHB}.

\subsubsection{\bf \boldmath$\alpha_{A}$ versus $\alpha_{ABH}$ \label{subsec:alA-alABH}}

Here again a numerical parametric scan over the various angles 
helps guessing the boundary of the $\alpha_{A}$ versus $\alpha_{ABH}$ domain. However, one still needs for that 
to admit {\sl ad hoc} that the whole boundary is obtained when 
$\sin \theta = 1$. The analytical proof is quite involved 
and is given in Appendix \ref{appendix:alA-alAHB} for completeness.
We find that the boundary is determined by the following:
\begin{equation}
\begin{aligned}
{\rm (VII):}& ~\alpha_{A}=1, \ \text{for} \; \alpha_{ABH}\in [-\sqrt 2 , +\sqrt 2],  \\
{\rm (VIII):}& ~\alpha_{A}=\frac12, \ \text{for} \; \alpha_{ABH}\in [-1 , +1],  \\
{\rm (IX):}& ~\alpha_{A} = 1 - \alpha_{ABH}^2 + \frac12 \alpha_{ABH}^4, \ 
\text{for} \; \alpha_{ABH}\in [-\sqrt 2 , -1] \cup [+1 , +\sqrt 2].
\label{eq:boundary-alA-alABH1-3}
\end{aligned}
\end{equation}
 
\begin{figure*}[!h]
\captionsetup{justification=raggedright,
singlelinecheck=false}
    \begin{center}
      {\includegraphics[width=0.60\textwidth, 
      height=0.3\textheight, keepaspectratio]{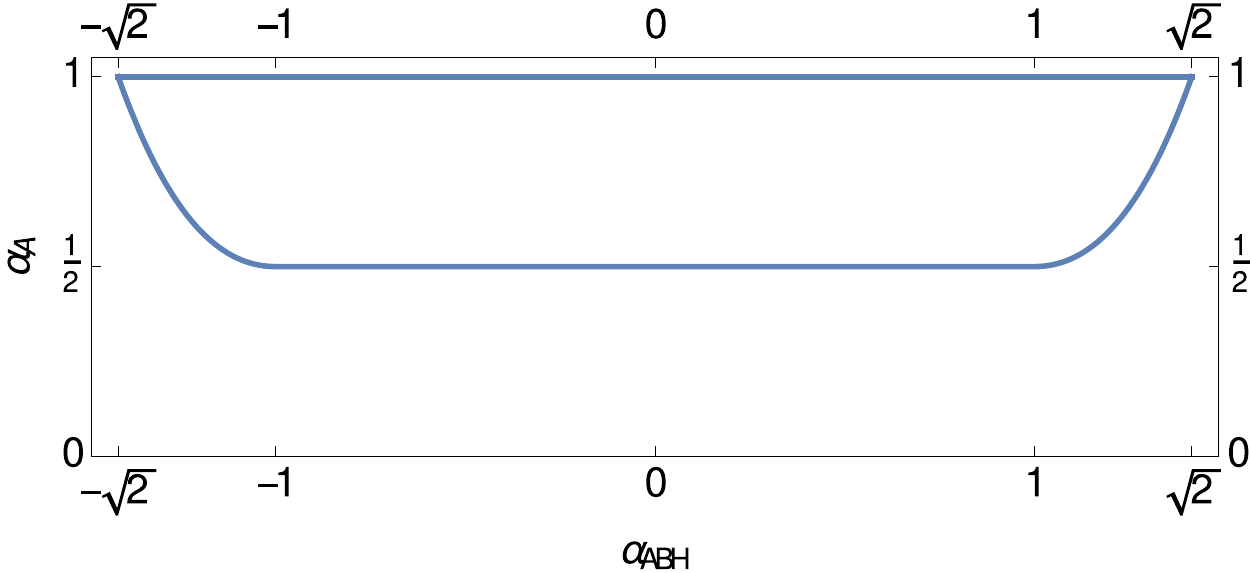}}
      \end{center}
\caption{\label{fig:alfa-alfabh} \small{Projection of the 
{\sl $\alpha$-parameters} domain onto the $(\alpha_{ABH}, 
\alpha_A)$ plane.}}
\end{figure*}

\subsubsection{\bf \boldmath$\alpha_{AB}$ versus $\alpha_{ABH}$}
\begin{figure*}[t]
\captionsetup{justification=raggedright,
singlelinecheck=false}
    \begin{center}
      {\includegraphics[width=0.60\textwidth, 
      height=0.3\textheight, keepaspectratio]{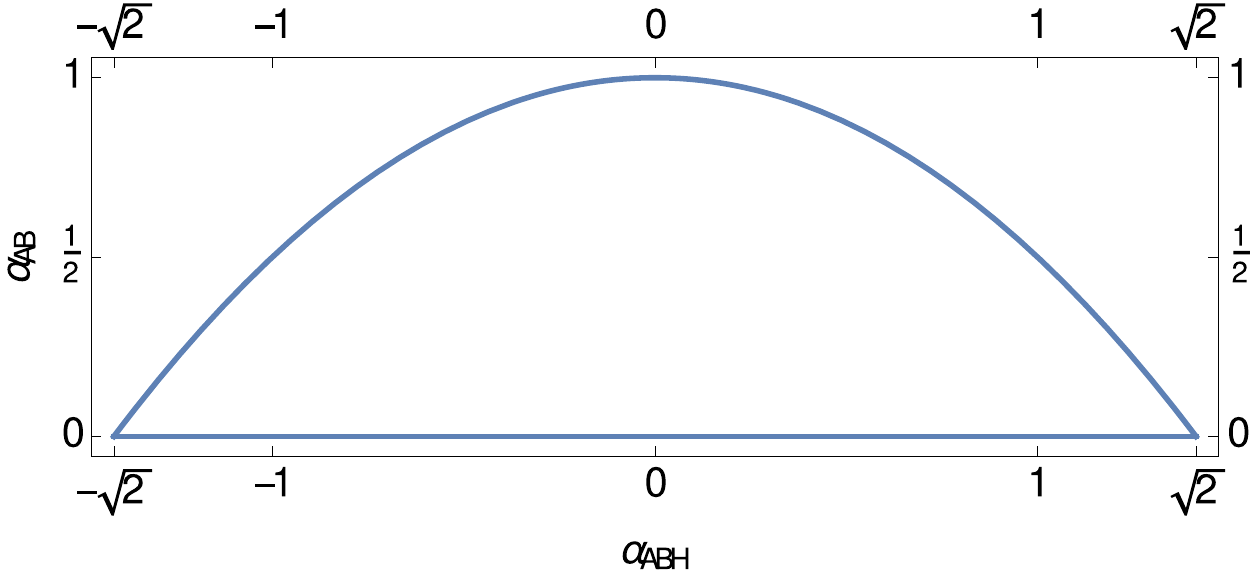}}
      \end{center}
\caption{\label{fig:alfab-alfabh} \small{Projection of the 
{\sl $\alpha$-parameters} domain onto the $(\alpha_{ABH}, 
\alpha_{AB})$ plane.}}
\end{figure*}

From Eqs.~(\ref{eq:alAB},\ref{eq:alABH},\ref{eq:angles-range},\ref{eq:alfaHABrange}),  one obtains readily
\begin{equation}
\alpha_{ABH}^2 =2 Y^2 ( 1 - \alpha_{AB}) \label{eq:rel-AB-ABH}
\end{equation}
where $Y$ (defined in Eq.~(\ref{eq:ZZ})) and  $\alpha_{AB}$ vary independently in the domain 
$[0,1]$. It is then clear that for each given value of 
$\alpha_{ABH}$, $\alpha_{AB}$ reaches its maximal value compatible
with Eq.~(\ref{eq:rel-AB-ABH}) when $Y^2=1$. Also the minimal value
$\alpha_{AB} = 0$ is reached for any value of $\alpha_{ABH}^2$.
The boundary of the allowed domain in the plane $\alpha_{AB}$
versus $\alpha_{ABH}$ is thus delimited by the two curves:
\begin{eqnarray}
&{\rm (X):}&\alpha_{AB}=0, \ \forall  \alpha_{ABH}\in [-\sqrt 2 , +\sqrt 2],\\
&{\rm (XI):}& \alpha_{AB} = 1 - \frac12 \alpha_{ABH}^2\ ,
\end{eqnarray}
as shown on Fig.~\ref{fig:alfab-alfabh}.

\subsubsection{\bf \boldmath$\alpha_{AH}$ versus $\alpha_{AB}$}
\begin{figure*}[!t]
\captionsetup{justification=raggedright,
singlelinecheck=false}
    \begin{center}
      {\includegraphics[width=0.50\textwidth, 
      height=0.50\textheight, keepaspectratio]{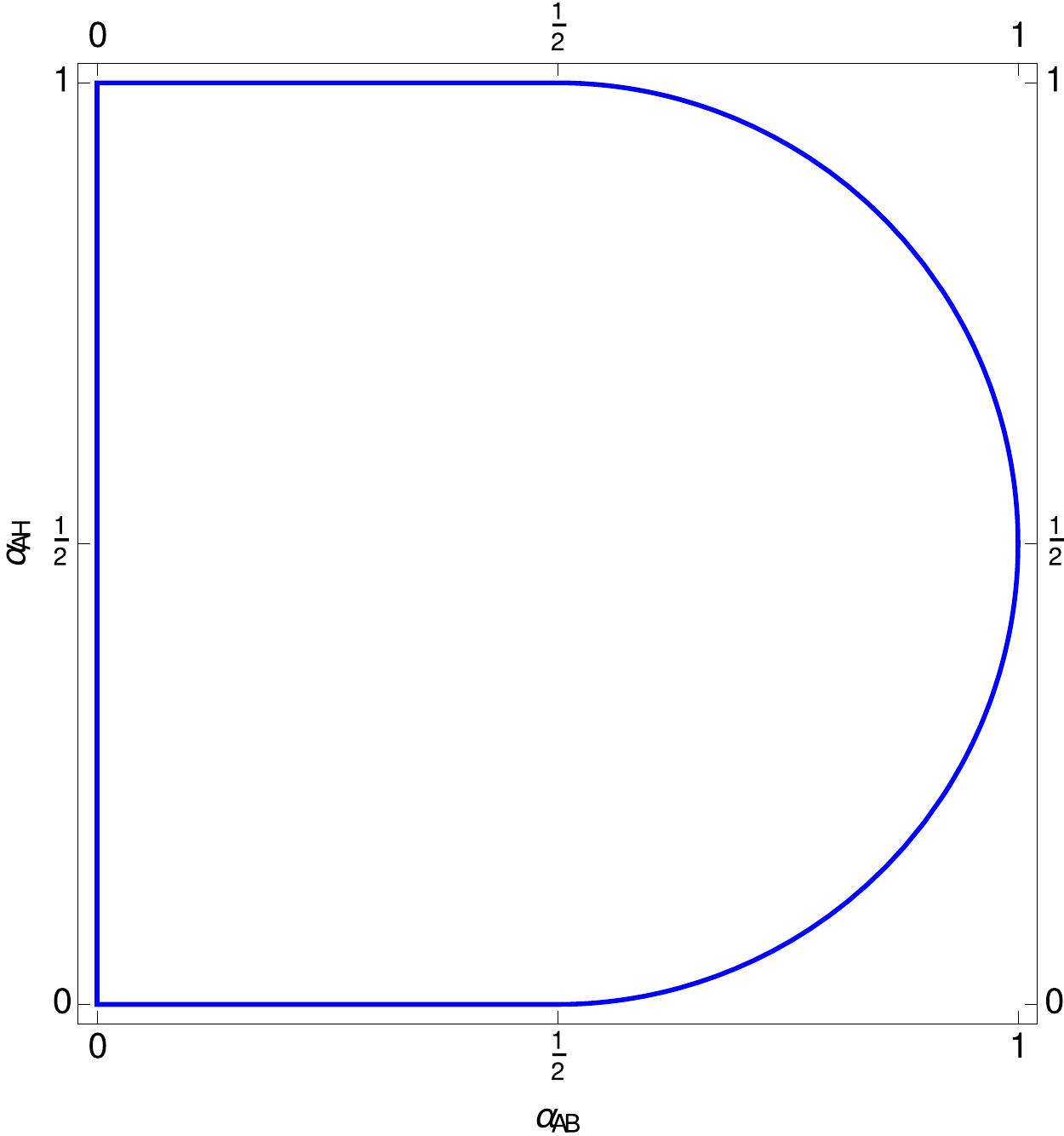}}
      \end{center}
\caption{\label{fig:alfab-alfah} \small{Projection of the 
{\sl $\alpha$-parameters} domain onto the $(\alpha_{AB}, 
\alpha_{AH})$ plane.}}
\end{figure*}

The boundary of the allowed domain in the 
$(\alpha_{AB},\alpha_{AH})$ plane is given by:
\begin{eqnarray}
&{\rm (XII):}&\alpha_{AB}=0, \ \forall  \alpha_{AH}\in [0, 1], \label{eq:alAB-alAH-bound1}\\
&{\rm (XIII):}&\alpha_{AH}=0, \ \forall  \alpha_{AB}\in [0 , 
\frac12],\\
&{\rm (XIV):}& \alpha_{AH} = 1  ,\ \forall  \alpha_{AB}\in [0 , 
\frac12], \label{eq:alAB-alAH-bound3}\\
&{\rm (XV):}& \left(\alpha_{AB}-\frac12\right)^2 +  \left(\alpha_{AH}-\frac12\right)^2= \frac14  ,\ \rm{for} \  \alpha_{AB}\in [\frac12,1] \ \rm{and} \
\alpha_{AH}\in [0, 1] , \label{eq:alAB-alAH-bound4}
\end{eqnarray}
see Fig.~(\ref{fig:alfab-alfah}). The proof strategy is similar to the one in Sec.~(\ref{subsec:alA-alAB}) albeit somewhat more involved, the convenient variable here to study the variation of $\alpha_{AH}$ being $x\equiv \cos 2 \psi$. (See Appendix \ref{appendix:alAB-alAH} for details.)

\subsubsection{\bf \boldmath The $4$D $\alpha$-potatoid \label{subsec:potatoid}}
The $2$D projections of the
{\sl $\alpha$-parameters} domain determined analytically in 
the previous subsections will allow, in some
cases, a fully analytical resolving of the BFB conditions on the $\lambda$'s.
Obviously generalizing beyond $2$D along the same lines becomes non-tractable analytically.
In principle one can then proceed numerically, scanning over 
part or all of  the seven angles entering Eqs.~(\ref{eq:alA}, \ref{eq:alAB}, \ref{eq:alAH}, \ref{eq:alABH}), to determine  
the $3$D projections as well as the true $4$D allowed domain of 
the {\sl $\alpha$-parameters}.  However, this would cut short
the possibility of further analytical resolving for the conditions
on the $\lambda$'s. 

We will proceed differently here by constructing an analytical approximation of the true {\sl $\alpha$-parameters} domain
from a back-projection using only six planes.  
Obviously any point in 
the true domain should have its projections on the six planes 
lying within the six domains determined above. 
This necessary condition can be characterized by 
the interior of a four dimensional convex domain that we will refer
to as the $4$D potatoid. To determine explicitly this 
$4$D potatoid 
we first express separately in the form of a logical (inclusive) disjunction each of 
the six domains of Figs.~\ref{fig:alfa-alfah}-- \ref{fig:alfab-alfah}, 
then form 
the logical conjunction of these disjunctions.  The resulting Boolean 
expression is somewhat involved but, interestingly enough, 
it eventually simplifies to the following form:   
\begin{eqnarray}
&\alpha_A \leq 1 \, \land \, \alpha_{AB} \geq 0 \, \land \,  
\alpha_{AB} \leq 
  1 - \displaystyle \frac12 \alpha_{ABH}^2  &\nonumber \\ 
&\, \land \,& \nonumber \\  
&  \displaystyle \alpha_{AH} \geq \frac12 \alpha_{ABH}^2  \, \land \,  
\alpha_A \geq 1 + 2 (\alpha_{AH} -1) \alpha_{AH} & \label{eq:potatoid}\\
 &  \, \land \,  &\nonumber \\
 & \displaystyle \left \{ \alpha_{AB} \leq \frac12 \, \lor \, \left \{ \alpha_A \leq \frac12 + 2 (1-\alpha_{AB}) \alpha_{AB} \, \land \,  
      \left(\alpha_{AB}-\frac12\right)^2 +  \left(\alpha_{AH}-\frac12\right)^2 \leq \frac14 \right \} \right \}& \nonumber 
\end{eqnarray}
This form is non-trivial in that it does not display explicitly
all six correlations among the four {\sl $\alpha$-parameters};
in particular, the correlation between $\alpha_A$ and 
$\alpha_{ABH}$ does not appear explicitly and, depending on $\alpha_{AB}$, either only three or five of the six correlations are explicitly needed. These features will prove useful
when resolving the constraints in Section~\ref{sec:part-resolv}. 
It is also informative to partially visualize the $4$D potatoid 
by considering 
its $3$D projections along each of the four directions. This amounts to combining the domains three by three which leads after some simplifications to:  
\begin{eqnarray}
\left\{ \alpha_A, \alpha_{AB}, \alpha_{AH}\right\} = \left\{
\begin{array}{c}
\displaystyle \frac12 \leq \alpha_A \leq 1 \, \land \,
\displaystyle 0 \leq \alpha_{AB} \leq \frac12 + \sqrt{\frac{1 - \alpha_A}{2}}   \\
\, \land \,    \\
\displaystyle  \frac12 \left(1 - \sqrt{2 \alpha_A- 1} \right) \leq \alpha_{AH} \leq \frac12\left(1 + \sqrt{2 \alpha_A- 1} \right), 
\end{array} \right.
\end{eqnarray}
\begin{eqnarray}
 \left\{ \alpha_A, \alpha_{AH}, \alpha_{ABH} \right\}=
1+2 (\alpha_{AH} -1 ) \alpha_{AH} \leq \alpha_A \leq 1 \, \land \, \alpha_{ABH}^2 \leq 2 \alpha_{AH} , 
\end{eqnarray}
\begin{eqnarray}
\left\{ \alpha_A, \alpha_{AB}, \alpha_{ABH}\right\} = \left\{
\begin{array}{c}
\displaystyle  \frac12 \leq \alpha_A \leq 1
\, \land \, 0 \leq \alpha_{AB} \leq 1 - \frac{\alpha_{ABH}^2}{2}  \\
\, \land \,\\
\displaystyle \left \{\alpha_{AB}\leq \frac12\, \lor \, 
\frac12 - \sqrt{\frac{1 - \alpha_A}{2}}\leq \alpha_{AB} \leq \frac12 + \sqrt{\frac{1 - \alpha_A}{2}}\right \}  \\
\, \land \,   \\
\left \{\alpha_{ABH}^2 \leq 1\, \lor \, 1 - \sqrt{2 \alpha_A- 1}  \leq \alpha_{ABH}^2 \leq 1 + \sqrt{2 \alpha_A- 1} \right \},
\end{array} \right. 
\end{eqnarray}
\begin{eqnarray}
\left\{ \alpha_{AB}, \alpha_{AH}, \alpha_{ABH}\right\} = \left\{
\begin{array}{c}
\displaystyle 0 \leq \alpha_{AB}\leq 1-\frac{\alpha_{ABH}^2}{2}\, \land \, \frac{\alpha_{ABH}^2}{2} \leq \alpha_{AH}\leq 1  \\
\, \land \,  \\
\displaystyle \left \{\alpha_{AB}\leq \frac12 \, \lor \, (1-2 \alpha_{AB})^2+(1-2 \alpha_{AH})^2\leq 1 \right \} 
\end{array} \right. 
\end{eqnarray}

\begin{figure*}[!b]
\captionsetup{justification=raggedright,
singlelinecheck=false}
\begin{center}
{\includegraphics[width=0.60\textwidth, 
height=0.3\textheight, keepaspectratio]{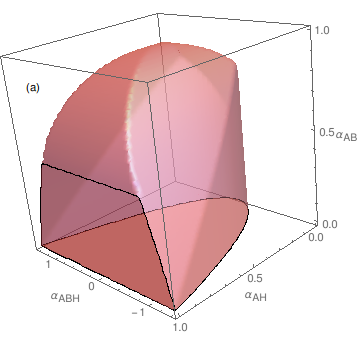}}
{\includegraphics[width=0.60\textwidth, 
      height=0.3\textheight, keepaspectratio]{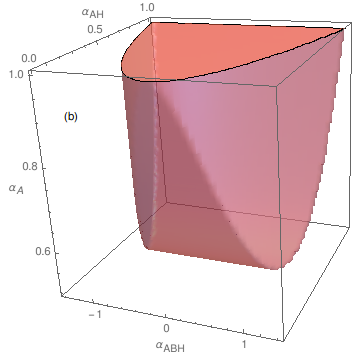}}
\end{center}
\begin{center}
{\includegraphics[width=0.60\textwidth, 
height=0.3\textheight, keepaspectratio]{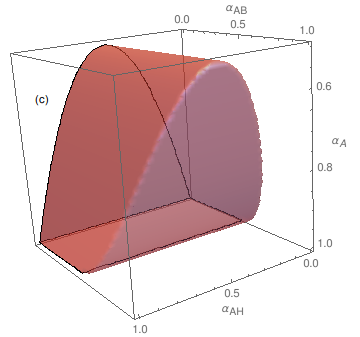}}
{\includegraphics[width=0.60\textwidth, 
height=0.3\textheight, keepaspectratio]{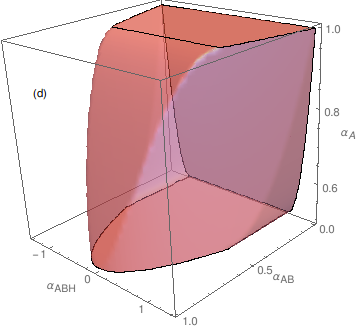}}
\end{center}
\caption{\label{fig:alfaFull}  Projection of the 
$\alpha$-potatoid along: (a) the $\alpha_A$ direction, (b) the $\alpha_{AB}$ direction, (c) the $\alpha_{ABH}$ direction, (d) the $\alpha_{AH}$ direction.}
\end{figure*}

Figure~\ref{fig:alfaFull} shows these $3$D projections. It is
easy to check by eye from this figure that further projection on the various planes reproduces the domains shown in Figs.~\ref{fig:alfa-alfah}-- \ref{fig:alfab-alfah}. However, the rounded 
(and even non-smooth) edges featured in Fig.~\ref{fig:alfaFull} hint at the fact that looking at projections is necessary but not sufficient to determine the true $4$D domain of the 
{\sl $\alpha$- parameters}. For instance a point lying just outside the chopped edge in Fig.~\ref{fig:alfaFull}~(a), that
is a point excluded for sure, 
would still project on the interior of
the domains of Figs.~\ref{fig:alfah-alfabh},\ref{fig:alfab-alfabh},\ref{fig:alfab-alfah}. Obviously this is not yet fully a counter example as the considered point might still project outside one of the three remaining $2$D domains. But on general grounds the potatoid determined by Eq.(\ref{eq:potatoid}), even though enclosing the true $4$D domain, is not necessarily identical
to it. Since relying on continuity arguments one does not 
expect holes in the interior of the true domain, that would 
leave no imprint in the projections on the six planes, one concludes that differences between the potatoid and the true domain should be located on the boundaries of the former.  
We defer a detailed study showing that this is indeed the case till section \ref{sec:peeling}. There we will make use of 
an interesting feedback on the 
issue from the more constrained Georgi-Machacek model.
 
\subsection{Resolved forms of the pre-custodial BFB conditions
\label{sec:resolved-p-c-BFB}}

For now  we ignore the above subtleties  
and exploit in the present section the domains of the {\sl $\alpha$- parameters}, as determined so far, to push as much as possible an explicit resolving of the conditions given by
Eqs.~(\ref{eq:p-c-BFB-1}, \ref{eq:p-c-BFB-2}) for the $\lambda$
parameters themselves. 


\subsubsection{Resolving $\mathfrak{a_0} > 0 \, \land \, \mathfrak{c_0} > 0$
\label{sec:resolv-a0-AND-c0}}
Resolving conditions (\ref{eq:p-c-BFB-1}) with respect to $t$, they become
equivalent to Eqs.~(\ref{eq:p-c-BFB-3}, \ref{eq:p-c-BFB-4}).
As stressed at the end of Section~\ref{sec:a0-AND-c0},
in order to 
fulfill the 'or' structure of Eq.~(\ref{eq:p-c-BFB-4}) one should
in principle consider all possible partitions into
subsets of the domain depicted in Fig.~\ref{fig:alfa-alfab}.
Since obviously the two inequalities could be simultaneously satisfied in some parts of the domain, the subsets
should be allowed to overlap. So, strictly speaking, we should
consider {\sl coverings} rather than {\sl partitions}.
More precisely:
\begin{itemize}
\item[] A set of values $( \lambda_{A}^{(1)}, \lambda_{A}^{(2)}, \lambda_{AB}^{(1)}, \lambda_{AB}^{(2)})$ will satisfy
Eq.~(\ref{eq:p-c-BFB-4}) $\forall \alpha_A, \alpha_{AB}$, 
{\it if and only
if there exists a covering of the $(\alpha_{A}, \alpha_{AB})$ domain consisting of a family of subsets of this domain for which Eq.~(\ref{eq:p-c-BFB-4})-(I) is satisfied on a collection of these
subsets, and Eq.~(\ref{eq:p-c-BFB-4})-(II) satisfied
on the complementary collection.}
\end{itemize}
The task can seem daunting since there are
a priori infinitely many ways of forming a covering of the domain. 
However, one can identify a clear procedure. Note first
the obvious fact that, for a given $( \lambda_{A}^{(1)}, \lambda_{A}^{(2)}, \lambda_{AB}^{(1)}, \lambda_{AB}^{(2)})$
in the $\lambda$--space, 
each of the two inequalities in Eq.~(\ref{eq:p-c-BFB-4}) defines
separately natural {\sl partitions} of the $(\alpha_{A}, \alpha_{AB})$ domain,
namely partitions formed by a collection made of subsets where the inequality is satisfied and subsets where it is not. Moreover, among all these natural partitions, one can show that a {\sl minimal} partition, made of the smallest possible number of subsets, is actually unique and made of at most two subsets.\footnote{This is an immediate consequence of the binary "yes/no" characterization of the subsets of the natural partitions defined above. Indeed, starting from a given natural partition and taking the union of all the "yes" subsets and the union of all the "no" subsets forms two subsets (including possibly an empty one) defining a minimal partition. The uniqueness proof then follows
easily: if $\{s_1, s_2\}$ and $\{s_1', s_2'\}$ are two minimal partitions, then at least one set $s_i$ and one set $s_j'$ should have a non-empty intersection $s_i \cap s_j'$, since the partitions cover the same domain. This implies the whole of $s_i$ and $s_j'$ to have the same "yes/no" characterization. But this contradicts the fact that the complementary of $s_i$ has by definition the opposite characterization, unless $s_i \cap s_j'= s_i= s_j'$. The two remaining subsets should thus be identical too, whence the uniqueness of the minimal partition.}  A clear strategy follows: For each given point in the 
$\lambda$--space, determine the two minimal partitions defined respectively by
Eq.~(\ref{eq:p-c-BFB-4})-(I) and Eq.~(\ref{eq:p-c-BFB-4})-(II),
call them $\{s_{yes}^I, s_{no}^I\}$ and $\{s_{yes}^{II}, s_{no}^{II}\}$;
then check whether their union forms a covering that satisfies 
the required property stated above in italics, that is check whether
\begin{equation}s_{no}^I \subset s_{yes}^{II}, \, \text{or equivalently,} \, s_{no}^{II} \subset s_{yes}^{I} ,\label{eq:covering}
\end{equation}
to select or reject the considered point in $\lambda$--space.

Given the linear dependence on $\alpha_{AB}$ in Eq.~(\ref{eq:p-c-BFB-4})-(I), the associated minimal partitioning corresponds  simply to cutting the $(\alpha_{A}, \alpha_{AB})$ domain into regions by a 
straight line going vertically across the domain, at $\displaystyle \alpha_{AB}=\alpha^*_{AB} \equiv - \frac{\lambda_{AB}^{(1)}}{\lambda_{AB}^{(2)}}$, as illustrated by the dashed line in Fig.~\ref{fig:alfa-alfab}. The inequality (\ref{eq:p-c-BFB-4})-(I) is then true for any $\alpha_{AB}$ in an entire interval of the form $[\alpha^*_{AB}, 1]$ or $[0,\alpha^*_{AB}]$, and false on their respective complement. This corresponds respectively to 
the two
minimal partitions where the "yes" assignment holds for the 
right side or the left side region. Moreover, since $\alpha_{AB}$ lives in $[0,1]$, 
the minimal partition reduces trivially to either $\{s_{yes}^I, \emptyset\}$ or $\{\emptyset, s_{no}^I\}$ if $\alpha^*_{AB} \notin [0,1]$. It is thus convenient to consider separately the NAS
conditions on $\lambda_{AB}^{(1)}$ and $\lambda_{AB}^{(2)}$ that 
correspond to each of these four configurations of the minimal partition. These NAS conditions are easy to write down given the monotonic dependence on $\alpha_{AB} \in [0,1]$ in  Eq.~(\ref{eq:p-c-BFB-4})-(I). They are given in Fig.~\ref{fig:resolved-p-c-BFB} with the labels (i), (ii), (iii) and (iv). Note that
(i) and (iv) correspond to the two extreme configurations, respectively $\{s_{yes}^I, \emptyset\}$, cf. Eq.~(\ref{eq:p-c-BFB-4-part}), and $\{\emptyset, s_{no}^I\}$, while
(ii) and (iii) are the two intermediate generic partitions. 

On the other hand, as can be easily seen
from the dependence on $\alpha_{A}, \alpha_{AB}$ in Eq.~(\ref{eq:p-c-BFB-4})-(II), the corresponding minimal partitions are determined by convex parabolae in the $(\alpha_{AB}, \alpha_{A})$ plane, illustrated by the black curves in Fig.~\ref{fig:alfa-alfab}. The middle and bottom figures in Fig.~\ref{fig:alfa-alfab}
show several possible configurations when $\alpha^*_{AB} \in [0,1]$. The middle-left illustrates a generic case where Eq.~(\ref{eq:covering}) can never be satisfied irrespective of the
"yes/no" configurations. The  middle-right and bottom-left figures,
and more generally when the solid and dashed lines do not cross, 
illustrate the necessary configurations to allow for Eq.~(\ref{eq:covering}), yet one still needs to examine the "yes/no" configurations for sufficiency. Finally the bottom-right figure where the two branches of the parabola cut through the domain, is another configuration for which Eq.~(\ref{eq:covering}) is impossible. Finally, when $\alpha^*_{AB} \notin [0,1]$, not represented on Fig.~\ref{fig:alfa-alfab}, the entire 
$(\alpha_{AB}, \alpha_{A})$ domain is contained either in the
non-empty subset of $\{s_{yes}^I, \emptyset\}$ or in the non-empty subset of $\{\emptyset, s_{no}^I\}$. In the latter case it is required to be entirely contained in the "yes" region determined by
the parabola. 

Putting everything together, the problem becomes equivalent to solving for the following complementary  conditions: 
\begin{itemize}
\item[(i)] Eq.~(\ref{eq:p-c-BFB-4})-(I) valid $\forall \alpha_{AB} \in [0, 1]$, partition $\{s_{yes}^I, \emptyset\}$
\item[(ii)] $\displaystyle \alpha^*_{AB}= - \frac{\lambda_{AB}^{(1)}}{\lambda_{AB}^{(2)}}$, 

Eq.~(\ref{eq:p-c-BFB-4})-(I) 
valid only $\forall \alpha_{AB} \in [0, \alpha^*_{AB}]$,

Eq.~(\ref{eq:p-c-BFB-4})-(II) should be
valid $\forall \alpha_{AB} \in [\alpha^*_{AB},1]$, i.e. $s_{no}^I \subset s_{yes}^{II}$
\item[(iii)] $\displaystyle \alpha^*_{AB}= - \frac{\lambda_{AB}^{(1)}}{\lambda_{AB}^{(2)}}$, 

Eq.~(\ref{eq:p-c-BFB-4})-(I) 
valid only $\forall \alpha_{AB} \in [\alpha^*_{AB},1]$,

Eq.~(\ref{eq:p-c-BFB-4})-(II) should be
valid $\forall \alpha_{AB} \in [0,\alpha^*_{AB}]$, i.e. $s_{no}^I \subset s_{yes}^{II}$

\item[(iv)] Eq.~(\ref{eq:p-c-BFB-4})-(I) false $\forall \alpha_{AB} \in [0, 1]$, partition $\{\emptyset, s_{no}^I\}$,

Eq.~(\ref{eq:p-c-BFB-4})-(II) should be valid $\forall \alpha_{AB} \in [0, 1]$, partition $\{s_{yes}^{II}, \emptyset\}$
\end{itemize}
where the numbering corresponds to that of Fig.~\ref{fig:resolved-p-c-BFB}. 

We can now derive in a fully analytical way  the resolved form of 
Eqs.~(\ref{eq:p-c-BFB-3}, \ref{eq:p-c-BFB-4}), or equivalently
of Eqs.~(\ref{eq:p-c-BFB-4}, \ref{eq:p-c-BFB-3-resolved}) 
  in conjunction with $\lambda_H > 0 \; \land \; \lambda_B > 0$.
The NAS conditions thus obtained on the $\lambda$'s have no 
residual dependence on $\alpha_{AB}$ and $\alpha_{A}$.
To retrieve these NAS conditions
we followed step-by-step the partitions described above and analyzed the non-monotonic dependence on $\alpha_{AB}$ in
Eq.~(\ref{eq:p-c-BFB-4})-(II) when applicable.

The details are very technical and will not be described here. 
We give the final result in Fig.~\ref{fig:resolved-p-c-BFB} where
we have defined the following Boolean expressions 
${}^{\ref{foot:Bool}}$: 
\begin{eqnarray}
&&{\mathcal B}_3 \Leftrightarrow  (2\lambda_{A}^{(1)} + \lambda_{A}^{(2)})\lambda_B > 12(\lambda_{AB}^{(1)} + \lambda_{AB}^{(2)})^2, 
\label{eq:Bool1}\\
&&{\mathcal B}_4 \Leftrightarrow  3(2\lambda_{A}^{(1)} + \lambda_{A}^{(2)}) {\lambda_{AB}^{(2)}}^2 + 2\lambda_{A}^{(2)}(\lambda_{A}^{(1)} + \lambda_{A}^{(2)})\lambda_B < 
12\lambda_{A}^{(2)}\lambda_{AB}^{(1)}(\lambda_{AB}^{(1)} + \lambda_{AB}^{(2)}) \nonumber\\
&&~~~~~~~~\, \lor \,  6\lambda_{AB}^{(2)}(\lambda_{AB}^{(1)} + \lambda_{AB}^{(2)}) + \lambda_{A}^{(2)}\lambda_B > 0, \\
&&{\mathcal B}_5 \Leftrightarrow 2(\lambda_{A}^{(1)} + \lambda_{A}^{(2)})\lambda_B > 3(2\lambda_{AB}^{(1)} + \lambda_{AB}^{(2)})^2, 
\label{eq:Bool3} \\
&&{\mathcal B}_6 \Leftrightarrow (2\lambda_{A}^{(1)} + \lambda_{A}^{(2)}) {\lambda_{AB}^{(2)}}^2 > 4 \lambda_{A}^{(2)}\lambda_{AB}^{(1)}(\lambda_{AB}^{(1)} + \lambda_{AB}^{(2)}) \, \lor \,
3 {\lambda_{AB}^{(2)}}^2 + \lambda_{A}^{(2)}\lambda_B < 0 \\
&&{\mathcal B}_7 \Leftrightarrow \lambda_B \min\left\{2(\lambda_{A}^{(1)} + \lambda_{A}^{(2)}),(2\lambda_{A}^{(1)} + \lambda_{A}^{(2)})\right\} > 12{\lambda_{AB}^{(1)}}^2  \label{eq:Bool5}.
\end{eqnarray}
 In writing this final form we used occasionally the fact that $\lambda_B >0$
to obtain compact expressions where 
Eqs.~(\ref{eq:p-c-BFB-3-resolved}) are implicitly taken into
account in Eqs.~(\ref{eq:Bool1}, \ref{eq:Bool3},\ref{eq:Bool5}). 

\begin{figure*}[!h]
\captionsetup{justification=raggedright,
singlelinecheck=false}
    \begin{center}
      {\includegraphics[width=0.60\textwidth, 
      height=0.3\textheight, keepaspectratio]{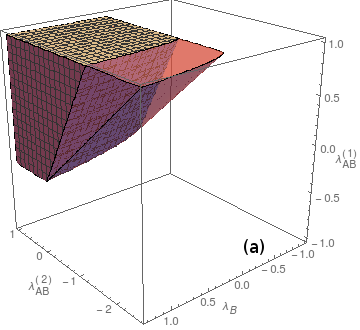}}{\includegraphics[width=0.60\textwidth, 
      height=0.3\textheight, keepaspectratio]{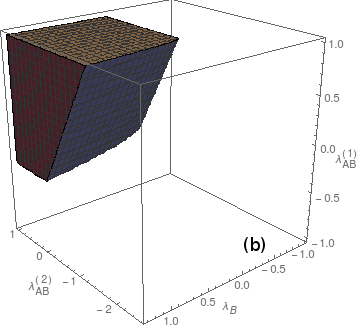}}
      \end{center}
    \begin{center}
      {\includegraphics[width=0.60\textwidth, 
      height=0.3\textheight, keepaspectratio]{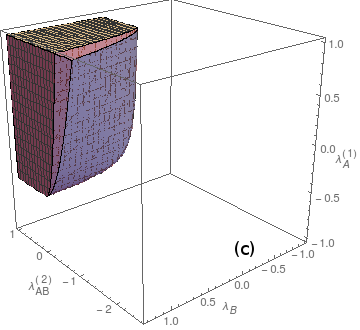}}{\includegraphics[width=0.60\textwidth, 
      height=0.3\textheight, keepaspectratio]{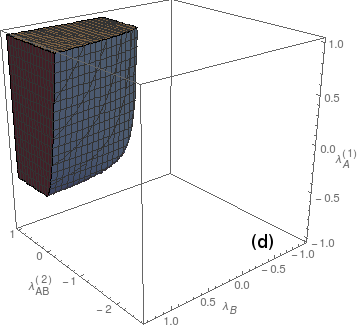}}
      \end{center}
\caption{\label{fig:resolved-scan} \small{Upper figures: 
the allowed domain
in the $\lambda_B, \lambda_{AB}^{(1)},\lambda_{AB}^{(2)}$ space
for $\lambda_{A}^{(1)}=-0.1, \lambda_{A}^{(2)} =1,  \lambda_H=1$;
(a) the brown domain corresponds to enforcing Eqs.~(\ref{eq:p-c-BFB-4}, \ref{eq:p-c-BFB-3-resolved}) for just the three sets of values,
$(\alpha_{AB}, \alpha_A) = (0, \frac12), (\frac12, \frac12), 
(1, \frac12)$; the light pink indicates the increased  
domain when replacing the last set by $(\frac12, 1)$;
(b) exact resolved conditions of Fig.~\ref{fig:resolved-p-c-BFB}.
Lower figures: the allowed domain in the $\lambda_B, \lambda_{A}^{(1)},\lambda_{AB}^{(2)}$ space, for  $\lambda_{AB}^{(1)}=-0.1, \lambda_{A}^{(2)} =1,  \lambda_H=1$; (c) as in (a), (d) as in (b).}}
\end{figure*}

For a cross-check of our results we have performed various 
numerical scans simultaneously on the 
$\lambda$'s, and on $\alpha_A, \alpha_{AB}$ in the domain defined 
by Eqs.~(\ref{eq:dom-alfa-alfab}-I -- \ref{eq:dom-alfa-alfab}-IV).
This amounts to checking the validity of the
conditions Eqs.~(\ref{eq:p-c-BFB-4}, \ref{eq:p-c-BFB-3-resolved}),
and comparing the Boolean output with that of the resolved 
conditions of Fig.~\ref{fig:resolved-p-c-BFB}.\footnote{Throughout the paper we rely
significantly on the Mathematica package \cite{Mathematica} for symbolic and numerical computations as well as for the generation of the plots.} One can take advantage of the fact that $\alpha_{AB}$ and 
$\alpha_{A}$ are not correlated in the square $[0, \frac12] \times [\frac12, 1]$ to replace for this part of the domain, and 
without loss of information, 
$\alpha_{AB}$ and $\alpha_{A}$ by their edge values in
Eqs.~(\ref{eq:p-c-BFB-4})-(I), -(II). The parabola-edged part of the domain (where
$\alpha_{AB} \in [\frac12, 1]$), is more tricky to treat.
If not sufficiently finely meshed, a numerical scan could miss 
some features depending on the configuration of the maximum/
minimum of the parabola. As an illustration we show in
Fig.~\ref{fig:resolved-scan} the allowed $3$D domains,    
for subsets of the $\lambda$ parameters, obtained from the resolved exact conditions of Fig.~\ref{fig:resolved-p-c-BFB} and compare them with the approximate ones obtained from requiring Eqs.~(\ref{eq:p-c-BFB-4}, \ref{eq:p-c-BFB-3-resolved}) to hold for just three sets of benchmark values of $\alpha_{AB}$ and $\alpha_{A}$ lying on the boundary of their allowed domain.
As expected, one of the benchmark sets leads to an approximate domain
 that is much less restrictive (the pink colored regions in Figs.~\ref{fig:resolved-scan}~(a),~(c)) than the exact domain 
shown in Figs.~\ref{fig:resolved-scan}~(b),~(d)). However, one finds that the other benchmark set (the brown colored regions in Figs.~\ref{fig:resolved-scan}~(a),~(c)) leads unexpectedly to an extremely good approximation of the exact domain. Obviously this accidental agreement could not have been guessed without the comparison and is not by itself a cross-check of the validity of the conditions given in Fig.~\ref{fig:resolved-p-c-BFB} \& Eqs~(\ref{eq:Bool1} --\ref{eq:Bool5}). For that we have performed large scans, 
$4 \times 10^6$ points on a regular grid in the $\lambda$-space in the configurations of Fig.~\ref{fig:resolved-scan}, or fixing only $\lambda_H=1$ and taking $1.3 - 2 \times 10^6$ points in the $\lambda$-space with much larger number of benchmark points, $60$ benchmark points within, or $30$ benchmark points on, the boundary of the $(\alpha_{AB}, \alpha_A)$ domain. Counting the hits where the Boolean values of the approximate and exact conditions are equal or different we found in all cases a difference of less than $2\%$ between the approximate and exact conditions. Another significant feature of the check is that the Boolean yield of the difference is found
in $100\%$ of the cases to be "approximate=True, exact=False". Only one hit with the reverse configuration would have meant the exact
conditions are wrong!

In summary, we have derived the NAS conditions for $\mathfrak{a_0} > 0 \, \land \, \mathfrak{c_0} > 0$ 
in a fully analytical resolved form. They are thus {\sl necessary} for the BFB of the general potential given by
Eqs.~(\ref{eq:V-p-c} -- \ref{eq:p-c-4}), and can be safely applied irrespective of the $A$, $B$ and $H$ field 
configurations.\footnote{Note that an alternative approach to obtain these results is to start from 
the third inequality in Eq.~(\ref{eq:a0c0b0p2sqrta0c0}) with
no 'OR' structure rather than from
Eq.~(\ref{eq:p-c-BFB-2}). Its advantage is to avoid the use of partitions and coverings but necessitates
the study of functions with square roots as in Appendix \ref{appendix:B} leading though to more compact conditions. We have checked the agreement of the two approaches. 
The partitions/coverings approach we developped will nevertheless
be unvoidable for the all-field-directions full analytical resolving of the pre-custodial model in the case
$\lambda_{ABH}=0$, not treated in the present paper.
} Further comments on these conditions are deferred to
Sections \ref{sec:peeling} and \ref{sec:disc}.

\begin{minipage}{40em}
\begin{numcases}{\ovalbox{\parbox[t]{8em}{$\lambda_H > 0 \, \land \, \lambda_B > 0$}} \, \land \, }
\begin{minipage}{18em}
\setlength{\fboxrule}{2pt}
\begin{center}
\framebox{\parbox[t]{16em}{
\vspace{-.2cm}
\begin{center} 
\ovalbox{\parbox[t]{12em}{ $\lambda_{AB}^{(1)} > 0 \ \land \ \lambda_{AB}^{(1)} + \lambda_{AB}^{(2)} > 0$}} $\leftarrow$ \text{(i)}\\
{$\mathord{\displaystyle\land}$}\\
\ovalbox{\parbox[t]{14em}{$\displaystyle 2 \lambda_{A}^{(1)} + \lambda_{A}^{(2)} > 0\ \land \ \lambda_{A}^{(1)} + \lambda_{A}^{(2)} > 0$}}
\end{center}}}\\[.2cm]
{$\mathord{\displaystyle\lor}$}\\[-.3cm]
\fbox{\parbox[t]{16em}{

\vspace{-.2cm}
\begin{center}
\ovalbox{\parbox[t]{12em}{$\displaystyle \lambda_{AB}^{(1)} > 0 \ \land \ \lambda_{AB}^{(1)} + \lambda_{AB}^{(2)} < 0$}} $\leftarrow$ \text{(ii)}\\

{$\mathord{\displaystyle\land}$}\\

\ovalbox{\parbox[t]{6em}{$\displaystyle  \lambda_{A}^{(1)} + \lambda_{A}^{(2)} > 0$}}\\

$\displaystyle \land$\\[-.8em]
\ovalbox{\parbox[t]{14em}{
\vspace{-.2cm}
\begin{center}$\displaystyle {\mathcal B}_3$ \\[-.8em]
$\displaystyle \land$\\[-.8em]
$\displaystyle \left(2\lambda_{AB}^{(1)} + \lambda_{AB}^{(2)} > 0  \, \lor \,  {\mathcal B}_4 \land {\mathcal B}_5\right)$ 
\end{center}}}
\end{center}}}\\[.2cm]      
\ $\mathord{\displaystyle\lor}$\\[-.3cm]
\fbox{\parbox[t]{16em}{
\vspace{-.2cm}
\begin{center}
\ovalbox{\parbox[t]{12em}{$\displaystyle \lambda_{AB}^{(1)} < 0 \ \land \ \lambda_{AB}^{(1)} + \lambda_{AB}^{(2)} > 0$}} $\leftarrow$ \text{(iii)}\\
{$\mathord{\displaystyle\land}$}\\[-.3cm]
\ovalbox{\parbox[t]{15em}{
\vspace{-.2cm}
\begin{center}$\displaystyle \left(2 \lambda_{AB}^{(1)} + \lambda_{AB}^{(2)} > 0 \,\lor\, {\mathcal B}_6\right) \,\land\, {\mathcal B}_7$
 \end{center}}} \end{center}}}\\[.2cm]
\ $\mathord{\displaystyle\lor}$\\[-.3cm]
\fbox{\parbox[t]{18em}{
\vspace{-.2cm}
\begin{center}
\ovalbox{\parbox[t]{12em}{$\displaystyle \lambda_{AB}^{(1)} < 0 \ \land \ \lambda_{AB}^{(1)} + \lambda_{AB}^{(2)} < 0$}} $\leftarrow$ \text{(iv)}\\
{$\mathord{\displaystyle\land}$}\\[-.3cm]
\ovalbox{\parbox[t]{17em}{
\vspace{-.2cm}
\begin{center}
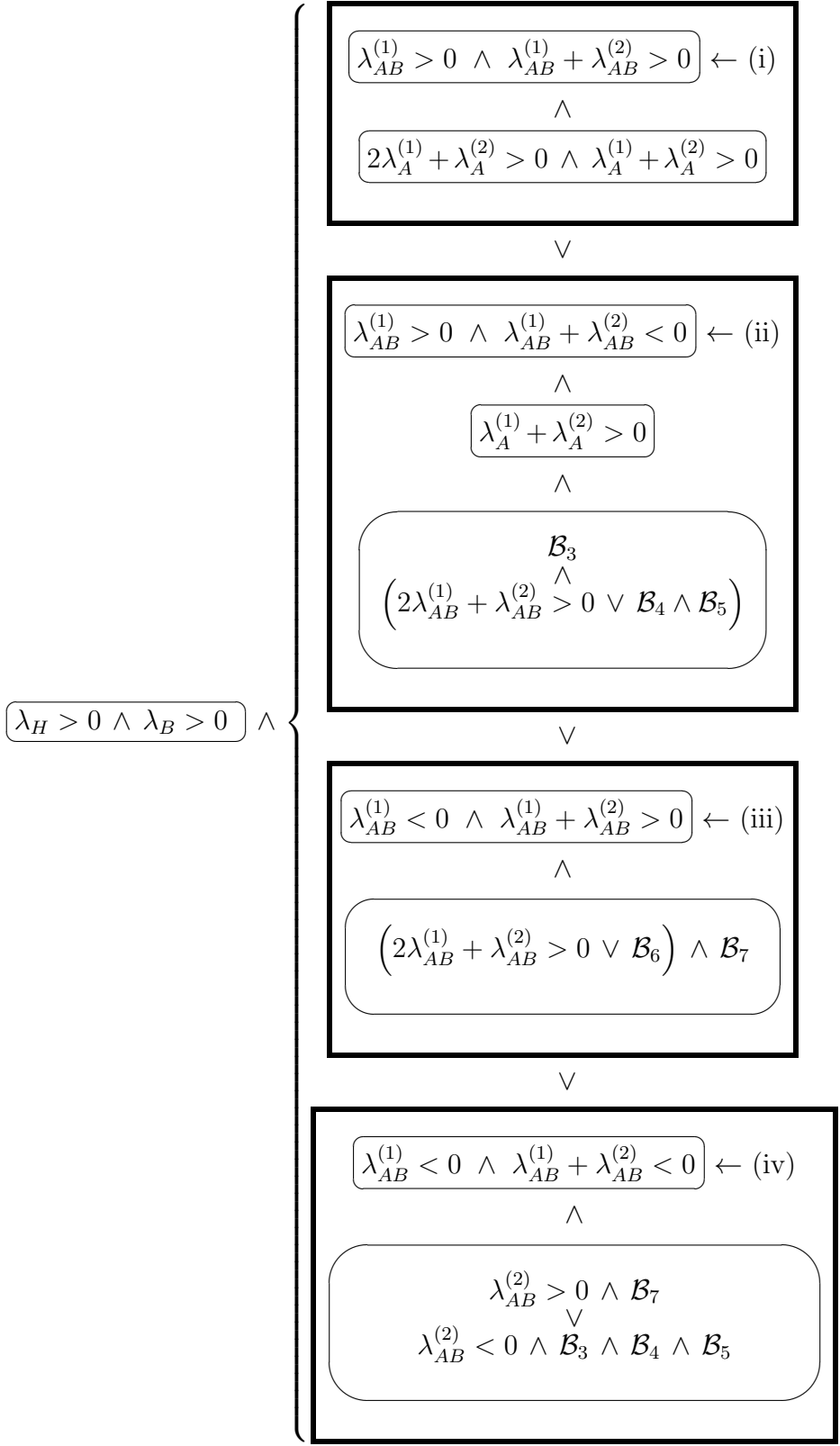
$\displaystyle \lambda_{AB}^{(2)} > 0 \,\land\, {\mathcal B}_7 $\\[-.8em] 
$\displaystyle\lor$\\[-.8em]
$\displaystyle \lambda_{AB}^{(2)} < 0 \,\land\, {\mathcal B}_3 \,\land \, {\mathcal B}_4 \,\land\, {\mathcal B}_5$ \end{center}}} \end{center}}}
\end{center}
\end{minipage}
\nonumber
\end{numcases}
\captionof{figure}{
Boolean flowchart of the fully resolved 
form, i.e. with no dependence on the fields, of the NAS conditions on $\lambda_B, \lambda_H, \lambda_{A}^{(1)}, \lambda_{A}^{(2)}, \lambda_{AB}^{(1)},\lambda_{AB}^{(2)}$ satisfying the inequalities 
given by Eqs.(\ref{eq:p-c-BFB-3}, \ref{eq:p-c-BFB-4}).}
\label{fig:resolved-p-c-BFB}
\end{minipage}

\subsubsection{Partial resolving of $\mathfrak{b_0} > 0 \,\lor  \, 4 \mathfrak{a_0} \mathfrak{c_0} -\mathfrak{b_0^2} > 0$ \label{sec:part-resolv}}

We investigate now Eq.~(\ref{eq:p-c-BFB-2}) that should be valid
$\forall Z, \alpha_A, \alpha_{AB}, \alpha_{AH}, \alpha_{ABH}$ in their allowed domains. (We 
use here the variable $Z$ defined in Eq.~(\ref{eq:defZ}) instead of
$t$, and refer the reader to Section~\ref{sec:b0-AND-4a0c0-b02} for
a discussion on the relevance of $Z$.)
As argued repeatedly in  
Sections~\ref{sec:a0-AND-c0}, \ref{sec:b0-AND-4a0c0-b02} and discussed in detail
in the previous subsection, 
the 'or' structure in Eq.~(\ref{eq:p-c-BFB-2}) implies that the
validity of the inequalities should be required
for all possible coverings of the $(Z, \alpha\text{-parameters})$
space. However, the situation is more complex here than in the 
previous subsection, since  $4 \mathfrak{a_0 c_0} -\mathfrak{b_0^2}$, cf. Eq.~(\ref{eq:c0}), involves simultaneously all four  $\alpha$'s {\sl and} is a complete quartic
polynomial in $Z$. Given the particularly involved NAS conditions
for quartic polynomials, Eqs.~(\ref{eq:conda} - \ref{eq:condd}), 
we do not expect to resolve
completely this case in an explicit form similar to that 
given in Fig.~\ref{fig:resolved-p-c-BFB}. The aim here is to 
proceed as far as possible towards an explicit resolving, then deal 
with the rest through mere numerical scans on the {\sl $\alpha$-parameters}
defined by Eq.~(\ref{eq:potatoid}), including some further refinements to be 
discussed in Sec.~\ref{sec:peeling}. To proceed let us first address the flowchart of the 
overall logic. This is sketched in Fig.~\ref{fig:partial-resolved-p-c-BFB}, together with the following definitions:

\begin{itemize}
\item ${\cal B}_8$ denotes the NAS conditions for $\mathfrak{b_0}$ to always 
have a constant sign,
\item ${\cal B}_{9}^{(a,b)}$ denotes the NAS conditions for $4 \mathfrak{a_0 c_0} -\mathfrak{b_0^2}$
to be positive when $Z$ is in the interval $(a,b)$
and the {\sl $\alpha$-parameters} satisfying Eq.~(\ref{eq:potatoid}).
\end{itemize}

The strategy underlying this flowchart is similar to the one adopted in the previous subsection (which the reader is referred
to for definitions and notations), and should be clear by now.
The upper left box of Fig.~\ref{fig:partial-resolved-p-c-BFB} corresponds to the $\lambda$--space points for which 
$\mathfrak{b_0} > 0$ defines two trivial minimal partitions, $\{s_{yes}^I, \emptyset\}$ or $\{\emptyset, s_{no}^I\}$, corresponding respectively to $\lambda_{BH}>0$ and $\lambda_{BH}<0$, while the lower left box corresponds to the $\lambda$--space points where $\mathfrak{b_0} > 0$ defines a generic minimal partition $\{s_{yes}^I, s_{no}^I\}$. The boxes to the right indicate the Boolean structure
including the minimal generic partition $\{s_{yes}^{II}, s_{no}^{II}\}$ defined by $4 \mathfrak{a_0} \mathfrak{c_0} -\mathfrak{b_0^2} > 0$ to satisfy Eq.~(\ref{eq:covering}). We now investigate how far the Boolean expressions ${\mathcal B}_{8}$ and
${\mathcal B}_{9}^(...)$ can be resolved analytically.

\begin{minipage}{40em}
\begin{center}
\begin{numcases}{\hspace{1.6cm}
\ovalbox{\parbox[h]{14em}{ \begin{center} constant $\sgn(\mathfrak{b_0})$, $\forall Z, \alpha\text{-params}$ \end{center} }}}
\hspace{-.85cm}
\begin{minipage}{17em}
\setlength{\fboxrule}{2pt}
\begin{center}
\framebox{\parbox[t]{12em}{
\vspace{-.2cm}
\begin{center} 
\ovalbox{\parbox[t]{1em}{ ${\mathcal B}_8$}} \\
{$\mathord{\displaystyle\land}$}\\
\ovalbox{\parbox[t]{10em}{
\begin{center}$\displaystyle \lambda_{BH} > 0  \lor   
{\mathcal B}_{9}^{\scriptscriptstyle (-\infty,+\infty)}$
\end{center}
}}
\end{center}}}\\[.2cm]
%
\end{center} 
\end{minipage}
\nonumber
\end{numcases}
\end{center}
\hspace{11.6cm} $\mathord{\displaystyle\lor}$\\[-.3cm]
\begin{numcases}{\ovalbox{\parbox[h]{17em}{\begin{center}$\exists Z, \alpha\text{-params}$ giving varying $\sgn(\mathfrak{b_0})$ 
\end{center}}} }
\begin{minipage}{18em} 
\setlength{\fboxrule}{2pt}   
\fbox{\parbox[t]{16em}{
\vspace{-.2cm}
\begin{center}
\ovalbox{\parbox[t]{2em}{$\displaystyle \neg {\cal B}_8$}} \\
{$\mathord{\displaystyle\land}$}\\[-.3cm]
\ovalbox{\parbox[t]{14em}{
\vspace{-.2cm}
\begin{center}
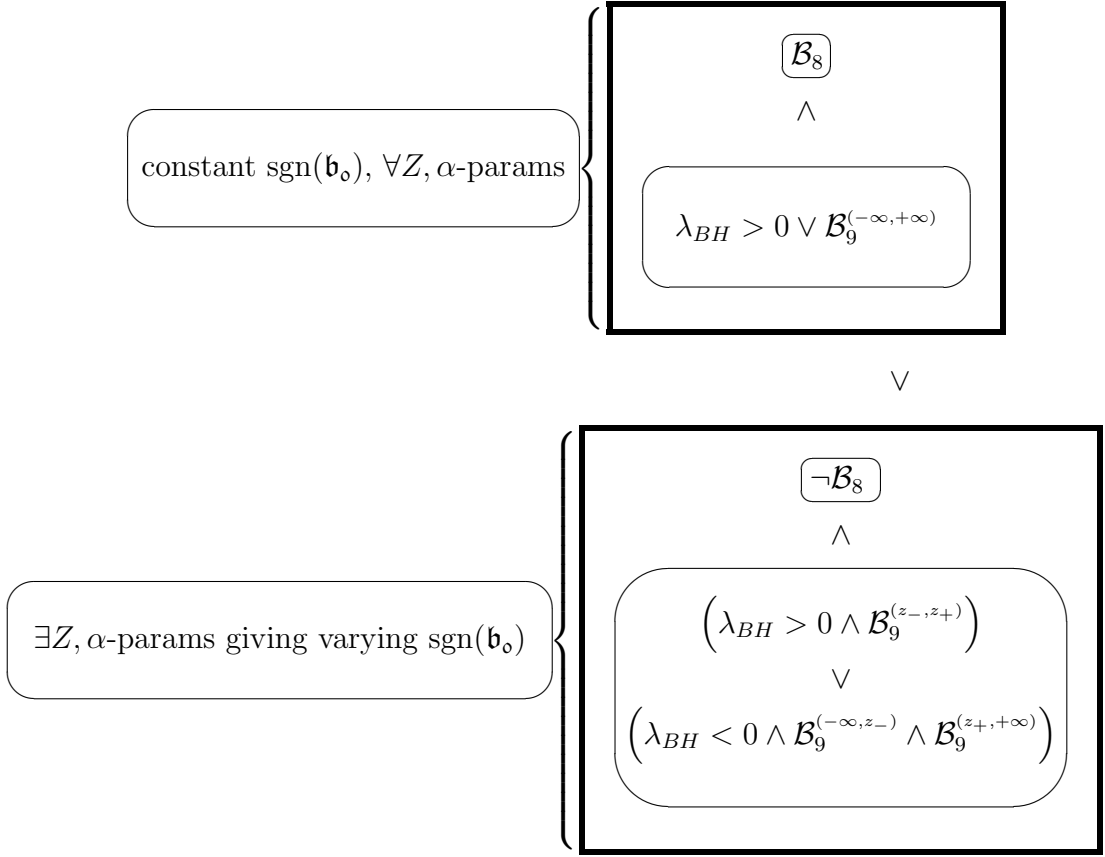

$\displaystyle \left(\lambda_{BH} > 0  \land  {\mathcal B}_{9}^{\scriptscriptstyle(z_{-},z_{+})}
\right)$ \\ {$\mathord{\displaystyle\lor}$}\\ $\left(\lambda_{BH} < 0  \land  
 {\mathcal B}_{9}^{\scriptscriptstyle(-\infty, z_{-})} \land  
 {\mathcal B}_{9}^{\scriptscriptstyle(z_{+},+\infty)} \right)$
 \end{center}
 }} \end{center}}}
\end{minipage}
\nonumber
\end{numcases}
\captionof{figure}{Boolean flowchart${}^{\ref{foot:Bool}}$ for the resolving of $\mathfrak{b_0} > 0 \,\lor  \, 4 \mathfrak{a_0} \mathfrak{c_0} -\mathfrak{b_0^2} > 0$. $z_\pm$ denote the two real-valued roots of $\mathfrak{b_0}(Z)$ when they exist. See text for the definitions of ${\cal B}_8$ and
${\cal B}_9$.}
\label{fig:partial-resolved-p-c-BFB}
\end{minipage}

\vspace{1cm}
\noindent
$\bullet {\cal B}_8$:~Viewing $\mathfrak{b_0}$, Eq.~(\ref{eq:c0}), as a quadratic
polynomial in $Z$, we denote by $z_\pm$ its two roots. Thus ${\cal B}_8$ corresponds to the NAS condition 
for which $z_\pm$ are not real-valued, that is to requiring the discriminant of this polynomial to
be negative, 
\begin{equation}
{\cal B}_8\equiv(\alpha_{ABH} \lambda_{ABH})^2 - 8 (\lambda_{AH}^{(1)} + \alpha_{AH} \lambda_{AH}^{(2)}) \lambda_{BH} \le 0
\label{eq:Bool8-0}
\end{equation}
{\sl for all $\alpha_{AH}, \alpha_{ABH}$ in the domain given by Eqs.~(\ref{eq:V}, \ref{eq:VI})}. Taking into account
the correlations at the boundary of this domain one can obtain condition ${\cal B}_8$  in a fully 
resolved analytical form.
After some non-trivial Boolean simplifications we find, 
\begin{equation}
 \left\{(\lambda_{ABH})^2 \le 4 (\lambda_{AH}^{(1)} + \lambda_{AH}^{(2)}) \lambda_{BH} \ \land \ 
   \left((\lambda_{ABH})^2 \ge 4 \lambda_{AH}^{(2)} \lambda_{BH} \ \lor \ \lambda_{AH}^{(1)} \lambda_{BH} \ge 0 \right) \right\}
  \Leftrightarrow {\cal B}_8 .
   \label{eq:Bool8}
\end{equation}
Clearly then, the NAS conditions for the sufficient condition $\mathfrak{b_0} >0$ read, see Fig.~\ref{fig:partial-resolved-p-c-BFB},
\begin{equation}
{\cal B}_8 \, \land \, \lambda_{BH}>0 . 
\end{equation}
However, as will be discussed later on in Sec.~\ref{sec:peeling}, the condition on the left-hand side of Eq.~(\ref{eq:Bool8}) is in fact
only sufficient to yield ${\cal B}_8 $.\\

\noindent
$\bullet {\cal B}_9^{\scriptscriptstyle (-\infty,+\infty)}$:~To obtain ${\cal B}_9^{\scriptscriptstyle (-\infty,+\infty)}$
we consider $4 \mathfrak{a_0 c_0} -\mathfrak{b_0^2}$ as a quartic
polynomial in $Z$ and thus require all the conditions given by
Eqs.~(\ref{eq:conda} - \ref{eq:condd}). The coefficients
$a_{i=0,...,4}$ are straightforwardly read from the combination 
$4 \mathfrak{a_0 c_0} -\mathfrak{b_0^2}$ upon use of Eqs.~(\ref{eq:c0}, \ref{eq:defZ}):
\begin{equation}
\begin{aligned}
&a_0 =  \gamma_0 -\delta_0^2 , a_1 = -2 \delta_0 \delta_1, a_2 = \gamma_1 -\delta_1^2 - 2 \delta_0 \delta_2 , a_3 = -2 \delta_1 \delta_2, a_4 =  \gamma_2 -\delta_2^2 , \label{eq:acoefs}\\
\end{aligned}
\end{equation}
where
\begin{equation}
\begin{aligned}
\delta_0 = \lambda_{AH}^{(1)} + \alpha_{AH} \lambda_{AH}^{(2)},\ &\delta_1 = \frac12 \lambda_{ABH},\ \delta_2 = \frac{\lambda_{BH}}{2 \alpha_{ABH}^2}, \\ \gamma_0 = \frac14 (\lambda_A^{(1)} + \alpha_A \lambda_A^{(2)}) \lambda_H,\  
&\gamma_1 = \frac{(\lambda_{AB}^{(1)} + \alpha_{AB} \lambda_{AB}^{(2)}) \lambda_H}{2 \alpha_{ABH}^2},\ \gamma_2 = \frac{\lambda_B \lambda_H}{24 \alpha_{ABH}^4}. \label{eq:acoefsannex}\\
\end{aligned}
\end{equation}

We provide here
explicitly the resulting first three conditions given by Eqs.~(\ref{eq:conda}):
\begin{eqnarray}
&4\, a_0 = (\lambda_A^{(1)} + \alpha_A \lambda_A^{(2)}) \lambda_H -
4 (\lambda_{AH}^{(1)} + \alpha_{AH} \lambda_{AH}^{(2)})^2 > 0,& 
\label{eq:condF30a1}\\ 
&24  \, \alpha_{ABH}^4 \, a_4 =  \lambda_B \lambda_H -6 \lambda_{BH}^2 > 0,& \label{eq:condF30a2}\\
&16 \, \alpha_{ABH}^4 \, \Delta_0=  \lambda_{ABH}^4\alpha_{ABH}^4   - 4  \lambda_{ABH}^2 (\lambda_H X_{AB} + 4 \lambda_{BH} X_{AH}) \alpha_{ABH}^2& \nonumber \\
&~~~~~+ 4 \, (\lambda_H X_{AB} - 2 \lambda_{BH} X_{AH})^2+ 8 a_0 (\lambda_B \lambda_H -6 \lambda_{BH}^2) > 0,& \label{eq:condF30a3}
\end{eqnarray}
where we defined
\begin{equation}
X_{AK} \equiv \lambda_{AK}^{(1)} + \alpha_{AK} \lambda_{AK}^{(2)}, \, (K=B, H).
\end{equation}
Condition (\ref{eq:condF30a1}) can be readily resolved: Being linear in $\alpha_A$, one requires it to hold simultaneously on
the upper and lower boundary lines of the $\alpha_A$ domain given by Eq.~(\ref{eq:dom-alfa-alfah2}). The resulting conditions depend only on $\alpha_{AH}$ quadratically and can be studied straightforwardly taking into account Eq.~(\ref{eq:dom-alfa-alfah1}). After several Boolean simplifications we find the following resolved form of Eq.~(\ref{eq:condF30a1}), adding also 
Eq.~(\ref{eq:condF30a2}),
\begin{eqnarray}
{\cal B}_9^{\scriptscriptstyle (-\infty,+\infty)} \supset &\lambda_{B} \lambda_H > 6 \lambda_{BH}^2& \label{eq:condF30a2bis} \\
 &\land& \nonumber \\
& (\lambda_{A}^{(1)} + \lambda_{A}^{(2)}) \lambda_H > 4 \max\left\{(\lambda_{AH}^{(1)})^2, (\lambda_{AH}^{(1)} + \lambda_{AH}^{(2)})^2\right\} & \nonumber \\
 &\land&  \label{eq:condF30a1bis}\\
 & \left(\lambda_{A}^{(2)} \lambda_H < 2 (\lambda_{AH}^{(2)})^2  \ \lor \  
      \lambda_{A}^{(2)} \lambda_H < 4 \max\left\{-\lambda_{AH}^{(1)} \lambda_{AH}^{(2)}, \lambda_{AH}^{(2)} (\lambda_{AH}^{(1)} + \lambda_{AH}^{(2)})\right \} \right.& \nonumber \\
  &    \ \lor \ \left. \lambda_{A}^{(2)} (2 \lambda_{A}^{(1)} + \lambda_{A}^{(2)}) \lambda_H > 
      4 ((\lambda_{A}^{(1)} + \lambda_{A}^{(2)}) (\lambda_{AH}^{(2)})^2 + 2 \lambda_{A}^{(2)} \lambda_{AH}^{(1)} (\lambda_{AH}^{(1)} + \lambda_{AH}^{(2)}))\right). & \nonumber
\end{eqnarray}

Condition (\ref{eq:condF30a3}) appears much less amenable to a 
resolved form as it involves all four {\sl $\alpha$-parameters}
simultaneously. One can however still resolve it partially but this will not be pursued
further here.\footnote{For 
instance, since it is biquadratic in $\alpha_{ABH}$ with a positive definite coefficient of $\alpha_{ABH}^4$, a {\sl sufficient} 
condition is then a negative discriminant. The latter has a simple 
form depending linearly on
$\alpha_A, \alpha_{AB}$ and quadratically on $\alpha_{AH}$.}
The remaining conditions corresponding to Eqs.~(\ref{eq:condb},\ref{eq:condc},\ref{eq:condd})
will be treated numerically. \\

\noindent
$\bullet$ ${\mathcal B}_{9}^{\scriptscriptstyle(-\infty, z_{-})}, 
 {\mathcal B}_{9}^{\scriptscriptstyle(z_{+},+\infty)}$:~To obtain these conditions one again considers $4 \mathfrak{a_0 c_0} -\mathfrak{b_0^2}$ as a quartic polynomial in $Z$. However, now the positivity is not required on all
 ${Z\!\in\!(-\infty,  +\infty)}$ and one needs to rely on the results derived in Appendix \ref{appendix:generalquarticplus}.
 Since the latter hold for $[0, +\infty)$, we first map one-to-one the domains $(-\infty, z_{-}]$ 
 and $[z_{+},+\infty)$ on $[0, +\infty)$ through the two changes of variable 
 \begin{equation}
 Z= z_{-} -\xi \ \text{and} \ Z= z_{+} +\xi \label{eq:varchange1}
 \end{equation}
  respectively, with $\xi \in [0, +\infty)$, then search for the conditions on the quartic polynomial in $\xi$ 
  satisfying criterion (\ref{statmnt:Rplus}).
 We note, however, two simplifcations due to the linear changes of variable: $a_4$, the coefficient of $Z^4$ given by Eq.~(\ref{eq:condF30a2}), is the same
  as that of $\xi^4$. It follows that the {\sl necessary} condition Eq.~(\ref{eq:condF30a2bis}) remains valid. 
  On the other hand, the coefficients $a_0$ are modified with respect to Eq.~(\ref{eq:condF30a1}) to, respectively, $a_0^-$ and
  $a_0^+$ given by:
  \begin{equation}
  a_0^{\mp}=\lambda_H \left(6 (\lambda_A^{(1)} + \alpha_A \lambda_A^{(2)}) + 12 (\lambda_{AB}^{(1)} + \alpha_{AB} \lambda_{AB}^{(2)}) z_{\mp}^2 + \lambda_B z_{\mp}^4 \right). \label{eq:a0plusminus}
  \end{equation}
  Interestingly, one can show that when combined with Eqs.~(\ref{eq:p-c-BFB-3}, \ref{eq:p-c-BFB-4}), the {\sl necessary} constraints
  $a_0^{\mp} >0$ as dictated by the first of Eqs.~(\ref{eq:conda}), will {\sl always} be satisfied by Eq.~(\ref{eq:a0plusminus}) irrespective of the values of 
  $z_{\mp}$! Indeed, given Eq.~(\ref{eq:p-c-BFB-3}),
  when Eq.~(\ref{eq:p-c-BFB-4})-(I) is satisfied then $a_0^{\mp} >0$ follows trivially, and when Eq.~(\ref{eq:p-c-BFB-4})-(II) is satisfied
  then $a_0^{\mp}$, taken as a quadratic equation in $z_{\mp}^2$, has no real-valued roots and thus again always positive.\\
  
  \noindent
$\bullet$ ${\mathcal B}_{9}^{\scriptscriptstyle(z_{-},z_{+})}$:~In this case a nonlinear change of variable 
\begin{equation}
\displaystyle Z=z_{-} + (z_{+} - z_{-}) \frac{\xi}{1 + \xi} \label{eq:varchange2}
\end{equation}
is used with $\xi \in [0, +\infty)$ before applying criterion (\ref{statmnt:Rplus}). 
Here too a simplication occurs for $a_0$ and $a_4$ after the change of variable. Up to a global positive definite denominator, 
they are expressed in terms of Eq.~(\ref{eq:a0plusminus}):
\begin{equation}
\begin{aligned}
a_0=a_0^{-},\\
a_4=a_0^{+},
\end{aligned}
\end{equation}
and are thus {\sl always} positive when combined with Eqs.~(\ref{eq:p-c-BFB-3}, \ref{eq:p-c-BFB-4}), as explained above.

To summarize, we have identified a subset of analytically 
resolved {\sl necessary} conditions in the various branches of Fig.~\ref{fig:partial-resolved-p-c-BFB} flowchart. One now should
combine these conditions with the other analytically resolved conditions given in Fig.~\ref{fig:resolved-p-c-BFB} and Eqs.~(\ref{eq:Bool1} - \ref{eq:Bool5}) and possibly also with those
given by Eqs.~(\ref{eq:BFB} -- \ref{eq:BFBnew5prime}).
This allows a quick determination of 
{\sl necessary} domains in the $\lambda$--space. Then adding the remaining {\sl necessary} conditions that can be
treated through numerical scans on the {\sl $\alpha$-parameters},
one delineates the NAS BFB conditions.
However, before doing so in Sec.~\ref{sec:disc}, we need to reexamine first the BFB conditions of the more constrained Georgi-Machacek model, as this will have some bearing on the general case.

\subsection{The Georgi-Machacek BFB conditions \label{sec:GM-BFB}}

In \cite{Hartling:2014zca} the authors provided a detailed study 
of the properties of the potential relying on a generalization
of the parameterization used in \cite{Arhrib:2011uy}. They 
identified the two parameters

\begin{eqnarray}
\hat\omega &=& \frac{Tr(\Phi^\dag \tau^a \Phi \tau^b) Tr(X^\dag t^a X t^b)}{Tr(\Phi^\dag \Phi) Tr(X^\dag X)}, \label{eq:omegaGM}\\
\hat\zeta &=& \frac{Tr(X^\dag X X^\dag X)}{(Tr(X^\dag X))^2},
\label{eq:zetaGM}
\end{eqnarray}
relevant to the study of the BFB conditions, 
writing $V^{(4)}_{\text{G-M}}$ in the form
\begin{equation}
V^{(4)}_{\text{G-M}}= \hat r^4\cos^4 \hat\gamma
\left( \hat\lambda_1 + (\hat\lambda_2 - \hat\omega \hat\lambda_5) \tan^2 \hat\gamma + (\hat\zeta \hat\lambda_3 + \hat\lambda_4) \tan^4 \hat\gamma\right) , \label{eq:G-M-4prime}
\end{equation}
with
\begin{eqnarray}
\hat r^2 &\equiv& Tr(\Phi^\dagger \Phi) + Tr(X^\dagger X), \\
\tan^2 \hat\gamma &\equiv& \frac{Tr(X^\dagger X)}{Tr(\Phi^\dagger \Phi)} .
\end{eqnarray}
Noting that $Tr(\Phi^\dagger \Phi) = 2 H^\dagger H$ and
$Tr(X^\dagger X) = 4 Tr(A A^\dagger)  + 2 Tr(B^2)$ one can relate
$\hat r$ and $\tan \hat\gamma$ to the parameters defined in
Eqs.~(\ref{eq:paramH} -- \ref{eq:paramr}) to obtain,
\begin{equation}
\tan^2 \hat{\gamma} = (1 + \cos^2 b) \ \tan^2a  , \; {\rm and} \; 
\hat r^2 \cos^2 \hat\gamma = 2 r^2 \cos^2a. 
\end{equation}
Then equating $V^{(4)}_{\text{G-M}}$, 
Eq.~(\ref{eq:G-M-4prime}), with $V_{\text{p-c}}^{(4)}$,  
Eq.~(\ref{eq:p-c-4}), and taking into account the above relations 
and Eqs.~(\ref{eq:correspond},
\ref{eq:correlate}),  one identifies $\hat\omega$ and $\hat\zeta$ as 
the coefficients of $-\hat \lambda_5 \tan^2\hat\gamma$ and
$\hat \lambda_3 \tan^4\hat\gamma$ which allows to relate 
them to the parameters defined in the pre-custodial case,
Eqs.~(\ref{eq:T} -- \ref{eq:def-alphas2}),  
as follows: 
\begin{eqnarray}
\hat\omega &=& -\frac{1 - 2 \alpha_{AH} - \sqrt{2} \, \alpha_{ABH} \, t}{2 (2 + t^2)}, \label{eq:omega-alphas}\\
\hat\zeta &=& \frac{6 - 4 \, \alpha_A + 4 \alpha_{AB} \, t^2 + t^4}{(2 + t^2)^2} . \label{eq:zeta-alphas}
\end{eqnarray}
As a cross-check of the validity of these relations, one can indeed retrieve from the fact that $t\in[0, +\infty)$ and the exact knowledge of the two domains given by Eqs.~(\ref{eq:dom-alfa-alfab}-I -- \ref{eq:dom-alfa-alfab}-IV) and Eqs.~(\ref{eq:V}, \ref{eq:VI}),  that $\hat\omega \in[-\frac14, \frac12]$ and
$\hat\zeta \in [\frac13, 1]$ as already found in \cite{Hartling:2014zca}.

\begin{figure*}[!h]
\captionsetup{justification=raggedright,
singlelinecheck=false}
    \begin{center}
      {\includegraphics[width=0.60\textwidth, 
      height=0.3\textheight, keepaspectratio]{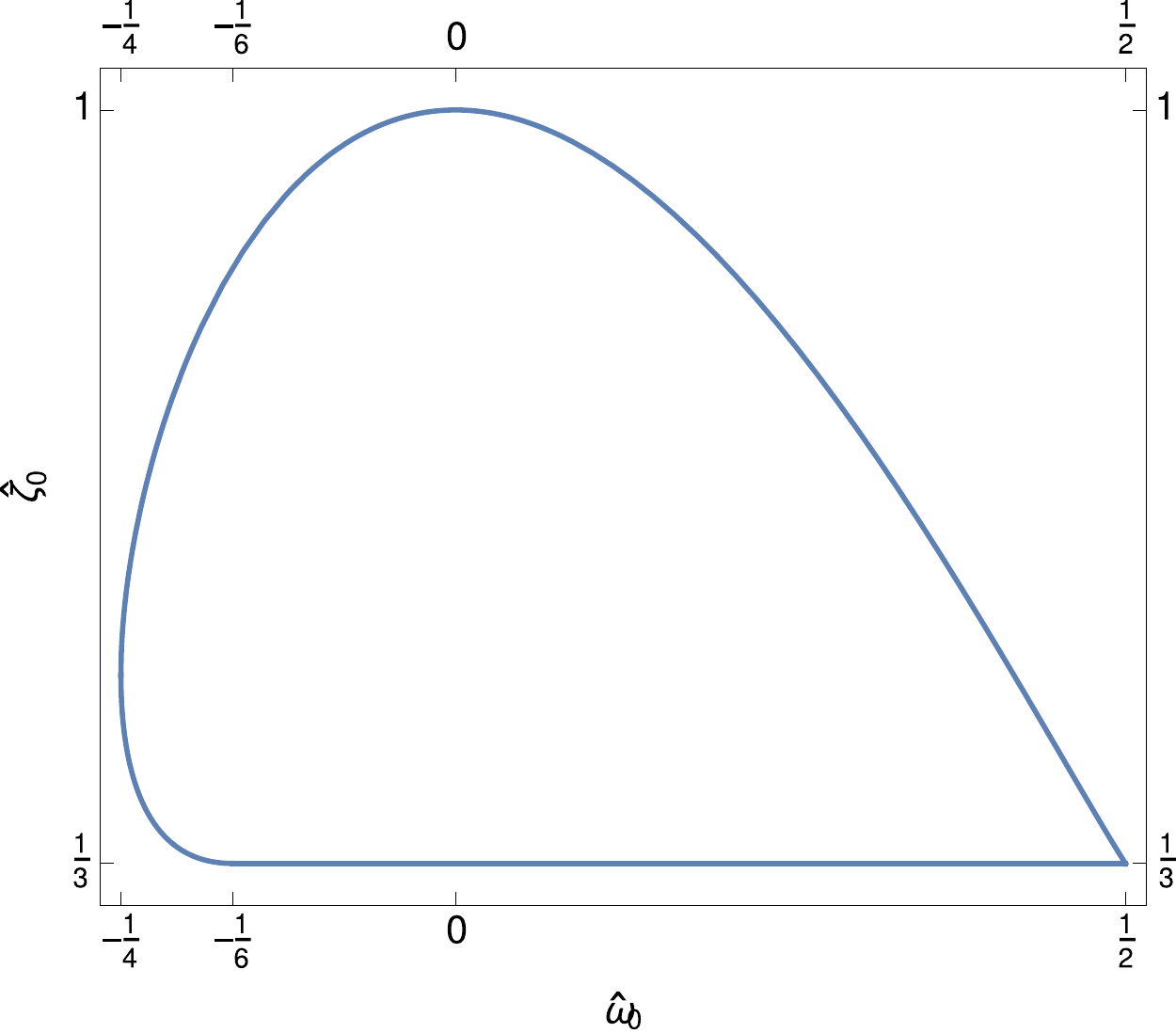}}
      \end{center}
\caption{\label{fig:omega-zeta} \small{The boundary in the $(\hat\omega_0, \hat\zeta_0)$ plane delimiting the allowed inner domain,  in the limit $u=0$. This agrees 
with reference \cite{Hartling:2014zca}. }}
\end{figure*}

The allowed domain in the $(\hat\omega, \hat\zeta)$ plane has been given 
in \cite{Hartling:2014zca}. This was done stating
that the boundary of the domain is obtained from the 
real valued components of the neutral field directions, that 
is keeping
only $\operatorname{Re} \chi^0$ and $\xi^0$ and zeroing all the others in 
Eqs.~(\ref{eq:omegaGM}, \ref{eq:zetaGM}). However, no justification
was given for this statement. The aim of the present section 
is to provide an explicit proof for the 
equation of the boundary 
of the $(\hat\omega, \hat\zeta)$ 
domain based on the symmetries of $V_{\text{G-M}}$. We choose to
use $SU(2)_R$  to rotate away the lower as well as
the imaginary part of the upper components of $H$, so that
\begin{equation}
\frac{Tr(\Phi^\dag \tau^a \Phi \tau^b)}{Tr(\Phi^\dag \Phi)} = \frac14 \delta_{a b},
\end{equation} 
(note that ref.~\cite{Hartling:2014zca} used $SU(2)_L$ instead), 
and use $SU(2)_L$ to rotate away for instance $\chi^{++}$ and 
the imaginary part of  $\chi^{+}$, bringing the bi-triplet X in the form

\begin{equation}
X = \left(
\begin{array}{ccc}
\chi^{0 *} & \xi^+ & 0\\
-u & \xi^0& u \\
0 & -\xi^{+ *}& \chi^0\\
\end{array}
\right)  \label{eq:Xchoice}
\end{equation}
\noindent
where $u (\equiv \operatorname{Re} \chi^+)$ denotes a real-valued scalar field.\footnote{One could be tempted to zero, 
on top of $\chi^{++}$, the (real-valued) $\xi^0$ entry rather than $\operatorname{Im} \chi^+$.
However one can show that this is not possible through a non infinitesimal
$SU(2)$ rotation. More generally, one cannot zero more than
two entries of $X$ through $SU(2)_L \times SU(2)_R$ rotations.} With this choice of gauge $\hat\omega$ and $\hat\zeta$ take the following form
 \begin{eqnarray}
\hat\omega &=& \frac14 \, \left(2 \sqrt{2} \cos \theta_0 \, \cos(\arg(\chi^0)) + \sin \theta_0 \right) \, \sin \theta_0 \,\sin^2 \theta_+ \label{eq:omega0}\\
&& + \frac12 \cos(\arg(\xi^+)) \, \cos\theta_+ \, \cot\theta_u +
{\cal O}(\cot^2 \theta_u)  \label{eq:omega1} \, ,\\
\hat\zeta &=& 1 - \sin^2 \theta_0 \sin^2 \theta_+ \left(1  + \frac14 (1 + 3 \cos 2 \theta_0)  \sin^2 \theta_+ \right)
\label{eq:zeta0} \\
&& -\sqrt{2} \,\cos(\arg(\xi^+) +\arg(\chi^0)) \, 
\cos\theta_+ \, \sin^2\theta_+ \,\sin 2 \theta_0 \cot\theta_u\,
+{\cal O}(\cot^2 \theta_u) \, , \label{eq:zeta1}
\end{eqnarray}
where we defined the polar angles by
\begin{eqnarray}
u &=& R \cos \theta_u, \\
|\xi^+| &=& R \cos \theta_+ \, \sin \theta_u, 
\label{eq:defteta+}\\
|\chi^0| &=& R \sin \theta_+ \, \sin \theta_0 \, \sin \theta_u, \\
\xi^0 &=& \sqrt{2} R \sin \theta_+ \, \cos \theta_0 \, \sin \theta_u,\\
R^2&=&\frac12 Tr(X^\dagger X) ,
\end{eqnarray}
with
\begin{eqnarray}
&0 \leq \arg(\chi^0), \arg(\xi^+) \leq 2 \pi , \label{eq:angrange1}\\ 
&0 \leq \theta_0, \theta_+ \leq \frac{\pi}{2}, \label{eq:angrange2}\\
&0 \leq \theta_u \leq \pi \label{eq:angrange3}.
\end{eqnarray} 
Note that due to the invariance 
of $V^{(4)}_{\text{G-M}}$ under $X \to -X$ one can always fix
uniquely either the sign of $\xi^0$ or that of $u$.
In our parameterization $\xi^0 >0$ while $u$ can take
either signs. In Eqs.~(\ref{eq:omega1}, \ref{eq:zeta1}) we kept
for simplicity only linear terms in $u$. We will come back to the 
exact contribution later on. Here we first concentrate on the 
$0^{th}$ order $u$ contributions to $\hat\omega$ and
$\hat\zeta$, i.e. Eqs.~(\ref{eq:omega0}, \ref{eq:zeta0}) which
we dub $\hat\omega_0$ and
$\hat\zeta_0$. In Appendix \ref{appendix:omega-zeta} we give a 
detailed proof for the determination of the boundary in the 
$(\hat\omega_0, \hat\zeta_0)$ domain, i.e. {\sl under the working assumption that $u=0 \, (=\cot\theta_u)$}. We find that this boundary is
 defined by the following upper and lower curves: 
\begin{equation}
\hat{\zeta_0}^{max}(\hat\omega_0)=\frac13 + \frac{2}{27} \left(1 - 2 \hat\omega_0 + 
2 \sqrt{(1- 2 \, \hat\omega_0)(1 + 4 \, \hat\omega_0)}
\right)^2, \; \; \text{for} \, \hat\omega_0 \in  [-\frac14, \frac12] \label{eq:upperbound}
\end{equation}
\begin{numcases}
{\hat{\zeta_0}^{min}(\hat\omega_0)=}\!\!\frac13 + \frac{2}{27}\left(1 - 2\hat\omega_0 - 
2\sqrt{(1- 2 \, \hat\omega_0)(1 + 4 \, \hat\omega_0)}
\right)^2\!\!\!\!, \text{for} \, \hat\omega_0 \!\in \! [-\frac14, -\frac16] \label{eq:lowerbound1} \\
\!\! \frac13, \; \;\; \;\; \;\; \;\; \; \; \;\; \;\; \;\; \;\; \;\; \;\; \;\; \;\; \;\; \;\; \;\; \;\; \;\; \;\; \;\; \;\; \;\; \;\; \;\; \;\; \;\; \;\; \;\; \;\; \;\; \;\; \;\; \;\text{for} \, \hat\omega_0 \!\in \! [-\frac16, \frac12] \label{eq:lowerbound2}
\end{numcases}
This reproduces exactly the boundary given in reference \cite{Hartling:2014zca} as illustrated in Fig.~\ref{fig:omega-zeta}
(note however that we deal with the inverse
function with respect to reference \cite{Hartling:2014zca}). 
As shown
in Appendix \ref{appendix:comments} the condition $\sin^2 \theta_+ =1$, i.e.  $\xi^+=0$, is sufficient and necessary for the determination of the $(\hat\omega_0, \hat\zeta_0)$ boundary.
In particular the necessity of this condition is a 
non-trivial result. From Eq.~(\ref{eq:omega0}) one sees that 
$\sin^2 \theta_+ =0$ could as well have defined a boundary. 
More importantly, the involved dependence on $\sin^2 \theta_+$
in $\hat\zeta_0$, Eq.~(\ref{eq:zeta0}), could in principle lead
to portions of the boundary with $\sin^2 \theta_+ <1$, since
we are interested in the projection on the $(\hat\omega_0, \hat\zeta_0)$ plane. (This was for instance the case for the 
$(\alpha_A, \alpha_{ABH})$ domain studied in Sec.~\ref{subsec:alA-alABH}.)
Moreover, this is not the end of the 
story because the boundary defined by Eqs.~(\ref{eq:upperbound} --
\ref{eq:lowerbound2}) is obtained in the case $u=0$. 
It remains to be seen whether 
$u\neq 0$ would possibly enlarge the allowed domain outside 
this boundary. We turn now to this point. The idea is to consider a subspace of the field space for which the boundary of $(\hat\omega_0, \hat\zeta_0)$ is reached and determine within this subspace the boundary of $(\hat\omega, \hat\zeta)$ allowing
for $u\neq 0$. As discussed above, such a subspace has necessarily  $\xi^+=0$, ($\sin^2 \theta_+ =1$). The bi-triplet
of Eq.~(\ref{eq:Xchoice}) becomes
\begin{equation}
X = \left(
\begin{array}{ccc}
\chi^{0 *} & 0 & 0\\
-u & \xi^0& u \\
0 & 0 & \chi^0\\
\end{array}
\right)  \label{eq:Xchoice1} .
\end{equation}
One then sees from Eqs.~(\ref{eq:omega1}, \ref{eq:zeta1})
that the $1^{st}$ order $u$ contributions
 vanish for any $u$ in this subspace,
indicating that the boundary is indeed unchanged when $u\neq 0$
at least if
$u$ remains sufficiently small. In fact this result remains true
in general beyond the first order as a consequence of an 
accidental symmetry: $Tr(X^\dag X)$ and $Tr(X^\dag t^a X t^a)$ (summation over $a$) are invariant under the substitution
$\chi^+ \leftrightarrow \xi^{+ *}, (\chi^{+ *} \leftrightarrow \xi^{+})$, and $Tr(X^\dag X X^\dag X)$ is invariant under the same
substitution supplemented by $\chi^{++} \leftrightarrow \chi^{++*}$. Thus $\hat\omega$ and $\hat\zeta$ are invariant under these
substitutions, in which case $X$ defined in Eq.~(\ref{eq:Xchoice1})
is replaced by
\begin{equation}
\tilde X = \left(
\begin{array}{ccc}
\chi^{0 *} & u & 0\\
0 & \xi^0& 0 \\
0 & -u & \chi^0\\
\end{array}
\right)  \label{eq:Xchoice2} .
\end{equation}
The key point is that the latter $\tilde X$ has the same form as $X$ given by Eq.~(\ref{eq:Xchoice}) with $u=0$. We are then brought back to the same
configuration that leads to the fact that the boundary is reached
for $\xi^+=0$ and is given by Eqs.~(\ref{eq:upperbound} -- \ref{eq:lowerbound2});
applied to the present case where $X$ is replaced by $\tilde X$ implies similarly that the boundary is reached for $u=0$ and
is given by the same Eqs.~(\ref{eq:upperbound} -- \ref{eq:lowerbound2}). This completes the proof that $u \neq 0$ in Eq.~(\ref{eq:Xchoice}) remains within the boundary obtained for $u=0$. Thus the full boundary in the $(\hat\omega, \hat\zeta)$ plane
is given by Eqs.~(\ref{eq:upperbound} -- \ref{eq:lowerbound2}):
\begin{equation}
\hat{\zeta_0}^{min}(\hat\omega)\leq \hat\zeta \leq \hat{\zeta_0}^{max}(\hat\omega) \label{eq:chips}
\end{equation}
 In the following we will refer to this domain as the $\omega$-$\zeta$--chips. 
  
\section{Peeling the potatoid with the chips \label{sec:peeling}}
As already announced at the end of Sec.~\ref{subsec:potatoid}, the knowledge of the
exact domain of the $\omega$-$\zeta$--chips of the Georgi-Machacek model  
will have a spin-off on the refinement
of the $4$D {\sl $\alpha$-parameters} potatoid in the general pre-custodial
model. That a model with an enlarged symmetry
would backreact on a less symmetric and more general model is
somewhat unusual. It can be understood as follows in the case at hand: The $SU(2)_L \times SU(2)_R$
symmetry of the Georgi-Machacek model has allowed regroup the four 
{\sl $\alpha$-parameters} and the $t$ parameter into just two relevant
parameters $\hat \omega$ and $\hat \zeta$ that are related to the former as given 
by Eqs.~(\ref{eq:omega-alphas}, \ref{eq:zeta-alphas}). However, the equations defining
the $\omega$-$\zeta$--chips, Eqs.~(\ref{eq:chips}, \ref{eq:upperbound} -- \ref{eq:lowerbound2}), were arrived at thanks to the gauge and
global symmetries, as well as to an accidental invariance of the quartic part of the Georgi-Machacek potential (see Sec.~\ref{sec:GM-BFB}
and Appendix~\ref{appendix:omega-zeta});
in this, Eqs.~(\ref{eq:omega-alphas}, \ref{eq:zeta-alphas}) played no role. The latter,  
in conjunction with Eq.~(\ref{eq:chips}), will thus lead to a supplementary correlation among the {\sl $\alpha$-parameters} and $t$ that should be valid
in the general pre-custodial model. It is in that sense that the Georgi-Machacek model informs about the more general model. 
Obviously, this information would have been redundant had we had beforehand a full knowledge of the exact $4$D {\sl $\alpha$-parameters} domain. This is however not the case as pointed out in Sec.~\ref{subsec:potatoid} regarding the $\alpha$-potatoid. 
Hence one can use the above information as a sufficient condition to exclude points in the  $\alpha$-potatoid as follows: Each set of {\sl $\alpha$-parameters} in the $\alpha$-potatoid defines, through Eqs.~(\ref{eq:omega-alphas}, \ref{eq:zeta-alphas}),
a unique trajectory $(\hat \omega_{\alpha}(t), \hat\zeta_{\alpha}(t))$ in the $(\hat \omega, \hat\zeta)$ plane, parameterized by $t \in [0, +\infty)$. {\sl If the trajectory
goes out of the $\omega$-$\zeta$--chips then the corresponding set of {\sl $\alpha$-parameters} values should be excluded.}

\begin{figure*}[h!]
\captionsetup{justification=raggedright,
singlelinecheck=false}
    \begin{center}
    \scalebox{.8}{
{\includegraphics[width=0.60\textwidth, 
height=0.3\textheight, keepaspectratio]{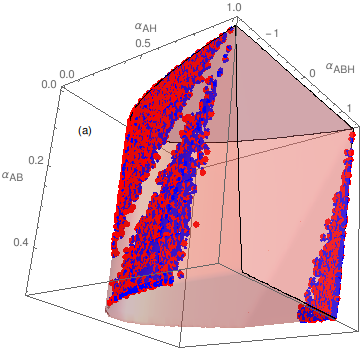}}\hspace{1cm} {\includegraphics[width=0.60\textwidth, 
height=0.3\textheight, keepaspectratio]{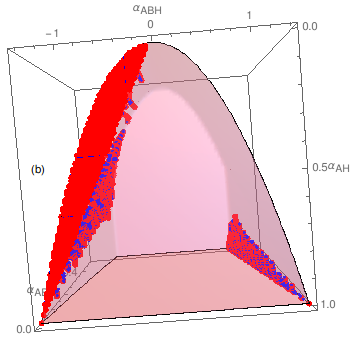}} }
\end{center}
\begin{center}
\scalebox{.8}{
{\includegraphics[width=0.60\textwidth, 
height=0.3\textheight, keepaspectratio]{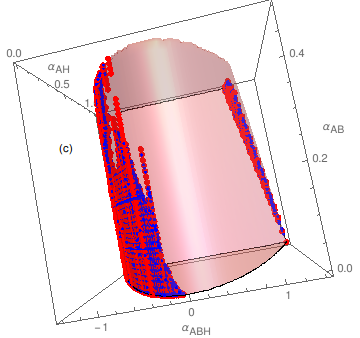}}{\includegraphics[width=0.60\textwidth, 
      height=0.3\textheight, keepaspectratio]{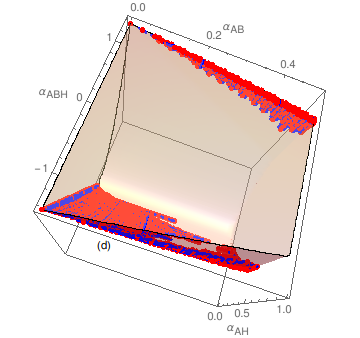}} }
      \end{center}
\caption{The $(\alpha_{AB}, \alpha_{AH}, \alpha_{ABH})$ $3$D-section of the $4$D $\alpha$-potatoid at $\alpha_A=1$, viewed from four different angles. The red and blue dots denote the regions excluded by the $\omega$-$\zeta$--chips. See text for further discussions.}
\label{fig:peeling-fixed-alfA}
\end{figure*}

\begin{figure*}[h!]
\captionsetup{justification=raggedright,
singlelinecheck=false}
    \begin{center}
    \scalebox{.8}{
{\includegraphics[width=0.60\textwidth, 
height=0.3\textheight, keepaspectratio]{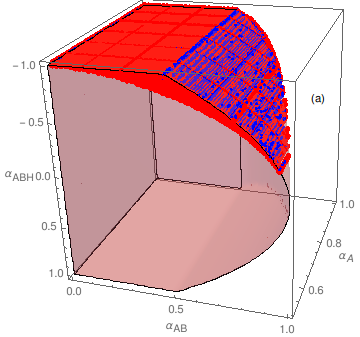}} \ \ {\includegraphics[width=0.63\textwidth, 
height=0.33\textheight, keepaspectratio,angle=0]{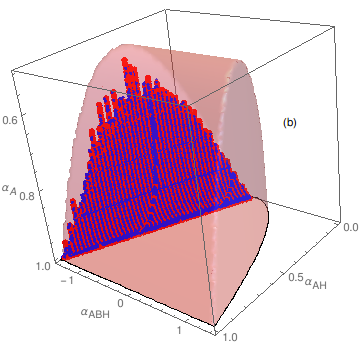}} }
    \end{center}
\caption{Two $3$D-sections of the $4$D $\alpha$-potatoid: (a) $(\alpha_{A}, \alpha_{AB}, \alpha_{ABH})$ at $\alpha_{AH}=\frac12$; 
(b)~$(\alpha_{A}, \alpha_{AH}, \alpha_{ABH})$ at $\alpha_{AB}=0$. The red and blue dots denote the regions excluded by the $\omega$-$\zeta$--chips. See text for further discussions.}
\label{fig:peeling-fixed-alfAH}
\end{figure*}

We show in Figs.~\ref{fig:peeling-fixed-alfA} \&\ref{fig:peeling-fixed-alfAH} numerical scans taking into account this exclusion criterion.\footnote{\label{foot:7} In practice this is achieved by scanning over the four {\sl $\alpha$-parameters} that satisfy Eq.~(\ref{eq:potatoid}) and following each trajectory $(\hat \omega_{\alpha}(t), \hat\zeta_{\alpha}(t))$ scanning over $0\leq u \leq \frac{\pi}{2}$ with $t=\sqrt{2} \tan u$.
Alternatively, one can use the exact $t$-resolved form for Eqs.~(\ref{eq:omega-alphas}, \ref{eq:zeta-alphas}), see Appendix \ref{subsec:resolved}, and scan only on the {\sl $\alpha$-parameters}. We used this latter alternative to cross-check our results.
} 
The red and blue dots delineate the somewhat convoluted regions of the $\alpha$-potatoid that are incompatible with the $\omega$-$\zeta$--chips. As anticipated in Sec.~\ref{subsec:potatoid} and visible from the different viewing angles in Fig.~\ref{fig:peeling-fixed-alfA}, the excluded portions lie only at the boundary of the $\alpha$-potatoid. Note that the domains (shown in pink) in Figs.~\ref{fig:peeling-fixed-alfA} \&\ref{fig:peeling-fixed-alfAH} are $3$D {\sl sections} of the $4$D $\alpha$-potatoid at fixed values of $\alpha_A$ or $\alpha_{AB}$
 or $\alpha_{AH}$ respectively; not to be confused with the $3$D {\sl projections} of the  $\alpha$-potatoid shown on Figs.~\ref{fig:alfaFull} (a), (b) and (d), with 
which it would not be possible to disentangle boundaries unambiguously. Moreover the choices of $\alpha_A=1$, $\alpha_{AH}=\frac12$ and
$\alpha_{AH}=0$ made in Figs.~\ref{fig:peeling-fixed-alfA}~\&\ref{fig:peeling-fixed-alfAH}
entail the inclusion, in the corresponding $3$D-sections, of the full $2$D domains Eq.~(\ref{eq:dom-alfa-alfab}), Fig.~\ref{fig:alfa-alfab}, and Eqs.~(\ref{eq:V}, \ref{eq:VI}), Fig.~\ref{fig:alfah-alfabh}, and  Eqs.~(\ref{eq:dom-alfa-alfah1}, \ref{eq:dom-alfa-alfah2}),
Fig.~\ref{fig:alfa-alfah} respectively.
These scans will thus allow to judge whether the resolved conditions on the
$\lambda$'s given by Fig.~\ref{fig:resolved-p-c-BFB}, or those given by Eq.~(\ref{eq:condF30a1bis}) or by Eq.~(\ref{eq:Bool8}), in which the pairs of 
parameters $(\alpha_A, \alpha_{AB})$, $(\alpha_A, \alpha_{AH})$ and $(\alpha_{AH}, \alpha_{ABH})$ have been eliminated respectively, are indeed necessary and sufficient or not. The answer will be yes for the first two and
no for the last:
\begin{itemize}
\item[--] One sees from Fig.~\ref{fig:peeling-fixed-alfAH} (a)
that for $\alpha_{ABH} \gtrsim -0.27$ there are no exclusions by the $\omega$-$\zeta$--chips. In particular, the $2$D section at 
$\alpha_{ABH}=0$ corresponds to the full $\alpha_A, \alpha_{AB}$ domain of Fig.~\ref{fig:alfa-alfab} which is thus not reduced
by the constraint from the $\omega$-$\zeta$--chips. In fact this result could be easily retrieved once noted that the $\alpha_A, \alpha_{AB}$ domain of Fig.~\ref{fig:alfa-alfab} corresponds indeed to the $2$D section of the $\alpha$-potatoid Eq.~(\ref{eq:potatoid})
at $\alpha_{AH}=\frac12, \alpha_{ABH}=0$.
 For these values imply $\omega=0$, cf. Eq.~(\ref{eq:omega-alphas}); and as seen from Fig.~\ref{fig:omega-zeta}, all points 
 $(\omega=0,$ $ \zeta)$ remain within the $\omega$-$\zeta$--chips $\forall \zeta \in [\frac13, 1]$. If follows that
the study in Sec.~\ref{sec:resolv-a0-AND-c0} that lead to the NAS conditions given by Fig.~\ref{fig:resolved-p-c-BFB} remains valid, at
least for the $\alpha_{AH}=\frac12, \alpha_{ABH}=0$ section. 
Moroever, since the domain of Fig.~\ref{fig:alfa-alfab} is not only a projection but corresponds as well to the latter section of the
$\alpha$-potatoid, then the above mentioned NAS conditions are  {\sl sufficient} conditions for all other sections at fixed $\alpha_{AH}, \alpha_{ABH}$ since by construction they all fall in the interior of the $\alpha_A, \alpha_{AB}$ domain of Fig.~\ref{fig:alfa-alfab}. 
Obviously this holds even if these sections have portions excluded by the $\omega$-$\zeta$--chips, e.g. when $\alpha_{ABH} < -0.27$ as seen from Fig.~\ref{fig:peeling-fixed-alfAH} (a), since sufficiency is more constraining. We can thus safely conclude that the conditions given by Fig.~\ref{fig:resolved-p-c-BFB} are NAS for the validity of Eq.~(\ref{eq:p-c-BFB-1}) in all the $\alpha$-potatoid.
\item[--] Along a similar line of thought, one deduces from Fig.~\ref{fig:peeling-fixed-alfAH} (b), where there are no exclusions by the 
$\omega$-$\zeta$--chips as soon as  $\alpha_{ABH} \gtrsim -0.06$, and from the fact that the projected domain shown in Fig.~\ref{fig:alfa-alfah} is also retrieved as a $2$D section at $\alpha_{AB}=\alpha_{ABH}=0$, 
that the conditions given by Eq.~(\ref{eq:condF30a1bis}) remain NAS for the validity of Eq.~(\ref{eq:condF30a1}) in all the $\alpha$-potatoid.
\item[--] The case of Eq.~(\ref{eq:Bool8}) is more involved. This condition resulted from eliminating $(\alpha_{AH}, \alpha_{ABH})$ based on the full domain of Fig.~\ref{fig:alfah-alfabh}. However, as seen from Fig.~\ref{fig:peeling-fixed-alfA} (b), a portion of this domain in the $-\sqrt{2} \leq \alpha_{ABH} \leq 0$ range is excluded by the $\omega$-$\zeta$--chips constraint.
Equation (\ref{eq:Bool8}) becomes thus only sufficient for the domain of Fig.~\ref{fig:alfah-alfabh} that corresponds furthermore to the
$2$D section at $\alpha_A=1, \alpha_{AB}=0$ on Figs.~\ref{fig:peeling-fixed-alfA} (a)--(d). It is thus also only sufficient for the full
$\alpha$-potatoid, again because the domain of Fig.~\ref{fig:alfah-alfabh} is the largest section. Note that one can do better by resolving
the NAS conditions for this largest section, taking into account the actual $\omega$-$\zeta$--chips constraint which is simply defined by a 
straight line joining the points $(\alpha_{AH}=0, \alpha_{AHB}=0)$ and $(\alpha_{AH}=1, \alpha_{AHB}=-\sqrt{2})$, 
see Fig.~\ref{fig:peeling-fixed-alfA} (b). The resulting truncated domain will however cease to be the largest section so that the obtained conditions are now only {\sl necessary} for an extended fraction of the $\alpha$-potatoid. As seen from Fig.~\ref{fig:peeling-fixed-alfA} (d), the maximal section taking into account the $\omega$-$\zeta$--chips constraint does exist somewhere inside the $3$D domain but would be difficult to determine analytically. 
\end{itemize}

We end this section by a comment concerning $\alpha_{ABH}$: as argued at the end of section \ref{sec:b0-AND-4a0c0-b02} the sign of
$\alpha_{ABH}$ is not expected to be correlated with the three other $\alpha$'s. If a given point $(\alpha_A, \alpha_{AB},
\alpha_{AH}, \alpha_{ABH})$ lies in the true $4$D {\sl $\alpha$-parameters} domain, i.e. not just in the $\alpha$-potatoid, then the point
$(\alpha_A, \alpha_{AB}, \alpha_{AH}, -\alpha_{ABH})$ lies also in this domain. This is best seen from Eq.~(\ref{eq:alABH}) which
is the only one that depends on the $B$ field (in the chosen gauge), and only through an arbitrary global sign. However, Eqs.~(\ref{eq:omega-alphas}) are not symmetrical under $\alpha_{ABH} \to -\alpha_{ABH}$, and as discussed above and shown in
Figs~~\ref{fig:peeling-fixed-alfA} \& \ref{fig:peeling-fixed-alfAH} the $\omega$-$\zeta$--chips peels the $\alpha$-potatoid
asymmetrically with respect to $\alpha_{ABH}$. This is not a contradiction because the $\omega$-$\zeta$--chips constraint is only sufficient but not necessary to exclude points. But given the general symmetry with respect to the sign flip of $\alpha_{ABH}$, 
it follows that {\sl for any domain excluded by the $\omega$-$\zeta$--chips one should also
 exclude the domain corresponding to the replacement $\alpha_{ABH} \to -\alpha_{ABH}$.}



\section{Putting everything together: A User's Guide \label{sec:disc}}
It is time to recapitulate the various results we arrived at and then provide
a roadmap for an optimal exploitation:

\begin{itemize}
\item While studying the general pre-custodial potential we were lead automatically
in sections \ref{sec:a0-AND-c0} and  \ref{sec:resolv-a0-AND-c0} to constraints that
involved only the $A$ and $B$ multiplets for which we provided the fully resolved NAS BFB conditions
in analytical form, see Fig.~\ref{fig:resolved-p-c-BFB} and Eqs.~(\ref{eq:Bool1} - \ref{eq:Bool5}).
As such they thus correspond to the NAS conditions for a reduced model having only two triplets. Nonetheless, 
they do provide robust {\sl necessary} BFB conditions for the full pre-custodial potential since they correspond to the
potential in the $H=0$ field direction.
\item In sections \ref{sec:b0-AND-4a0c0-b02} and \ref{sec:part-resolv} we addressed the parts of the constraints that involve
simultaneously the three sectors $H,A$ and $B$. The sign of $\lambda_{BH}$ turned out to be critical, but again the BFB conditions  
that we obtained in a fully resolved analytical form correspond to field sub-sectors, namely $H,B$ or $H,A$, cf. Eqs.~(\ref{eq:condF30a2bis}, \ref{eq:condF30a1bis}), and are thus {\sl necessary} for the full model. It is noteworthy that
Eq.~(\ref{eq:condF30a1bis}) reproduces Eqs.~(\ref{eq:BFB} -- \ref{eq:BFBnew5prime}) of the Type-II seesaw model\footnote{\label{foot:10} with the correspondence $\Delta=A$, $\lambda = \lambda_H, \lambda_2 = \lambda_A^{(1)}/4, \lambda_3 = \lambda_A^{(2)}/4, \lambda_1 = 
\lambda_{AH}^{(1)}$ and $\lambda_4 = \lambda_{AH}^{(2)}$,}, that we had arrived at following a different path in section \ref{subsec:BFB},
a significant cross-check. Moreover, from the flowchart of Fig.~\ref{fig:partial-resolved-p-c-BFB} and the properties of 
${\cal B}_{9}^{(a,b)}$
one finds that the constraint Eq.~(\ref{eq:condF30a2bis}) should be applied whenever $\lambda_{BH} <0$, thus retrieving the fully resolved
NAS BFB conditions for the SM extended by one real $SU(2)$ triplet. 
\item We give in Table \ref{tab:user} a roadmap for a user's implementation of the constraints following two alternative roads each made of
two steps. Step  {\bf \large{\textcircled{\normalsize{1}}}} is common and corresponds to the fully resolved necessary constraints that are
also NAS if restricted to the $A,B$ or $H,A$ sectors. Note that these constraints are already stricter than the ones given in  
\cite{Blasi:2017xmc} under the assumption of two nonvanishing
complex fields at once or the ones extended to the ``custodial'' direction in \cite{Krauss:2017xpj}, as they are NAS in {\sl all directions} within $A,B$ or $H,A$. Also specifying to the Georgi-Machacek case we do retrieve the conditions found in \cite{Hartling:2014zca}.
Steps {\bf \large{\textcircled{\normalsize{2}}}} and {\bf \large{\textcircled{\normalsize{2'}}}} are two technically different but theoretically equivalent ways to complete the NAS
conditions. Note first that in both cases branches {\bf \large{\textcircled{\normalsize{a}}}} and 
{\bf \large{\textcircled{\normalsize{b}}}} approximate
Eq.(\ref{eq:Bool8}) as being necessary for the positivity of $\mathfrak{b_0}$ (on top of it being sufficient). Despite the issue discussed
in Sec.~\ref{sec:peeling}, this approximation is valid for all practical purposes, which we checked numerically by scanning over several
tens of thousands of points in the {\sl $\alpha$-parameters} space and verified that Eqs.~(\ref{eq:Bool8-0}) and (\ref{eq:Bool8}) 
delineated indeed the same $(\lambda_{AH}^{(1,2)},\lambda_{ABH})$-space regions.\footnote{This should not come as a surprise since the
further refinement discussed in Sec.~\ref{sec:peeling} concerns only boundaries of the $\alpha$-potatoid that would require much finer
scans as shown in Figs.~\ref{fig:peeling-fixed-alfA}, \ref{fig:peeling-fixed-alfAH}}. Then the {\bf \large{\textcircled{\normalsize{a}}}} branches with $\lambda_{BH} \geq 0$ are complete and provide fully resolved NAS BFB conditions. When $\lambda_{BH} < 0$, both {\bf \large{\textcircled{\normalsize{a}}}} and {\bf \large{\textcircled{\normalsize{b}}}} lead to the same fully resolved extra constraint
involving the $B,H$ sector, plus different sets of partially resolved constraints: In step {\bf \large{\textcircled{\normalsize{2}}}}
as well as in step {\bf \large{\textcircled{\normalsize{2'}}}}-{\bf \large{\textcircled{\normalsize{a}}}} with $\lambda_{BH} < 0$,
the latter constraints are resolved only with respect to the $T$ and $t$ parameters but still need a scan over the {\sl $\alpha$-parameters}
 (including optionally the refinements of Sec.~\ref{sec:peeling}). In contrast, {\bf \large{\textcircled{\normalsize{2'}}}}-{\bf \large{\textcircled{\normalsize{b}}}} is resolved only with respect to $T$ and a supplementary scan is still required on $t$. 
 Note also the different Boolean meanings in the last columns of {\bf \large{\textcircled{\normalsize{2}}}}-{\bf \large{\textcircled{\normalsize{b}}}} and {\bf \large{\textcircled{\normalsize{2'}}}}-{\bf \large{\textcircled{\normalsize{b}}}}. In the former one needs to find {\sl at least one} set of values $(u,v,c)$ satisfying a set of inequalities
 while the latter requires {\sl all} values of $t$ to satisfy one inequality. 
\end{itemize}

\begin{figure*}[!h]
\captionsetup{justification=raggedright,
singlelinecheck=false}
\begin{center}
{\includegraphics[width=0.60\textwidth, 
height=0.3\textheight, keepaspectratio]{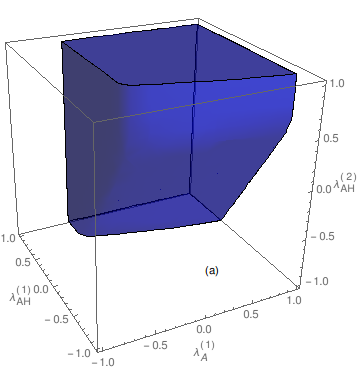}}
\ \ \ \ {\includegraphics[width=0.60\textwidth, 
height=0.3\textheight, keepaspectratio]{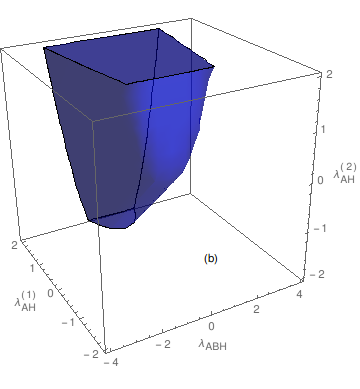}}
\end{center}
\begin{center}
{\includegraphics[width=0.60\textwidth, 
      height=0.3\textheight, keepaspectratio]{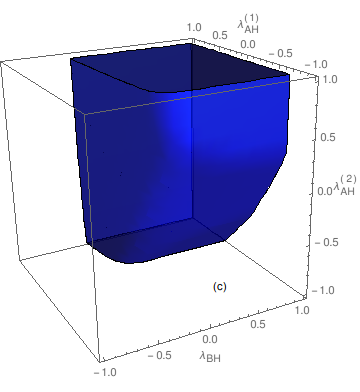}}
\ \ \ \ {\includegraphics[width=0.60\textwidth, 
height=0.3\textheight, keepaspectratio]{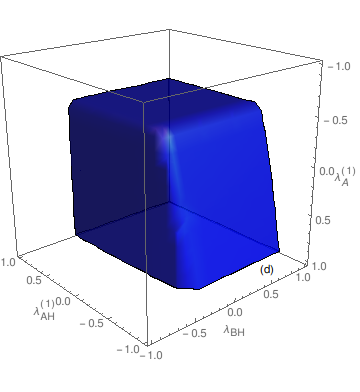}}
\end{center}
\caption{\label{fig:lambda-AH12} Necessary and sufficient $3$D $\lambda$-domains that ensure BFB of the pre-custodial
 potential Eq.~(\ref{eq:p-c-4}) illustrated for fixed 
$\lambda_B=\lambda_H=\lambda_A^{(2)}= \lambda_{AB}^{(1)}= \lambda_{AB}^{(2)}=1$,
(a) the $(\lambda_A^{(1)}, \lambda_{AH}^{(1)}, \lambda_{AH}^{(2)})$ domain with $\lambda_{BH}=-\frac{1}{10}$, $\lambda_{ABH}=1$;
(b) the $(\lambda_{AH}^{(1)}, \lambda_{AH}^{(2)},\lambda_{ABH})$ domain with $\lambda_{BH}=-\frac{1}{10}$, $\lambda_{A}^{(1)}=1$; 
(c) the $(\lambda_{AH}^{(1)}, \lambda_{AH}^{(2)},\lambda_{BH})$ domain with $\lambda_{A}^{(1)}=\lambda_{ABH}=1$;
 (d) the $(\lambda_A^{(1)}, \lambda_{AH}^{(1)},\lambda_{BH})$ domain with $\lambda_{A}^{(2)}=\lambda_{ABH}=1$. 
 }
\end{figure*}

We give in Fig.~\ref{fig:lambda-AH12} an illustration of allowed $\lambda$ domains following road 
{\bf \large{\textcircled{\normalsize{1}}}}-{\bf \large{\textcircled{\normalsize{2'}}}}.
A typical expectation is that the constraints are more stringent for negative values of the couplings associated with
the positive definite operators that are present in the potential.
This is indeed seen in Figs.~\ref{fig:lambda-AH12} (a), (c) and (d). 
 In contrast Fig.~\ref{fig:lambda-AH12} (b) shows that 
$\lambda_{ABH}$ can be in equally sized negative or positive regions since this coupling corresponds to the only operators that is 
not positive definite (cf. Eq.~\ref{eq:alfaHABrange}).\\

\begin{table}[t!]
\scalebox{.8}{
\begin{tabular}{c|c|c|c|c|c|l}
  \multicolumn{1}{ r }{\bf \large{\textcircled{\normalsize{1}}}} \\ \cline{1-2}
\multicolumn{1}{ |c|  }{Eqs.(\ref{eq:Bool1}-\ref{eq:Bool5}), Fig.\ref{fig:resolved-p-c-BFB} \Cmark } & 
Eqs.(\ref{eq:BFB}-\ref{eq:BFBnew5prime})${}^{\ref{foot:10}}$ \Cmark  \\ \cline{1-2}

  \multicolumn{1}{ r }{\bf \large{\textcircled{\normalsize{2}}}} \\ \cline{1-2}
\multicolumn{1}{ |c  }{{\bf \large{\textcircled{\normalsize{a}}}} \multirow{2}{*}{ Eq.(\ref{eq:Bool8}) \Cmark} } &
\multicolumn{1}{ |c| }{$\lambda_{BH} \geq 0$}  \\ \cline{2-4} 
\multicolumn{1}{ |c  }{}                        &
\multicolumn{1}{ |c| }{{$\lambda_{BH} < 0$}} & Eq.(\ref{eq:condF30a2bis}) \Cmark & Eqs.(\ref{eq:condF30a3}, \ref{eq:condb}-\ref{eq:condd}) \Cmark, with Eqs.(\ref{eq:acoefs}, \ref{eq:acoefsannex})  \\ \cline{1-3}
\multicolumn{1}{ |c }{} &  &  & and $\forall \alpha$-params, $\omega, \zeta$ Eqs.(\ref{eq:potatoid}, \ref{eq:chips}, \ref{eq:resolved}) \\ \cline{1-4}
\multicolumn{1}{ |c  }{{\bf \large{\textcircled{\normalsize{b}}}} \multirow{2}{*}{ Eq.(\ref{eq:Bool8}) \xmark} } &
\multicolumn{1}{ |c| }{\multirow{1}{*}{$\lambda_{BH} \geq 0$}} &  \textemdash & $4 \mathfrak{a_0 c_0} -\mathfrak{b_0^2}$ Eqs.(\ref{eq:c0},\ref{eq:defZ},\ref{eq:varchange2}) $\to a_i$,  (\ref{statmnt:Rplus}) \Cmark  \\ \cline{2-4}
\multicolumn{1}{ |c  }{}                        &
\multicolumn{1}{ |c| }{\multirow{1}{*}{$\lambda_{BH} < 0$}} & Eq.(\ref{eq:condF30a2bis}) \Cmark & $4 \mathfrak{a_0 c_0} -\mathfrak{b_0^2}$ Eqs.(\ref{eq:c0},\ref{eq:defZ},\ref{eq:varchange1}) $\to a_i$,  (\ref{statmnt:Rplus}) \Cmark     \\ \cline{1-3}
\multicolumn{1}{ |c }{} &  &  & and $\forall \alpha$-params, $\omega, \zeta$ Eqs.(\ref{eq:potatoid}, \ref{eq:chips}, \ref{eq:resolved}) \\ \cline{1-4}
  \multicolumn{1}{ r }{\bf \large{\textcircled{\normalsize{2'}}}} \\ \cline{1-2}
\multicolumn{1}{ |c  }{{\bf \large{\textcircled{\normalsize{a}}}}\multirow{2}{*}{ Eq.(\ref{eq:Bool8}) \Cmark} } &
\multicolumn{1}{ |c| }{$\lambda_{BH} \geq 0$}  \\ \cline{2-4} 
\multicolumn{1}{ |c  }{}                        &
\multicolumn{1}{ |c| }{{$\lambda_{BH} < 0$}} & Eq.(\ref{eq:condF30a2bis}) \Cmark & same as in \bf \large{\textcircled{\normalsize{2}}}  \\ \cline{1-4}
\multicolumn{1}{ |c  }{{\bf \large{\textcircled{\normalsize{b}}}} \multirow{2}{*}{ Eq.(\ref{eq:Bool8}) \xmark} } &
\multicolumn{1}{ |c| }{\multirow{1}{*}{$\lambda_{BH} \geq 0$}} &  \textemdash & $\mathfrak{b_0} + 2 \sqrt{\mathfrak{a_0 \mathfrak{c_0}}} > 0$ Eq.(\ref{eq:a0c0b0p2sqrta0c0}) \Cmark  \\ \cline{2-3}
\multicolumn{1}{ |c  }{}                        &
\multicolumn{1}{ |c| }{\multirow{1}{*}{$\lambda_{BH} < 0$}} & Eq.(\ref{eq:condF30a2bis}) \Cmark &   $\forall t \in [0, +\infty), \alpha$-params, $\omega, \zeta$ Eqs.(\ref{eq:potatoid}, \ref{eq:chips}, \ref{eq:resolved})   \\ \cline{1-4}
\end{tabular}}
\caption{ A roadmap for the complete NAS-BFB conditions for the pre-custodial model. Check/Cross marks following an equation number indicate that the equation should be satisfied/violated. See text for a detailed description.}
    \label{tab:user}
\end{table}





\begin{figure*}[!h]
\captionsetup{justification=raggedright,
singlelinecheck=false}
\begin{center}
{\includegraphics[width=0.60\textwidth, 
height=0.3\textheight, keepaspectratio]{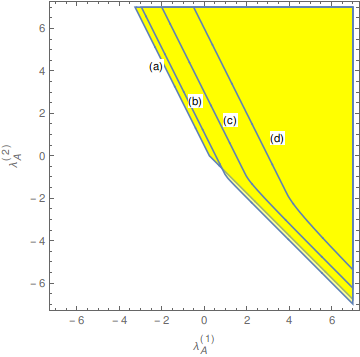}}
\end{center}
\caption{\label{fig:unitarity-tension} Trend of the $(\lambda_{A}^{(1)},\lambda_{A}^{(2)})$ necessary domains in yellow, as dictated by the 
necessary BFB conditions of Fig.~\ref{fig:resolved-p-c-BFB} \& Eqs~(\ref{eq:Bool1} --\ref{eq:Bool5}). The allowed domain lies to the right of each line, illustrated for:
$\lambda_B=1, \lambda_H=\frac12$ and (a) $\lambda_{AB}^{(1)}=-\frac15, \lambda_{AB}^{(2)}= \frac12$, 
(b) $\lambda_{AB}^{(1)}= \frac15$, $\lambda_{AB}^{(2)}=-\frac12$, (c) $\lambda_{AB}^{(1)}=-\frac15, \lambda_{AB}^{(2)}= -\frac{3}{10}$, and (d) $\lambda_B=\frac12$, $\lambda_H=\frac12, \lambda_{AB}^{(1)}=-\frac15, \lambda_{AB}^{(2)}= -\frac{3}{10}$.}
\end{figure*}

Let us close this section with an outlook on some issues related to the subject of the present paper but lying beyond its scope:\\

\noindent
\underline{\sl perturbative unitarity constraints.} They typically bind the absolute magnitudes of the $\lambda$ couplings
and some of their combinations from above. 
These constraints should eventually be studied  for the general pre-custodial model (see however \cite{Krauss:2017xpj}) and be 
combined with the NAS BFB conditions derived in this paper. 
Here we just note an interesting tension that might arise from such a combination, due to the form of conditions 
${\mathcal B}_3, {\mathcal B}_5,{\mathcal B}_7$. 
The relatively large numerical factors appearing in these inequalities, see Eqs.~(\ref{eq:Bool1}, \ref{eq:Bool3}, \ref{eq:Bool5}),
can easily force $|\lambda_A^{(i=1,2)}|$ or $\lambda_B$ to be
(much) larger than one even for $|\lambda_{AB}^{(i=1,2)}|, |\lambda_H| \lesssim 1$. At
least one among the conditions ${\mathcal B}_3, {\mathcal B}_5$ and ${\mathcal B}_7$ is active in cases (ii), (iii) or (iv) of the flowchart of 
Fig.~\ref{fig:resolved-p-c-BFB}.
We illustrate a few such configurations on Fig.~\ref{fig:unitarity-tension}. The domains shown in the figure are necessary but not
sufficient; they can be reduced further when adding the rest of the NAS BFB conditions. Note that such a potential tension disappears 
in the limit of decoupling
between the two triples ($\lambda_{AB}^{(i=1,2)} \to 0$) in accordance with the unitarity/BFB conditions found in \cite{Arhrib:2011uy}. \\

\noindent
\underline{\sl quantum corrections.} They affect the tree-level constraints in various ways: --they modify the form
of the constraints, introduce a notion of scale at which they should be satisfied and criteria for the validity of perturbativity,
as treated for instance in \cite{Staub:2017ktc}, \cite{Krauss:2017xpj,Krauss:2018orw} --however, it is not often appreciated that  combining perturbative-unitarity and stability requirements beyond the tree-level
needs some further care because the physical meaning of the running couplings becomes different in these two classes of constraints. 
Since unitarity is related to scattering processes the proper objects are the Green's functions. The scale appearing in the 
running couplings (and masses) of the renormalization group improved Green's functions encodes the way the scattering amplitudes scale
 with energy. In contrast, stability issues are expressed in terms of the renormalization group improved effective potential where
 now the scale on which depend the running couplings, masses, and fields, is in fact a combination of the fields themselves and encode
 the modification of the shape of the potential (see for instance \cite{Bando:1992np,Ford:1992mv}\footnote{where it was also stressed
 that even an additive constant becomes field dependent beyond~tree-level.}). It thus appears that, in so far as replacing the tree-level
 couplings by their runnings in the tree-level conditions is a good approximation, the potential stability conditions
 need not be required {\sl at all} 'scales', from the electroweak scale all the way up to some very high cut-off $\Lambda$ (e.g. $M_{GUT}$ 
 or $M_{Planck}$) as often done in the literature \cite{Ghosh:2017pxl,Blasi:2017xmc,Bonilla:2015eha}, but only at that scale $\Lambda$
 which represents the largest value of the fields. 
 Barring Landau poles, there is indeed no physical reason to require the improved quartic part of the potential to remain positive for 
 intermediate values of the fields. (Obviously this is at variance with the unitarity constraints that should be satisfied already at the
 energy scale of a given scattering experiment.) Furthermore, a longstanding issue is how to improve the effective potential in the presence of several scalar fields (see \cite{Chataignier:2018aud} for a recent reappraisal, and references therein). As concerns the
NAS BFB conditions of Table \ref{tab:user}, they can be used beyond the tree-level in two different ways: {\sl i)} The quartic part, $V_{\text{p-c}}^{(4)}$, Eq.~(\ref{eq:p-c-4}), of the pre-custodial potential has the same form as the general counterterms needed to renormalize the Georgi-Machacek model accounting for a deviation from the tree-level correlations Eq.~(\ref{eq:correlate})
due to the custodial symmetry breaking loop effect of the $U(1)_Y$ gauge couplings
 \cite{Gunion:1990dt}, \cite{Blasi:2017xmc}. One is thus guaranteed that the ten $\lambda$ couplings of $V_{\text{p-c}}^{(4)}$ will
 absorb the one-loop corrections of the Georgi-Machacek effective potential up to field dependent factors of the form
 $\log({\cal M}(\phi_i)^2/Q^2)  -c$, where ${\cal M}$ is typically a binomial function of the fields, $Q$ is some renormalization scale and 
 $c$ a renormalization scheme dependent constant. It follows that satisfying the conditions of Table \ref{tab:user} on the $\lambda$'s
 that absorb the one-loop induced quartic couplings, will also guarantee the stability of the full one-loop Georgi-Machacek effective potential at large field values with ${\cal M}(\phi_i)^2 \gg Q^2$. {\sl ii)} Table \ref{tab:user} can also obviously be used as a seed for
 the loop corrected stability conditions of the pre-custodial model itself, relying on whatever renormalization group improvement
 approaches quoted above. The main difference with {\sl i)} will reside essentially in the renormalization conditions not enforcing
 the custodial symmetry of the potential at a given scale.\\
 
\noindent
\underline{\sl comparison with other methods.} As already recalled in the introduction, several methods have been considered to 
address the boundedness from below conditions. Depending on the extended scalar sectors, the symmetries of the
potential and the multiplet representations, the different methods can have varying practical applicability.
An important case is when discrete symmetries are not imposed, thus allowing for odd powers of the scalar fields 
to occur in 4-dim gauge invariant operators of the (most general) renormalizable scalar potential. This was considered for the
two-Higgs-doublet model \cite{Maniatis:2006fs}, for
extensions 
with two real scalar singlets 
\cite{Kannike:2016fmd}, or for the most general potential in Left-Right symmetric models \cite{Chakrabortty:2013zja}. It is also
the case for the pre-custodial model with $\lambda_{ABH} \neq 0$ studied in this paper. The copositivity method ceases to be efficient
in this case since one cannot write the potential in a bilinear form with vectors of dimension $3$ or more having positive definite independent components. (The fact that a two-dimensional such a form is still possible is not helpful as one hides the complexity
of the conditions in the dependence on angles). In contrast, our approach remains
applicable albeit with an extended set of the {\sl $\alpha$-parameters}. It would provide further insight into the all-directions
NAS-BFB conditions for example in the study of Left-Right symmetric models, unlike in 
\cite{Chakrabortty:2016wkl} which relied on \cite{Chakrabortty:2013mha,Chakrabortty:2013zja} where specific directions were considered.
A distinctive feature in this case is the appearance of positivity conditions for full quartic polynomials as was found for the model
with two singlets discussed in \cite{Kannike:2016fmd} and in the present study with two triplets.

\section{Conclusion \label{sec:conclusion}}

We carried out in this paper a comprehensive study of tree-level {\sl necessary and sufficient} conditions for a bounded from below
potential in extensions
of the SM with one or two $SU(2)_L$ triplet scalar fields. We derived for the first time the complete set of such conditions
in the case of the general pre-custodial model having one complex and one real triplets. A fully resolved 
analytical form involving only the couplings was obtained for parts of these conditions. This could be achieved thanks to
a parameterization of the $13$-dimensional field space reducing the degrees of freedom to a small set of relevant gauge invariant variables.
We determined precisely the compact domains in which most of these variables live, thus allowing a well defined procedure for the
other parts of the conditions that remained in a partially resolved form. It would be interesting to see how the more general methods
quoted in the introduction would perform in the presence of triplets. In particular the fully resolved form we found in the purely
two triplets sector may lend itself to a generalization to multiple fields.  
In the course of the study we were lead to review some of the known results for the type-II seesaw and Georgi-Machacek models
providing complete proofs that were missing in the literature for key properties. The latter were important to settle on a firm basis
in relation with an unexpected feedback of the Georgi-Machacek reduced variables on those of the pre-custodial model. Furthermore,
we demonstrated the existence of simplified criteria for the positivity of a general quartic polynomial that can be used for any model
with a renormalizable potential. The pre-custodial BFB conditions on the couplings have to be fulfilled for any consistent tree-level 
phenomenological analysis of the model. They find also their motivation as a pattern for the one-loop BFB conditions in the Georgi-Machacek model.

\acknowledgements
We would like to thank Abdesslam Arhrib for his fruitful collaboration at an early stage of this work, and Michele Frigerio as well as
Michel Talon for profitable discussions. We also acknowledge an inspiring exchange with Kaladi Babu.
\appendix

\section{Proof of the properties of $\xi$ and $\zeta$ \label{appendix:A}}

In the following we give the proof of Eq.~(\ref{eq:xizetaranges}), then
establish Eq.~(\ref{eq:xizetaenv}) and the ensuing correlations.

\subsubsection{$0 \le \xi \le 1$ \label{appendix:xi}}

\noindent
First we note that $\Delta$ being traceless implies the identity
\begin{equation}
\Delta\Delta^{\dagger}  + \Delta^{\dagger}\Delta=
{\bf 1} \times Tr\Delta{\Delta}^{\dagger}, \label{eq:identity1}
\end{equation}
(see also Eq.~(\ref{eq:identity4})), 
from which follows immediatly
\begin{equation}
H^\dagger\Delta\Delta^{\dagger}H  + 
H^\dagger\Delta^{\dagger}\Delta H=
H^\dagger{H} Tr\Delta{\Delta}^{\dagger} \ . \label{eq:rel1}
\end{equation}
Since $H^\dagger\Delta^{\dagger}\Delta H$ is positive definite one
then has
\begin{equation}
 H^\dagger{H} Tr\Delta{\Delta}^{\dagger} -
 H^\dagger\Delta\Delta^{\dagger}H  \geq 0 
\end{equation}
and thus
\begin{equation}
\xi \equiv \frac{ H^\dagger\Delta\Delta^{\dagger}H}{H^\dagger{H} 
Tr\Delta{\Delta}^{\dagger} }  \leq 1
\end{equation}
Furthermore $\xi$ is trivially greater than zero since it is the
ratio of two positive definite quantities. Finally the two values
$0$ and $1$ are effectively reached respectively when 
$H^\dagger\Delta=0$ and $\Delta H =0$, which is always possible
for some given configurations of the $H$ and $\Delta$ field
components provided that $Det \Delta =0$ when $H\neq 0$. Thus 
\begin{equation}
0 \le \xi \le 1  \ . \label{eq:xirange}
\end{equation}

\subsubsection{$\frac12 \le \zeta \le 1$\label{appendix:zeta}}

\noindent
$\Delta \Delta^\dagger$ being a $2 \times 2$ matrix one has
\begin{equation}
\frac12 (Tr\Delta{\Delta}^{\dagger})^2 - \frac12 Tr(\Delta{\Delta}^{\dagger})^2  = Det  \Delta{\Delta}^{\dagger} \label{eq:rel2}
\end{equation}
Then, using $Det  \Delta{\Delta}^{\dagger} \equiv |Det \Delta|^2\geq 0$ implies straightforwardly from
Eq.~(\ref{eq:rel2}) that
\begin{equation}
\zeta \equiv \frac{Tr (\Delta{\Delta}^{\dagger})^2}{(Tr \Delta{\Delta}^{\dagger})^2} \leq 1
\end{equation}
Note that the value $1$ is indeed reached when 
$\Delta \Delta^\dagger$ has one zero and one non-zero eigenvalues,
which is always possible to find for some configurations of
the $\Delta$ field components. 

\noindent 
Also, we trivially have $\zeta \geq 0$ since it is the ratio
of two positive definite quantities. However, the value $0$ cannot
be trivially reached, since if the 
numerator of $\zeta$ vanishes then the denominator should vanish
as well! In fact $\zeta$ cannot go below $1/2$. To see this
we rewrite $\zeta$ in terms of $M_1^2, M_2^2$ the two (real and positive) 
eigenvalues of $\Delta \Delta^\dagger$,
\begin{equation}
\zeta = \frac{M_1^4 + M_2^4}{(M_1^2 + M_2^2)^2} \label{eq:zetarotated}
\end{equation}  
It is now easy to study the function
$\zeta(x) = (1+x^2)/(1+x)^2$ where $x\equiv M_1^2/M_2^2\geq 0$, to show
that it has a minimum of $\zeta=1/2$ at $x=1$, that is when
$\Delta \Delta^\dagger$ has degenerate eigenvalues. One also retrieves the fact that $\zeta(x) \leq 1$ and reaches $1$ for
$x \to 0$ or $x \to \infty$.  Thus  
\begin{equation}
\displaystyle \frac12 \le \zeta \le 1 \ . \label{eq:zetarange}
\end{equation}

\subsubsection{Correlation between $\xi$ and $\zeta$\label{appendix:xizetacorr}}
\noindent
Since from Eqs.~(\ref{eq:paramzeta}, \ref{eq:paramxi})
$\zeta$ depends solely on $\Delta$ while $\xi$ depends on both $H$ and $\Delta$, one could be tempted to assume
that $\zeta$ and $\xi$ can reach independently their extrema given by
Eqs.~(\ref{eq:xirange}, \ref{eq:zetarange}), by varying independently $H$ and $\Delta$. 
This is however not true as one can see easily from the fact that $\xi$ reaches 
its two extrema under the generic condition $Det \Delta =0$ as discussed above Eq.~(\ref{eq:xirange}).
But then  $Det \Delta =0$ together with Eq.~(\ref{eq:rel2}) imply necessarily $\zeta =1$ so that
$\zeta =\frac12$ can never be reached when $\xi$ takes its extremal values 0 or 1. 
 
We use now the invariance under the general gauge transformation
$H \to {\cal U}(x) H$, $\Delta \to {\cal U}(x) \Delta {\cal U}^\dag(x)$,
where ${\cal U}(x)$ denotes any element of $SU(2)_L \times U(1)_Y$, of 
the potential Eq.~(\ref{eq:Vpot}) and of the parameters
defined in Eqs.~(\ref{eq:paramr} - \ref{eq:paramxi}). Since ${\cal U}(x)$
is unitary and $\Delta \Delta^\dag$ hermitian, we can always find, for any given field
configuration $\Delta$,  a gauge transformation
 ${\cal U}_{\Delta}(x)$ that
diagonalizes $ \Delta \Delta^\dag$. Then $\zeta$ takes the form given in
Eq.(\ref{eq:zetarotated}) and  $\xi$ reads 
\begin{equation}
\xi = 
    \frac{(M_2^2 |\widetilde{\phi^0}|^2 + M_1^2 |\widetilde{\phi^+}|^2)}{
     (M_1^2 + M_2^2) (|\widetilde{\phi^0}|^2 + |\widetilde{\phi^+}|^2)}
     \label{eq:xirotated}
\end{equation}
where the tilde denotes the components of the transformed doublet $\widetilde{H} = {\cal U}_{\Delta}(x) H$.
It is then natural to define
\begin{eqnarray}
&\displaystyle c_{{}_\Delta}^2 \equiv \frac{M_1^2}{M_1^2 + M_2^2}&
, \;\; s_{{}_\Delta}^2 \equiv  1 - c_{{}_\Delta}^2 \label{eq:anglesDelta}\\
&\displaystyle c_{{}_H}^2 \equiv  \frac{|\widetilde{\phi^+}|^2}{|\widetilde{\phi^0}|^2 + |\widetilde{\phi^+}|^2}&, 
\;\; s_{{}_H}^2 \equiv 1 - c_{{}_H}^2 . \label{eq:anglesH}
\end{eqnarray}
with their obvious range of variation $c^2_{\Delta}, c_{{}_H}^2 \in [0,1]$.
Equations~(\ref{eq:xizetacorr1}, \ref{eq:xizetacorr2}) follow then straightforwardly
from Eqs.~(\ref{eq:zetarotated}, \ref{eq:xirotated} - \ref{eq:anglesH}):
\begin{eqnarray}
\xi &=& c_{{}_\Delta}^2 c_{{}_H}^2 + s_{{}_\Delta}^2 s_{{}_H}^2 = \frac12(1 + c_{{}_{2H}} \, c_{{}_{2\Delta}}),  
\label{eq:xizetacorr1bis}\\ 
\zeta &=& c_{{}_\Delta}^4 + s_{{}_\Delta}^4 = \frac12(1 + c^2_{{}_{2\Delta}}),
\label{eq:xizetacorr2bis}
\end{eqnarray}
where we have defined $c_{{}_{2H}} = c_{{}_H}^2 - s_{{}_H}^2$,  $c_{{}_{2\Delta}} = c_{{}_H}^2 - s_{{}_H}^2$. It is crucial that these cosines vary
{\sl independently from each other} in their allowed domains $c_{{}_{2H}} \in [-1,1]$,  
$c_{{}_{2\Delta}} \in [-1,1]$. That they indeed scan independently all their allowed domain is obvious
from the definitions Eqs.~(\ref{eq:anglesDelta}, \ref{eq:anglesH}) and the fact
that  ${\cal U}_{\Delta}(x)$ is invertible: Indeed one can always choose the magnitudes of
$M_1^2, M_2^2, |\widetilde{\phi^0}|, |\widetilde{\phi^+}| $ to reach any value of 
$c_{{}_H}^2, c_{{}_\Delta}^2\in [0,1]$; this will correspond to the domain of all field configurations
obtained by gauge transforming $\widetilde{H} \equiv (\widetilde{\phi^+}, 
\widetilde{\phi^0})^T $ and $\widetilde{\Delta} \equiv {\rm diag}(e^{i\theta_1} M_1, e^{i\theta_2} M_2)$
with an arbitrary ${\cal U} \equiv {\cal U}_{\Delta}^{-1}$.   

Eliminating $c^2_{{}_{2\Delta}}$ in Eqs.~(\ref{eq:xizetacorr1bis}, \ref{eq:xizetacorr2bis}) one finds
\begin{equation}
 2 \xi^2 - 2 \xi + 1 + \frac{(c_{{}_{2H}}^2-1)}{2}  = c_{{}_{2H}}^2 \zeta \ .
\end{equation}
This allows to determine the lower envelope in the $\xi, \zeta$ plane, i.e. when saturating the inequality in
Eq.~(\ref{eq:xizetaenv}) as discussed in \cite{Bonilla:2015eha}. We will however rely directly on 
Eqs.~(\ref{eq:xizetacorr1bis}, \ref{eq:xizetacorr2bis}) when determining the BFB conditions in the next section.

. 
   
%

   
%

\section{the BFB conditions for the Type-II seesaw model \label{appendix:B}}
\subsubsection{The new necessary and sufficient BFB conditions}
We give here a detailed proof of the NAS-BFB conditions Eqs.~(\ref{eq:BFB}, \ref{eq:BFBnew5prime}).
The condition $\lambda_2+ \zeta \lambda_3 \geq 0$ of Eq.~(\ref{eq:NAS-BFB}) depends only
on $\zeta$ so that the correlations given by Eqs.~(\ref{eq:xizetacorr1bis}, \ref{eq:xizetacorr2bis}) 
are not relevant here. It is thus equivalent
to replacing $\zeta$ by its two extreme values due to the monotonic
dependence on $\zeta$. Thus the first two condition of Eq.~(\ref{eq:NAS-BFB}) become
\begin{equation}
\lambda > 0 \;\;\land  \;\; \lambda_2+\lambda_3 \geq 0  \;\;\land  \;\;\lambda_2+\frac{\lambda_3}{2} \geq 0 \label{eq:BFBnew1prime}
\end{equation}
as was initially found in \cite{Arhrib:2011uy}. As for 
$\lambda_1+ \xi \lambda_4 + \sqrt{\lambda(\lambda_2+\zeta \lambda_3)} > 0$
of Eq.~(\ref{eq:NAS-BFB}), we first rewrite it in terms of
$c_{{}_{2H}}$ and $c_{{}_{2\Delta}}$ according to  
Eqs.~(\ref{eq:xizetacorr1bis}, \ref{eq:xizetacorr2bis}), as 
\begin{equation}
F(c_{{}_{2\Delta}}, c_{{}_{2H}}) > 0, \;\; \forall  c_{{}_{2\Delta}}, c_{{}_{2H}} \in[-1, 1] \ ,  
\end{equation}
where we defined 
\begin{equation}
F(c_{{}_{2\Delta}}, c_{{}_{2H}}) \equiv \lambda_1 + (1 + c_{{}_{2H}} \, c_{{}_{2\Delta}})\frac{\lambda_4}{2} + \sqrt{\lambda \big(\lambda_2 + (1 + c^2_{{}_{2\Delta}}) \frac{\lambda_3}{2}\big)} \label{eq:Fdef} \ .
 \end{equation}
 
\noindent 
However, since 
$c_{{}_{2H}}$ and $c_{{}_{2\Delta}}$ are mutually independent, the monotonic dependence
on $c_{{}_{2H}}$ in $F(c_{{}_{2\Delta}}, c_{{}_{2H}})$ allows again to replace the above positivity condition 
equivalently by two positivity conditions corresponding to the two extreme values $c_{{}_{2H}} = \pm1$. We will thus 
replace once and for all the condition 
$\lambda_1+ \xi \lambda_4 + \sqrt{\lambda(\lambda_2+\zeta \lambda_3)} > 0$
by
\begin{equation}
F(c_{{}_{2\Delta}}, +) > 0 \;\; {\rm and} \;\; F(c_{{}_{2\Delta}}, -) >0, \;\; \forall  c_{{}_{2\Delta}} \in[-1, 1] \ ,
\label{eq:NAS-BFBintermediate} 
\end{equation}
where we use the shorthand notation $F(c_{{}_{2\Delta}}, \pm) \equiv F(c_{{}_{2\Delta}}, \pm 1)$. 

At this point a careful study is needed, as the dependence on $c_{{}_{2 \Delta}}$ is not trivially monotonic so that
{\sl a priori} one does not necessarily have, 
\begin{equation}
Eq.~(\ref{eq:NAS-BFBintermediate})  \Leftrightarrow \big\{ F(+1, \pm) > 0 \;\; \land \;\; F(-1, \pm) > 0 \big\}\ .
\label{eq:NAS-BFBintermediate1} 
\end{equation}

It is nonetheless noteworthy that this equivalence does hold in half of the parameter space region where $\lambda_3 < 0$ despite
the non-monotonicity of $F$ in $c_{{}_{2\Delta}}$, as we will see in a moment.
Irrespective of the sign of $\lambda_3$ the first and second derivatives of $F$ read, 
\begin{eqnarray}
&&\displaystyle F'(c_{{}_{2\Delta}}, \pm) = \frac12\big(\frac{c_{{}_{2\Delta}} \lambda \lambda_3}{\sqrt{\lambda (\lambda_2 + 
\frac12(1 + c_{{}_{2\Delta}}^2) \lambda_3)}} \pm  \lambda_4\big)  \label{eq:dF}\\
&& F''(c_{{}_{2\Delta}}, \pm) = \frac{\lambda_3 (2 \lambda_2 + \lambda_3)}{(2 \lambda_2 + \lambda_3(1 + 
c_{{}_{2\Delta}}^2) )^{3/2}}  \label{eq:d2F}
\end{eqnarray}
In the sequel we will assume without further reference the conditions
given in Eq.~(\ref{eq:BFBnew1prime}).
It then immediately follows from Eq.~(\ref{eq:d2F}) that $F''<0, \, \forall c_{{}_{2\Delta}} \in[-1, 1]$, whenever $\lambda_3 <0$.
This implies that if $F$ admits an extremum it will be  necessarily a {\sl maximum} so that Eq.~(\ref{eq:NAS-BFBintermediate1}) is
valid, since in this case the value of $F$ at one of the two boundaries of $[-1,1]$ is necessarily the smallest value it can take. 
On the other hand, if $F$ does not admit an extremum then Eq.~(\ref{eq:NAS-BFBintermediate1}) is obviously valid as well, and 
one retrieves  Eqs.~(\ref{eq:BFBgen2}, \ref{eq:BFBgen3}). 

{\sl We thus conclude that the BFB conditions Eqs.~(\ref{eq:BFBgen2}, \ref{eq:BFBgen3}) initially found in \cite{Arhrib:2011uy} are necessary and sufficient, and thus complete, when $\lambda_3 <0$.}

The situation is quite different when $\lambda_3 >0$. In this case $F''$ is non-negative over the full domain $[-1,1]$. 
Thus if $F(c_{{}_{2\Delta}}, \pm)$ 
admit extrema in the domain, they will be necessarily minima. On the other hand, $F'(c_{{}_{2\Delta}}, +)$ and 
$F'(c_{{}_{2\Delta}}, -)$ cannot vanish simultaneously (except in the special cases where $\lambda_4=0$ or $2 \lambda_2 + \lambda_3=0$), 
but rather at two opposite values of $c_{{}_{2\Delta}}$, as can be seen from Eq.~(\ref{eq:dF}). This occurs for
\begin{equation}
c_{{}_{2\Delta}}^{(\pm)} = \pm |\lambda_4| \sqrt{\frac{(2 \lambda_2 + \lambda_3)}{\lambda_3 (2 \lambda \lambda_3 - \lambda_4^2)}}
\label{eq:cDeltamin}
\end{equation}
with the consistency condition   $0 \leq (c_{{}_{2\Delta}}^{(\pm)} )^2 \leq 1$. The latter condition reads
\begin{equation}
2 \lambda \lambda_3 - \lambda_4^2 > 0 \;\; \land \;\; 
\lambda_4^2 (2 \lambda_2 + \lambda_3) \leq \lambda_3 ( 2 \lambda \lambda_3 - \lambda_4^2) \ .
\end{equation} 
Note that the second of these two inequalities always implies the first due to the case assumption $\lambda_3 >0$ and the 
validity of Eq.~(\ref{eq:BFBnew1prime}). Moreover this second inequality can be rewritten equivalently as
\begin{equation}
\sqrt{\lambda} \lambda_3 \geq \sqrt{(\lambda_2 + \lambda_3) \lambda_4^2} \label{eq:consistency}
\end{equation}
where we again relied on the case assumption $\lambda_3 >0$. Thus Eq.~(\ref{eq:consistency}) is {\sl  necessary and sufficient for 
the existence of minima within the domain} $[-1, 1]$.
In this case one of the two functions 
$F(c_{{}_{2\Delta}}, +)$, 
$F(c_{{}_{2\Delta}}, -)$ will have a minimum at $c_{{}_{2\Delta}}^{(+)}$ and the other at $c_{{}_{2\Delta}}^{(-)}$. Moreover,
the values of the two $F$ functions at these minima turn out to be the {\sl same}, given by,
\begin{equation}
F_{\rm min}= \lambda_1 +
\frac{\lambda_4}{2} + \sqrt{\lambda (\lambda_2 + \frac{\lambda_3}{2}) \big(1 - \frac{\lambda_4^2}{2 \lambda \lambda_3} \big)}
\label{eq:Fmin} \ .
\end{equation}
[To determine $F_{\rm min}$ some care should be taken  by considering the sign of $\lambda_4$ and noting that
the $\pm$ in Eq.~(\ref{eq:cDeltamin}) refer neither to the sign of $\lambda_4$ nor to the two functions
$F$.] In fact the uniqueness of $F_{\rm min}$ is a direct consequence of the symmetry property 
$F(c_{{}_{2\Delta}}, - c_{{}_{2H}}) = F(-c_{{}_{2\Delta}}, c_{{}_{2H}})$, cf. Eq.~(\ref{eq:Fdef}).
This symmetry is also responsible for the fact that $c_{{}_{2\Delta}}^{(+)}$ and $c_{{}_{2\Delta}}^{(-)}$
are the opposite of each other so that when Eq.~(\ref{eq:consistency}) is satisfied they both remain in the domain $[-1,1]$. 

It follows that even though the two functions $F(c_{{}_{2\Delta}}, +)$ and $F(c_{{}_{2\Delta}},-)$ do not reach their  
minimum for the same value of $c_{{}_{2\Delta}}$, requiring 
\begin{equation}
F_{\rm min} > 0  \label{eq:NAS-BFBintermediate2}
\end{equation}
when Eq.~(\ref{eq:consistency}) is satisfied, will be equivalent to Eq.~(\ref{eq:NAS-BFBintermediate}). Note in particular that Eq.~(\ref{eq:NAS-BFBintermediate2}) should
imply $F(\pm 1, \pm) > 0$ and $F(\mp 1, \pm) > 0$, that is,
\begin{equation}
\lambda_1 + \frac{\lambda_4}{2} + \sqrt{\lambda (\lambda_2 + \frac{\lambda_3}{2}) \big(1 - \frac{\lambda_4^2}{2 \lambda \lambda_3} \big)} >0
\Rightarrow  \lambda_1+ \sqrt{\lambda(\lambda_2+\lambda_3)} >0 \;\; \land \;\; \lambda_1+ \lambda_4 +\sqrt{\lambda(\lambda_2+\lambda_3)} >0 \label{eq:implication}
\end{equation}
which is indeed the case.\footnote{This is due to the inequality $\displaystyle \sqrt{\lambda(\lambda_2+\lambda_3)} - \sqrt{\lambda (\lambda_2 + \frac{\lambda_3}{2}) \big(1 - \frac{\lambda_4^2}{2 \lambda \lambda_3} \big)} > \pm \frac{\lambda_4}{2}$ being valid whenever $\lambda_3 >0$ and Eq.~(\ref{eq:BFBnew1prime}) valid and thus consistently also $2 \lambda \lambda_3 - \lambda_4^2 > 0$.} Finally, when Eq.~(\ref{eq:consistency}) is not satisfied, but still $\lambda_3 >0$, then either $c_{{}_{2\Delta}}^{(\pm)}$ are not real-valued
 or they lie outside of the $[-1, 1]$ domain. In both cases the two functions $F(c_{{}_{2\Delta}}, \pm)$ are monotonic on $[-1, 1]$ and Eq.~(\ref{eq:NAS-BFBintermediate1}) applies, which is similar to the previously
discussed case of $\lambda_3 < 0$. Putting everything together one can summarize the conditions that are equivalent to
$\lambda_1+ \xi \lambda_4 + \sqrt{\lambda(\lambda_2+\zeta \lambda_3)} > 0$ (or Eq.~(\ref{eq:NAS-BFBintermediate})), as follows:
\begin{itemize}
\item if $\sqrt{\lambda} \lambda_3 < \sqrt{(\lambda_2 + \lambda_3) \lambda_4^2}$ then  $F(\pm 1, \pm) > 0$ and $F(\mp 1, \pm) > 0$.
\item if $\sqrt{\lambda} \lambda_3 \geq \sqrt{(\lambda_2 + \lambda_3) \lambda_4^2}$ then $F_{\rm min} > 0$.
\end{itemize}
Adding Eq.~(\ref{eq:BFBnew1prime}) to these conditions, we obtain the Boolean form of the necessary and sufficient BFB conditions 
as given by Eqs.~(\ref{eq:BFB}, \ref{eq:BFBnew5prime}). 

\subsubsection{The old conditions \label{appendix:almostNAS}}

We recall here for further reference the sufficient and almost necessary BFB conditions \cite{Arhrib:2011uy}:
\begin{eqnarray}
&& \lambda > 0 \;\;\land\;\; \lambda_2+\lambda_3 > 0  \;\;\land  \;\;\lambda_2+\frac{\lambda_3}{2} > 0 
\label{eq:BFBgen1} \\
&& \land \;\;\lambda_1+ \sqrt{\lambda(\lambda_2+\lambda_3)} > 0 \;\;\land\;\;
\lambda_1+ \sqrt{\lambda(\lambda_2+\frac{\lambda_3}{2})} > 0  \label{eq:BFBgen2}\\
&& \land \;\; \lambda_1+\lambda_4+\sqrt{\lambda(\lambda_2+\lambda_3)} > 0 \;\; \land \;\; 
\lambda_1+\lambda_4+\sqrt{\lambda(\lambda_2+ \frac{\lambda_3}{2})} > 0  \label{eq:BFBgen3}
\end{eqnarray}

\section{The pre-Custodial potential \label{appendix:pre-c}}
We give hereafter some elements that can help define a systematic
procedure to construct the pre-custodial potential 
Eqs.~(\ref{eq:p-c-2-3}, \ref{eq:p-c-4}) from a minimal set of independent operators. They can be useful as well for the construction of extended models with several $SU(2)$ triplet and singlet fields.

Note first the following two general identities, valid for
any $2\times2$ matrices  $M$ and $N$:
\begin{eqnarray}
 M + \sigma^2 M^\top \sigma^2  &=& {\bf 1}  Tr M, \label{eq:identity3}\\ 
M N + N M &=& {\bf 1}  (Tr M N - Tr M \, Tr N) + M Tr N + N Tr M
. \label{eq:identity4}
\end{eqnarray}
 
The fundamental representation of 
$SU(2)$ is pseudo-real. In particular, any of its elements ${\cal U}$ satisfies
\begin{equation}
\sigma^2 {\cal U} \sigma^2 = {\cal U}^* \label{eq:pseudo-real},
\end{equation}
where $\sigma^2$ denotes the second Pauli matrix.
From this and Eq.~(\ref{eq:p-c-transfos}) it follows that
\begin{eqnarray}
&&\sigma^2 H^* \sim H,  \\
&& H^\top\sigma^2 \sim H^\dag, \\
&&\sigma^2 A^{\top} \sigma^2 \sim \sigma^2 A^* \sigma^2 \sim A
\sim A^\dag, \label{eq:translikeA} \\
&&\sigma^2 B^{\top} \sigma^2 \sim  B, \label{eq:translikeB}
\end{eqnarray}
where the symbol $\sim$ stands for "...transforms like... under $SU(2)$".
To systematize further the discussion it is useful to define
the $2 \times 2$ matrices
\begin{eqnarray}
&&\mathbb{H}_0 = H H^\dag, \\
&&\mathbb{H}_2 = H H^\top\sigma^2 
\end{eqnarray}
$\mathbb{H}_0$ is hermitian and transforms like $B$ under $SU(2)_L \times U(1)_Y$
but has a non-vanishing trace, while $\mathbb{H}_2$ is traceless
and transforms like $A$ under $SU(2)_L \times U(1)_Y$.
From the tracelessness of $\mathbb{H}_2, A$ and $B$, Eq.~(\ref{eq:identity3}) implies
\begin{equation}
\sigma^2 \mathbb{H}_2^{\top} \sigma^2 = -\mathbb{H}_2, \, \sigma^2 \mathbb{H}_2^{*} \sigma^2 = -\mathbb{H}_2^\dag,\, \sigma^2 A^{\top} \sigma^2 = -A, \, \sigma^2 A^{*} \sigma^2 = -A^\dag,  \, \sigma^2 B^{\top} \sigma^2 =-B, \label{eq:sig2Asig2}
\end{equation}
thus trivializing Eqs.~(\ref{eq:translikeA}, \ref{eq:translikeB}).

Similarly, Eq.~(\ref{eq:identity4}) leads to,
\begin{equation}
(\mathbb{H}_2)^2 = \frac12 {\mathbf 1} Tr (\mathbb{H}_2)^2, \, A^2 = \frac12 {\mathbf 1} Tr A^2, \, B^2 = \frac12 {\mathbf 1} Tr B^2 . \label{eq:A2B2}
\end{equation}
For the sake of conciseness we do not write here other useful relations resulting from Eqs.~(\ref{eq:identity3}, \ref{eq:identity4}), 
involving $\mathbb{H}_0$ or products involving $A,B$ (generalizing Eq.~(\ref{eq:identity1})).
We have now all the ingredients to show that the 
 $SU(2)_L \times U(1)_Y$ invariant
operators in Eqs.~(\ref{eq:p-c-2-3}, \ref{eq:p-c-4}) form a complete and independent set: Any such operator is necessarily
either in the form of a trace of a $2\times2$ matrix operator that is neutral under $U(1)_Y$ and
constructed from a product of fields that transform similarly under $SU(2)_L$,   
or in the form of a product of such traces that are separately either neutral or charged under  $U(1)_Y$.
(Recall that the other invariant quantity, the determinant, is always expressible in terms of traces).
We sketch hereafter the main steps with some examples.

\begin{itemize}
\item[dim-2:] the list of all $U(1)_Y$ neutral operators is $\mathbb{H}_0,B^2, A A^\dag, A^\dag A$; recall that $Tr\mathbb{H}_0 = H^\dag H$.
\item[dim-3:] the exhaustive list of representative $U(1)_Y$ neutral operators is
 $B^3, \mathbb{H}_0 B, \mathbb{H}_2 A^\dag, A A^\dag B$. 
 Only the first one drops out after tracing, since $TrB^3=0$ as an immediate consequence of Eq.~(\ref{eq:A2B2}) and $Tr B=0$.
 All other neutral dim-3 operators obtained from the above list by arbitrary permutations
 of the fields or by substituting a field by its transpose or complex conjugate are, upon tracing, related to this list. 
 This is obtained by successive use of Eqs.~(\ref{eq:sig2Asig2}, \ref{eq:A2B2}) and the like. E.g.
 $Tr A A^\dag B^*$ is forbidden since $B^*$ does not transform like $A$, while $Tr A^* A^\top B^*$ is allowed but redundant:
 $Tr A^* A^\top B^* =(Tr A A^\dag B)^* = -(Tr A^\dag A B)^*= -Tr A^\top A^* B^*= + Tr A^\top A^* \sigma^2 B \sigma^2 = +Tr A A^\dag B$,
 where we used Eq.~(\ref{eq:identity4}) and the tracelessness of $A,B$ for the second equality, and Eq.~(\ref{eq:sig2Asig2}) for the last
 two equalities. 
\item[dim-4:] the exhaustive list of representative $U(1)_Y$ neutral operators is $\mathbb{H}_0 \mathbb{H}_0, \mathbb{H}_2 \mathbb{H}_2^\dag, \mathbb{H}_0 B^2 ,\mathbb{H}_0 A A^\dag, \mathbb{H}_2 A^\dag B,B^4, A B A^\dag B, A A^\dag A A^\dag$. Note that products of two dim-2 traced operators should also be added. Thus a systematic strategy would be to reduce in the above list the traces of the product of 
four matrices to products of two traces, whenever possible. This is done using the same tricks as illustrated for dim-3. E.g.,
$Tr B^4 = Tr (\frac12 {\mathbf 1} Tr B^2) B^2 = \frac12 (Tr B^2)^2$  as a consequence of Eq.~(\ref{eq:A2B2}), or
$Tr A B A^\dag B = -  Tr A A^\dag B^2 + 
Tr A (Tr A^\dag B) B  = -  \frac12 (Tr A A^\dag )(Tr B^2) +   (Tr A B) (Tr A^\dag B)$.
Note that $Tr A A^\dag A A^\dag$ can be transformed similarly but we chose not to do so in Eq.~(\ref{eq:p-c-4}) so as to keep close
to the notations in the literature. Finally, it is immediate from the list above, that there exists only one operator containing $\sigma^2$
up to complex conjugation, $Tr \mathbb{H}_2 A^\dag B$.
\end{itemize}

\section{Proofs of properties of~the~{\sl $\alpha$-parameters}\label{appendix:alphas}}

\subsubsection{$\alpha_{AB} \in[0,1]$}
Thanks to gauge invariance and to the fact that $B$ is self-adjoint one can always find, 
for each given value of $\alpha_{AB}$, an $SU(2)_L$ transformation ${\cal U}_L$ that diagonalizes
$B$ leading to
\begin{equation}
\alpha_{AB} \equiv \frac{Tr \tilde{A} B_d \, Tr \tilde{A}^\dagger B_d}{Tr \tilde{A}^\dagger \tilde{A} \, Tr (B_d^2)}
\end{equation}
where
\begin{eqnarray}
\tilde{A}={\cal U}_L A {\cal U}_L^\dagger~~ {\rm and}~~ B_d &= {\cal U}_L B {\cal U}_L^\dagger \equiv b_d
                  \displaystyle \left(
\begin{array}{cc}
1 & 0 \\
0 & -1\\
\end{array}
\right)  \ .
\end{eqnarray}
Then all dependence on $B$ drops out from $\alpha_{AB}$, and one is then left with
\begin{equation}
\alpha_{AB} = \frac{|\tilde{a}^+|^2}{|\tilde{a}^0|^2 + |\tilde{a}^+|^2 + |\tilde{a}^{++}|^2}, \label{eq:alfaABdiag}
\end{equation}
 from which Eq.~(\ref{eq:alfaABrange}) follows immediately when $\tilde{A}$ scans all its
field space values. It is to be stressed that appealing to gauge invariance is essential for the proof;
indeed, without gauge invariance, one would still be at liberty to 
choose the $B$-field space direction such as 
$b^+ =0$, leading through Eq.~(\ref{eq:alfaABdiag}) to the same result, however this would be no proof that
$\alpha_{AB}$ remains in the $[0,1]$ domain in other field directions. This is similar to the reason why we believe
the determination of the BFB conditions in \cite{Hartling:2014zca} for the Georgi-Machacek model lacks
a complete proof, a version of which we give in Sec.~\ref{sec:GM-BFB}.  

\subsubsection{$\alpha_{ABH} \in [-\sqrt 2 , +\sqrt 2]$}
Again, in the gauge where the real field $B=B_d$ (and denoting  the components of the gauge transformed $H$ and $A$ fields
with a tilde),  $\alpha_{ABH}$ defined in Eq.~(\ref{eq:def-alphas2}) takes the form:
  \begin{equation}
\alpha_{ABH} = \sqrt{2} \,\sgn(b) \, \frac{ \operatorname{Re} ( \tilde{a}^{++} (\tilde{\phi}^{+*})^2  -\tilde{a}^0 (\tilde{\phi}^{0*})^2 )}
    { \sqrt{|\tilde{a}^0|^2 + |\tilde{a}^+|^2 + |\tilde{a}^{++}|^2} (|\tilde{\phi}^0|^2  +|\tilde{\phi}^+|^2)}.
\end{equation} 
Using the fact that $-|z| \leq \operatorname{Re}(z) \leq |z|$ for any complex number $z$, one immediately finds
\begin{equation}
-  \frac{ |\tilde{a}^{++}||\tilde{\phi}^{+}|^2  + |\tilde{a}^0||\tilde{\phi}^{0}|^2 }
    { \sqrt{|\tilde{a}^0|^2 + |\tilde{a}^+|^2 + |\tilde{a}^{++}|^2} (|\tilde{\phi}^0|^2  +|\tilde{\phi}^+|^2)}\leq \frac{\alpha_{ABH}}{\sqrt{2}} \leq \frac{ |\tilde{a}^{++}||\tilde{\phi}^{+}|^2  + |\tilde{a}^0||\tilde{\phi}^{0}|^2 }
    { \sqrt{|\tilde{a}^0|^2 + |\tilde{a}^+|^2 + |\tilde{a}^{++}|^2} (|\tilde{\phi}^0|^2  +|\tilde{\phi}^+|^2)},
\end{equation}
where the upper (lower) bound is effectively reached in the field directions where $\operatorname{arg}(\tilde{a}^{++} (\tilde{\phi}^{+*})^2)=0 (\pi)$, $\operatorname{arg}(\tilde{a}^0 (\tilde{\phi}^{0*})^2)=\pi (0)$. 
Moreover, since $\sqrt{|\tilde{a}^0|^2 + |\tilde{a}^+|^2 + |\tilde{a}^{++}|^2} \geq \sqrt{|\tilde{a}^0|^2  + |\tilde{a}^{++}|^2}$, $\alpha_{ABH}$ scans a larger domain 
in the direction $\tilde{a}^+=0$, namely 
\begin{equation}
- \sqrt{2} \frac{ |\tilde{a}^{++}||\tilde{\phi}^{+}|^2  + |\tilde{a}^0||\tilde{\phi}^{0}|^2 }
    { \sqrt{|\tilde{a}^0|^2 + |\tilde{a}^{++}|^2} (|\tilde{\phi}^0|^2  +|\tilde{\phi}^+|^2)}\leq \alpha_{ABH} \leq \sqrt{2} \frac{ |\tilde{a}^{++}||\tilde{\phi}^{+}|^2+ |\tilde{a}^0||\tilde{\phi}^{0}|^2 }
    { \sqrt{|\tilde{a}^0|^2 +  |\tilde{a}^{++}|^2} (|\tilde{\phi}^0|^2  +|\tilde{\phi}^+|^2)}.  
\end{equation}
Defining $x =  |\tilde{\phi}^{+}|/|\tilde{\phi}^{0}|$ and $y=|\tilde{a}^{++}|/|\tilde{a}^{0}|$, 
the above domain is rewritten as
\begin{equation}
-f(x,y) \leq \alpha_{ABH} \leq +f(x,y),
\end{equation}
with 
\begin{equation}
f(x,y)= \sqrt{2} \frac{1+y x^2 }{(1+x^2) \sqrt{(1+y^2)} }.
\end{equation}
Noting that $f(x, y) = f(1/x, 1/y)$, a straightforward study of the function $f(x,y)$ in the domain $x,y \in[0, +\infty)$ shows that it possesses a saddle point at $x=y=1$ and reaches 
a global maximum at $x=y=0$ and at $x=y \to +\infty$ given by $f(0,0)= f_{max}= \sqrt{2}$, whence Eq.~(\ref{eq:alfaHABrange}). Incidentally, 
we note that the
ill-defined point $H=0, A=0$ in $\alpha_{ABH}$ is automatically accounted for through the behavior of $f$. 

\subsubsection{Boundary of the $(\alpha_{AH},\alpha_{ABH})$
domain \label{appendix:alAH-alAHB}}

To prove that Eq.~(\ref{eq:VI}) gives the lower boundary
we show hereafter that 

$\delta \equiv \alpha_{ABH}^2 -2 \, \alpha_{AH}$
is either negative or vanishing.
This combination is of the form 
\begin{equation}
\delta(x) = -1 + a x^2 + b x \sqrt{1 - x^2}, \label{eq:delta}
\end{equation}
with $x \equiv \sin \theta$ and $a,b$ easily read
from Eqs.~(\ref{eq:alABH}, \ref{eq:alAH}),
\begin{eqnarray}
&&a= -\cos 2\varphi \, \cos 2\psi + 2 (c_1 \, \cos\varphi \cos^2\psi -
c_2 \sin\varphi \sin^2\psi)^2, \label{eq:a} \\
&&b= - \sqrt{2} \, (c_3 \cos \varphi + c_4 \sin \varphi) \sin 2 \psi,
\label{eq:b}
\end{eqnarray}
and $c_i \equiv \cos \theta_i$.
The study of the $\delta(x)$ function 
shows that it reaches only one stationary point 
\begin{equation}
\delta^{stationary}= \delta(x_0) = -1 + \frac{b}{2} \ (r + \sqrt{1 + r^2}) \ . \label{eq:deltaextrem}
\end{equation}
for 
\begin{equation}
x =x_0= \frac{1}{\sqrt{2}} \ \sqrt{1 + \frac{r}{\sqrt{1 + r^2}}}
\, \in \left[0, 1\right],
\end{equation}
where we took into account the fact that $0 \leq x \leq 1$, cf.
Eq.~(\ref{eq:angles-range}), and defined $r=a/b$. 

One also finds
\begin{equation}
\frac{d^2 \delta(x)}{d x^2} |_{x=x0}=
-4 \, b \, (1 + r^2) (r + \sqrt{1 + r^2}) \ ,
\end{equation}
so that it is only the sign of $b$ that dictates whether this stationary point is a maximum ($b >0$) or a minimum ($b<0$). 
Note that $b$ as defined  in Eq.~(\ref{eq:b}) can take
either signs since $c_3,c_4 \in [-1,1]$. To proceed we consider
the two cases:

\begin{itemize}
\item $b <0$, $\delta(x_0)$ is a minimum: In this case 
one has $\delta(x) \leq 0$ $\forall x \in[0,1]$, if and only if
$\delta(0) \leq 0$ and $\delta(1) \leq 0$. The first condition is trivially satisfied. The second is equivalent to
$a \leq 1$. Since $\varphi, \psi \in [0, \frac{\pi}{2}]$ and $c_1,c_2$
vary independently in $[-1, 1]$ it follows from Eq.~(\ref{eq:a})
that, for fixed $\varphi$ and $\psi$, $a$ reaches a maximum
when $c_1=-c_2= \pm 1$. One thus has
\begin{eqnarray}
a &\leq& -\cos 2\varphi \, \cos 2\psi + 2 (\cos\varphi \cos^2\psi +
\sin\varphi \sin^2\psi)^2 = \frac14 (3 + \cos 4 \psi \, + \, 
2 \sin 2 \varphi \, \sin^2 2 \psi) \nonumber \\
  &\leq& \frac14 (3 + \cos 4 \psi \, + \, 
2  \, \sin^2 2 \psi) = 1,  \label{eq:aleq1}
\end{eqnarray}  
and $\delta(1) \leq 0$ as required.
\item $b >0$, $\delta(x_0)$ is a maximum: In this case 
one has $\delta(x) \leq 0$ $\forall x \in[0,1]$, if and only if
$\delta(x_0) \leq 0$. 
From Eq.~(\ref{eq:deltaextrem}) and the fact that $b >0$, the condition $\delta(x_0) \leq 0$ can be rewritten as
\begin{equation}
\sqrt{a^2 + b^2} \leq 2 - a . \label{eq:ineqbpos}
\end{equation}
This in turn is  {\sl equivalent} to
\begin{equation}
4 (1 - a) - b^2  \geq 0 , 
\end{equation}
since $a \leq 1$, Eq.~(\ref{eq:aleq1}) being valid 
independently of the sign of $b$. Expressing  $\sin \psi$ and
$\cos \psi$ in terms of $T\equiv \tan \psi$, the above inequality 
is equivalently rewritten as
\begin{equation}
  a_0+ a_2 \, T^2 + a_4 T^4 \geq 0 , \label{eq:ineqbpos1}
\end{equation}
with $T \in [0, +\infty)$ and 
\begin{eqnarray}
&&a_0=(1-c_1^2) \cos^2\varphi , \\
&&a_2=1  + c_1 c_2 \sin 2 \varphi - (c_3 \cos\varphi + c_4 \sin \varphi)^2
, \\
&&a_4= (1-c_2^2) \sin^2 \varphi .
\end{eqnarray}
We can now examine the NAS positivity conditions for 
the biquadratic polynomial in $T$, namely
 \begin{equation}
 a_0 \geq 0 \, \land \, a_4 \geq 0 \, \land \,  a_2 + 2 \sqrt{a_0 a_4} \geq 0. \label{eq:NAScondT}
 \end{equation} 
 The first two are trivially satisfied. To prove the third
 we should take into account the correlation $\theta_1= \theta_2 + 
 \theta_3 + \theta_4$ (modulo multiples of $2 \pi$), 
 see Eq.~(\ref{eq:t1-t4}). Moreover, since
 $\sin\theta_1$ and $\sin \theta_2$ can take either signs, we
 can include the two cases by simply using
 the inequality $\sqrt{(1-c_1^2)(1-c_2^2)}~\geq~\sin\theta_1 
 \, \sin \theta_2$ to write:
 \begin{eqnarray}
 a_2 + 2 \sqrt{a_0 a_4} &\geq& 1 + \sin 2 \varphi \, \cos (\theta_1 - \theta_2) - (\cos\theta_3 \cos\varphi + \cos\theta_4 \sin \varphi)^2 .
 \end{eqnarray}
 Using $\theta_1= \theta_2 + \theta_3 + \theta_4$ the right-hand
 side of the above inequality simplifies to
 \begin{equation}
 1 + \sin 2 \varphi \, \cos (\theta_3 + \theta_4) - (\cos\theta_3 \cos\varphi + \cos\theta_4 \sin \varphi)^2 = (\sin\theta_3 \cos\varphi - \sin \theta_4 \sin \varphi)^2 .
 \end{equation} 
 Thus the third NAS positivity condition in Eq.~(\ref{eq:NAScondT})
 is valid for all
 values of the angles $\varphi$ and $\theta_i$. This implies that
  Eq.~(\ref{eq:ineqbpos1}) is satisfied for all $T \geq 0$ thus for
  all values of $\psi$. Eq.~(\ref{eq:ineqbpos}) then holds for all
  the values of the angles, in particular those compatible with
  $b>0$; we have thus proven that $\delta(x) \leq 0$ $\forall x \in[0,1]$ in this case too. 
\end{itemize}
This ends the proof that $\alpha_{ABH}^2 -2 \, \alpha_{AH} \leq 0$
holds for all field directions and that Eq.~(\ref{eq:VI}) gives the lower boundary in the $(\alpha_{ABH}, \alpha_{AH})$ plane.

\subsubsection{Boundary of the $(\alpha_{A},\alpha_{ABH})$
domain \label{appendix:alA-alAHB}}

We rewrite Eq.~(\ref{eq:alABH}) as 
\begin{equation}
 -c_1 \sqrt{1-x^2} ( 1 - y^2) + c_2 x y^2 = \alpha_{ABH}
 \label{eq:alABH-template}
\end{equation}
with the obvious notations, $x= \sin \varphi, y = \sin \psi, 
c_1 = \sqrt{2} \cos \theta_1 \sin \theta, c_2 = \sqrt{2} \cos \theta_2 \sin \theta$.
We seek the conditions on $\alpha_{ABH}, c_1$ and $c_2$ that
ensure the existence of at least one value for $x \in [0, 1]$ for each
value of $y^2 \in [0, 1]$ and vice versa. This can be worked out by
solving for $y^2$ and considering the (relative) signs of $c_1$ and $c_2$. One
finds:
\begin{itemize}
\item When $c_1 \times c_2 \geq 0$, $\alpha_{ABH}$ can be of any sign, with $x$ and $\alpha_{ABH}$ satisfying 
\begin{equation}
0 \leq \frac{\alpha_{ABH}}{c_2} \leq x,
\; {\rm or} \;
x^2 \leq 1 - \frac{\alpha_{ABH}^2}{c_1^2} \; \left({\rm when} \; \frac{\alpha_{ABH}}{c_2} \leq 0 \right) \ . \label{eq:cond1}
\end{equation}
Thus upper and/or lower parts of the
$[0, 1]$ domain for $x$ will not be reached $\forall y^2 \in [0, 1]$, 
unless \begin{equation} \alpha_{ABH}=0 . \end{equation}
\item When $c_1 \times c_2 \leq 0$, $\alpha_{ABH}$ and $c_2$ should
have the same sign, with $x$ and $\alpha_{ABH}$ satisfying
\begin{equation}
0 \leq x^2 \leq \min\left\{\frac{\alpha_{ABH}^2}{c_2^2}, {1 - \frac{\alpha_{ABH}^2}{c_1^2}}, \frac{ c_1^2}{{c_1^2 + c_2^2}}\right\}
\; {\rm or} \; \max\left\{\frac{\alpha_{ABH}^2}{c_2^2}, {1 - \frac{\alpha_{ABH}^2}{c_1^2}}, \frac{ c_1^2}{{c_1^2 + c_2^2}}\right\}
\leq x^2 \leq 1 \ . \label{eq:cond2}
\end{equation}
Thus intermediate parts of the 
$[0, 1]$ domain for $x$ will not be reached $\forall y^2 \in [0, 1]$, 
unless 
\begin{equation}
\alpha_{ABH} = (\alpha_{ABH})_{crit} \equiv \frac{ |c_1| c_2}{\sqrt{c_1^2 + c_2^2}} \label{eq:many-to-many-condition}.
\end{equation}
\end{itemize}
Since $|c_1|, |c_2| \in[0, \sqrt{2}]$, the maximum value for
$|\alpha_{ABH}|$ from Eq.~(\ref{eq:many-to-many-condition}) is 
obtained when $|c_1| = |c_2| = \sqrt{2}$, and corresponds to the
maximal critical value 
$|(\alpha_{ABH})_{crit}^{max}| =1$, not $\sqrt{2}$ ! A direct consequence
is the absence of correlations between $\alpha_{ABH}$ and $x$
or $\alpha_{ABH}$ and $y^2$ in the domain $\alpha_{ABH} \in[-1, 1]$, i.e the square $[-1, 1] \times [0, 1]$ is totally filled in both
cases. Indeed, for any given $\alpha_{ABH} \in [-1, 1]$ 
one can always find $c_1$ and $c_2$ of opposite signs satisfying 
Eq.~(\ref{eq:many-to-many-condition}) so that to any $x$ 
corresponds at least one $y^2$ and vice versa, thus varying freely 
in $[0, 1]$. Note also that $|c_1| = |c_2| = \sqrt{2}$ entails 
maximizing $|\cos \theta_1|, |\cos \theta_2|$ and $\sin \theta$ to 
$1$. 

We can study now the allowed domain in the plane $(\alpha_{ABH}, \alpha_{A})$. 
We first  determine the allowed 
$(\alpha_{ABH}, \alpha_{A})$ sub-domain corresponding to $\sin \theta=1$, then show that all sub-domains that correspond to $\sin \theta < 1$ are necessarily 
within that sub-domain, which thus turns out to be the full $(\alpha_{ABH}, \alpha_{A})$ domain. 

When $\sin \theta=1$ the dependence on $\cos \rho$ drops out from  Eq.~(\ref{eq:alA}) and one can easily solve for 
$x(=\sin \varphi)$ as a function of $\alpha_{A}$,
\begin{equation}
x_\pm = \sqrt{\frac12 \left( 1 \pm \sqrt{2 \alpha_{A} - 1}\right)}
\ . \label{eq:x-alA}
\end{equation} 
The two $\pm$ solutions should be kept in the discussion 
as their union scans the full 
$[0, 1]$ domain of $x$ allowing $\alpha_{A}$ to scan all its
allowed domain $[\frac12, 1]$. Similarly,
since $\sin \theta = 1$,   $\alpha_{ABH}$ will scan
all its allowed domain $[-\sqrt{2}, +\sqrt{2}]$ by varying
$x$, $y$, $c_1$ and $c_2$. Let us choose a couple of values
$(\alpha_{ABH}, \alpha_{A})$ in their respective domains.
\begin{itemize}
\item[-]If $|\alpha_{ABH}| \leq 1$ then, relying on what was demonstrated after Eq.~(\ref{eq:many-to-many-condition}), one can always find
$c_1, c_2$ (or equivalently $\cos \theta_1, \cos \theta_2$) with
opposite signs and the sign of $c_2$ being that of $\alpha_{ABH}$, in such a way that $\alpha_{ABH} = (\alpha_{ABH})_{crit}$. It
follows that for any $x \in[0,1]$ there exists $y^2 \in[0,1]$ consistent with the given value of $\alpha_{ABH}$. In particular this is true
for the values of $x$ corresponding, through Eq.~(\ref{eq:x-alA}),
to any given value of $\alpha_{A}$. There is thus no obstruction
on the independent choice of the values of $\alpha_{ABH}$ and $\alpha_{A}$ as long as $|\alpha_{ABH}| \leq 1$. {\sl It follows that the entire square $[-1,1] \times [\frac12, 1]$ is allowed
in the  $(\alpha_{ABH}, \alpha_{A})$ plane.}
\item[-]If $|\alpha_{ABH}| > 1$, one has to examine separately 
the conditions given by Eqs.~(\ref{eq:cond1}, \ref{eq:cond2}).
Note also that since $|\alpha_{ABH}| > 1$ the $\min$ and $\max$
in Eqs.~(\ref{eq:cond2}) become uniquely defined, equaling  respectively 
$1 - \frac{\alpha_{ABH}^2}{c_1^2}$ and $\frac{\alpha_{ABH}^2}{c_2^2}$. Plugging $x$ as given by Eq.~(\ref{eq:x-alA}) in the four
inequalities, it is clear that a necessary condition in each case
obtains when $c_1$ and $c_2$ take their extreme values $\pm \sqrt{2}$. Taking consistently into account the various sign conditions
in each case as well as the $\pm$ in Eq.~(\ref{eq:x-alA}) one determines the necessary condition relating $\alpha_A$ and $\alpha_{ABH}$. One finds exactly the same inequality in the four cases, namely $\alpha_{A} \geq 1 - \alpha_{ABH}^2 + \frac12 \alpha_{ABH}^4$.  
Moreover, this conditions is also sufficient since it allows
at least the extremal values of $c_1, c_2$. Thus
\begin{equation}
\alpha_{A} = 1 - \alpha_{ABH}^2 + \frac12 \alpha_{ABH}^4 = \frac12 \left(1 + (\alpha_{ABH}^2-1)^2 \right)\ 
\label{eq:boundary-alA-alABH}
\end{equation}
{\sl gives the lower boundary for  $\alpha_{A}$ when $\alpha_{ABH} >1$.}
\end{itemize}
This completes the proof that when $\sin \theta =1$, the allowed domain in the 
$(\alpha_{ABH},\alpha_{A})$ plane is as defined by Eqs.~(\ref{eq:boundary-alA-alABH1-3})
 and illustrated in Fig.~\ref{fig:alfa-alfabh}.

Since $\theta_1$ and $\theta_2$ appear only in 
$\alpha_{ABH}$, 
they can be safely chosen without biasing the correlations between
$\alpha_A$ and $\alpha_{ABH}$, as long as they maximize the allowed
domain of the latter. 
The angle $\theta$ is however common to 
$\alpha_A$ and $\alpha_{ABH}$. One should then be careful that
the value $\sin \theta = 1$ does not miss points in the allowed 
domain.  A necessary condition for this not to happen is that 
$\sin \theta = 1$ still allows $\alpha_A$ and $\alpha_{ABH}$ to 
take
any value in their respective domains as given by 
Eqs.~(\ref{eq:alfaArange}, \ref{eq:alfaHABrange}). This is indeed 
the case
as one can check from Eqs.~(\ref{eq:alA}, \ref{eq:alABH}) by 
varying all the other angles at fixed $\sin \theta = 1$. 

However
this is not sufficient. One should still check that for $\sin \theta$
strictly smaller than one  
there exists no set of values for the remaining  angle variables 
 giving a point in the $(\alpha_{ABH}, \alpha_A)$ plane that is {\sl outside} the
 domain defined by Eqs.~(\ref{eq:boundary-alA-alABH1-3}). To show this it suffices 
 to prove (cf. Eq.~(\ref{eq:boundary-alA-alABH})) that
\begin{equation}
2 \alpha_{A} - 1 - \left(\alpha_{ABH}^2-1 \right)^2 \geq 0, \ \forall \sin \theta ,
\label{eq:inequality-alA-alABH}
\end{equation}
whenever 
\begin{equation}
|\alpha_{ABH}| > 1 \ . \label{eq:ABHcond}
\end{equation}
Rewriting Eq.~(\ref{eq:alABH}) as
\begin{equation}
\alpha_{ABH}= \sqrt{2} Y \sin \theta,
\end{equation}
where 
\begin{equation}
 Y= \sin \varphi\,\sin^2 \psi\,\cos \theta_2 -\cos \varphi\,\cos^2 \psi\,\cos \theta_1 \label{eq:ZZ}
 \end{equation}
and $Y \in [-1,1]$, condition (\ref{eq:ABHcond}) implies
\begin{equation}
|Y| \geq \frac{1}{\sqrt{2}} \ \text{and} \ \sin \theta \geq \frac{1}{\sqrt{2}} , \label{eq:Zsinbound}
\end{equation}
since none of $|Y|$ and $\sin \theta$ can exceed one. We can thus replace Eq.~(\ref{eq:ABHcond})
by
\begin{eqnarray}
1 \geq &|Y|& \geq \frac{1}{\sqrt{2}} \label{eq:Zbound} \\
&\text{and}& \nonumber \\
1 \geq &\sin \theta& \geq \frac{1}{\sqrt{2} |Y|} .\label{eq:sinbound}
\end{eqnarray}
On the other hand, as seen from Eqs.~(\ref{eq:alA}, \ref{eq:alABH}), the only dependence on the angle
$\rho$ in Eq.(\ref{eq:inequality-alA-alABH}) is linear in $\cos \rho$ and with a positive coefficient:
\begin{equation}
\frac14 \sin 2\varphi \,\sin^2 2 \theta \, \cos \rho\, + ... \geq 0 \ .
 \end{equation}
 Condition (\ref{eq:inequality-alA-alABH})
 is thus equivalent to the one where $\cos \rho$ takes its minimal value $\cos \rho = -1$, in which case
 (\ref{eq:inequality-alA-alABH}) can be recast in the form
\begin{equation}
 a_4 \tau^4 + a_2 \tau^2 -1 \geq 0
\label{eq:inequality-alA-alABH1}
\end{equation}
with
\begin{eqnarray}
a_2&=&2 \left(2 Y^2 - \sin 2 \varphi \right), \\
a_4&=&-4 \left(Y^2 -\cos^2 \varphi \right) \left(Y^2 -\sin^2 \varphi \right),
\end{eqnarray}
where we defined $\tau = \tan \theta$ and dropped out a positive denominator. 
The coefficients of $\tau^2$ and $\tau^4$ in Eq.~(\ref{eq:inequality-alA-alABH1}) both satisfy 
\begin{eqnarray}
a_2 &\geq& 0, \label{eq:a2pos}\\
a_4 &\geq& 0 . \label{eq:a4pos}
\end{eqnarray}
as a consequence of the lower bound in Eq.~(\ref{eq:Zbound}). The first is immediate to establish.
The positivity of $a_4$ is less obvious. Rewriting $|Y| \geq 1/\sqrt{2}$ and $a_4$ respectively as
\begin{eqnarray}
0 &\leq& (Y - \frac{1}{\sqrt{2}})(Y + \frac{1}{\sqrt{2}}), \label{eq:Zbound1}\\
a_4 &=& -4 (Y - \cos \varphi) (Y + \cos \varphi) (Y - \sin \varphi) (Y + \sin \varphi), \label{eq:a_4}
\end{eqnarray}
and noting that $Y$ is linear in $\cos \theta_1$ and $\cos \theta_2$, cf. Eq.~(\ref{eq:ZZ}),
one can easily study the sign of $a_4$ when Eq.~(\ref{eq:Zbound1}) is satisfied, in terms of a bundle of six
parallel straight lines with slope $\cot \varphi \cot^2\psi$ in the $(\cos \theta_1, \cos \theta_2)$ plane; 
the sign alternates each time one of
these lines is crossed. Moreover, since they are all parallel it suffices to study the change of sign
along a given axis in the $(\cos \theta_1, \cos \theta_2)$ plane, say the axis defined by $\cos \theta_2 =0$. 
On this axis the inequality Eq.~(\ref{eq:Zbound1})
is satisfied if and only if
\begin{equation}
\frac{1}{\sqrt{2} \cos \varphi \cos^2 \psi}  \leq \cos \theta_1 \leq 1 \;\; \text{or} \;\; 
-1 \leq \cos \theta_1 \leq-\frac{1}{\sqrt{2} \cos \varphi \cos^2 \psi} . \label{eq:Zbound2}
\end{equation}
This implies
\begin{equation}
\cos \varphi \geq \frac{1}{\sqrt{2} \cos^2 \psi} \geq \frac{1}{\sqrt{2}},
\end{equation}
thus
\begin{equation}
\sin \varphi \leq \frac{1}{\sqrt{2}} \; \text{and} \; \tan \varphi \leq 1 . \label{eq:sintan}
\end{equation}
On the other hand, it is easily seen from Eqs.~(\ref{eq:ZZ}, \ref{eq:a_4}), (with $\cos \theta_2 =0$),
 that $a_4$ is positive
if and only if $\cos \theta_1$ is between $1/\cos^2\psi$ and $\tan\varphi/\cos^2\psi$ 
or between $-1/\cos^2\psi$ and $-\tan\varphi/\cos^2\psi$, and negative otherwise.
Using Eq.~(\ref{eq:sintan}) these conditions read,
\begin{equation}
\frac{\tan\varphi}{\cos^2\psi} \leq \cos \theta_1 \leq \frac{1}{\cos^2\psi} \; \text{or} \; 
- \frac{1}{\cos^2\psi}\leq \cos \theta_1 \leq -\frac{\tan\varphi}{\cos^2\psi} . \label{eq:a4poscond}
\end{equation}
And, again from Eq.~(\ref{eq:sintan}),
\begin{equation}
\frac{\tan\varphi}{\cos^2\psi} \leq \frac{1}{\sqrt{2} \cos \varphi \cos^2 \psi},
\end{equation}
which shows that Eq.~(\ref{eq:a4poscond}) is satisfied whenever Eq.~(\ref{eq:Zbound2}) (or equivalently Eq.~(\ref{eq:Zbound})), 
is satisfied. 
Thus condition (\ref{eq:Zbound}) impliques $a_4 \geq 0$.  It is easy to see that this property remains true even when $\cos \theta_2 \neq 0$. Indeed if (\ref{eq:Zbound}) is satisfied for a given point $(\cos \theta_1, \cos \theta_2)$, then it remains true on all the straight
line with slope $\cot \varphi \cot^2\psi$ going through this point, in particular for the point intersecting the axis $\cos \theta_2=0$, 
and we are brought back to the known case.  

Now back to Eq.~(\ref{eq:inequality-alA-alABH1}):  The domain of variation of $\tau^2$
corresponding to Eq.~(\ref{eq:sinbound}) is given by 
\begin{equation}
\frac{1}{2 Y^2 -1} \leq \tau^2 < +\infty \ .
\end{equation}
Moreover, the quadratic function
in $\tau^2$ is a monotonically increasing function as can be seen from its derivative and Eqs.~(\ref{eq:a2pos}, \ref{eq:a4pos}).
Its minimum is thus reached for $\tau^2_{min} = \frac{1}{2 Y^2 -1}$ and is given by
\begin{equation}
a_4 \tau^4_{min} + a_2 \tau^2_{min} -1 = \frac{2 (1 - \sin 2 \varphi)\;( 4 Y^2 + \sin 2 \varphi -1 )}{(1 - 2 Y^2)^2},
\end{equation}
which is obviously positive when Eq.~(\ref{eq:Zbound}) is satisfied. Thus Eq.~(\ref{eq:inequality-alA-alABH1}) is always satisfied
whenever Eqs.~(\ref{eq:Zbound}, \ref{eq:sinbound}). This completes the proof that Eq.~(\ref{eq:inequality-alA-alABH}) is satisfied
whenever Eq.~(\ref{eq:ABHcond}) holds and that  the full allowed domain in the 
$(\alpha_{ABH},\alpha_{A})$ plane is given by Eqs.~(\ref{eq:boundary-alA-alABH1-3}).  

\subsubsection{Boundary of the $(\alpha_{AB},\alpha_{AH})$
domain \label{appendix:alAB-alAH}}
From Eq.~(\ref{eq:alAH}) one sees that $\alpha_{AH}$ is of the form
\begin{equation}
\alpha_{AH}(x)= \frac12 + a x + b \sqrt{1 - x^2}, \ \text{with} \ x \in [-1, +1],
\end{equation}
where we defined $x\equiv \cos 2 \psi$, and $a,b$ are readily obtained from Eqs.~(\ref{eq:alAB},\ref{eq:alAH}),
\begin{eqnarray}
a &=& \frac12 (1- \alpha_{AB}) \cos 2 \varphi  , \label{eq:anew}\\
b &=& \frac{1}{\sqrt{2}} \ (\cos \theta_3 \cos \varphi + \cos \theta_4 \sin \varphi) \sqrt{(1- \alpha_{AB}) \alpha_{AB}} \label{eq:bnew} \ .
\end{eqnarray}
It is easy to study the structure of maxima and minima of $\alpha_{AH}(x)$ at fixed $a,b$. One finds that
it always has only one stationary point, at $\displaystyle x=\frac{a \sgn{b}}{\sqrt{a^2 + b^2}} \in [-1, +1]$, given by
\begin{equation}
\alpha_{AH}^{stationary} = \frac12 + \sqrt{a^2 + b^2} \sgn{b}.
\end{equation}
Moreover, this stationary point is found to be a minimum (resp. maximum) when $b<0$ (resp. $b>0$), 
and thus with a corresponding maximum (resp. minimum) of $\alpha_{AH}$
given by $\max\{\alpha_{AH}(\pm 1)\}$ (resp. $\min\{\alpha_{AH}(\pm 1)\}$). This leads to:
\begin{eqnarray}
 \frac12 - \sqrt{a^2 + b^2} \leq &\alpha_{AH}& \leq \frac12 + |a|, \ \rm{iff} \ b \leq 0, \\
 \frac12 - |a| \leq &\alpha_{AH}& \leq \frac12 + \sqrt{a^2 + b^2} , \ \rm{iff} \ b \geq 0 .
\end{eqnarray}
The parameter $b$ as defined by Eq.~(\ref{eq:bnew}) can take either signs when all the angles are varied (since $\cos \theta_3, \cos \theta_4 \in [-1,1]$ and $\displaystyle \varphi \in [0, \frac{\pi}{2}]$, cf. Eq.~(\ref{eq:angles-range})). It is thus more relevant to
combine the $\alpha_{AH}$ domains given above, reducing them for fixed $a$ and $|b|$ to
\begin{eqnarray}
 \frac12 - \sqrt{a^2 + b^2} \leq \alpha_{AH}  \leq \frac12 + \sqrt{a^2 + b^2} , \label{eq:alAH-bounds}
\end{eqnarray}
or equivalently to
\begin{eqnarray}
 \left(\alpha_{AH} -\frac12 \right)^2 \leq a^2 + b^2 . \label{eq:alAH-bounds1}
\end{eqnarray}
Given Eq.~(\ref{eq:bnew}), the domain in Eq.~(\ref{eq:alAH-bounds1}) is obviously maximized for
$\cos \theta_3 = \cos \theta_4 = \pm 1$. Assuming these values
we now show that $\alpha_{AH}$ will scan its full allowed domain $[0, 1]$, i.e. that 
$a^2 + b^2$ will reach $\displaystyle \frac14$, only when $\displaystyle 0 \leq \alpha_{AB} \leq \frac12$.
We first note from Eqs.~(\ref{eq:anew}, \ref{eq:bnew}) that
$a^2 + b^2$ can be recast in the form,
\begin{equation}
 a^2 + b^2 =
-\frac14 \left(1 + \sin 2 \varphi \right)^2 \ \left(\alpha_{AB} - \frac{\sin 2 \varphi}{1 + \sin 2 \varphi}\right)^2
+\frac14 . \label{eq:a2pb2}
\end{equation}
Since $\sin 2 \varphi \in [0, 1]$, it is clear that $a^2 + b^2$ reaches $\displaystyle \frac14$ iff $\alpha_{AB} = \frac{\sin 2 \varphi}{1 + \sin 2 \varphi} \in [0, \frac12]$ . It then follows from Eq.~(\ref{eq:alAH-bounds}) that all the $\alpha_{AH}$ domain $[0, 1]$ is allowed
when $\alpha_{AB} \in [0, \frac12]$, whence the boundaries given in Eqs.~(\ref{eq:alAB-alAH-bound1} - \ref{eq:alAB-alAH-bound3}). 

Finally, when $\frac12 \leq \alpha_{AB} \leq 1$ the study of 
$a^2 + b^2$ as a function of $\sin 2 \varphi$ in Eq.~(\ref{eq:a2pb2}) shows that $a^2 + b^2$  reaches its 
maximum for $\sin 2 \varphi=1$, given
by 
\begin{equation}
a^2 + b^2 |_{max}= \frac14 - \left(\alpha_{AB}- \frac12\right)^2 .
\end{equation}
Plugging this back in Eq.~(\ref{eq:alAH-bounds1}), gives the largest allowed domain
\begin{equation}
\left(\alpha_{AH} - \frac12\right)^2 + \left(\alpha_{AB}- \frac12\right)^2 \leq \frac14 ,
\end{equation}
whence the half-circle boundary Eq.~(\ref{eq:alAB-alAH-bound4}).

\section{The $(\hat\omega_0, \hat\zeta_0
)$ domain \label{appendix:omega-zeta}}

To simplify the presentation we define: 
\begin{equation}
x= \cos 2 \theta_0,\; \; y= \cos arg(\chi^0),
\label{eq:canxydef}
\end{equation}
so that $\sin 2 \theta_0 = +\sqrt{1 - x^2}$ and 
$-1 \leq x, y \leq 1$, cf. Eqs.~(\ref{eq:angrange1}, \ref{eq:angrange2}). We also define 
\begin{equation}
{w}(x,y)= \frac18 \left(1 - x + 2 \sqrt{2(1 - x^2)} \, y \right)
\label{eq:wdef}
\end{equation}
so that the  $0^{th}$ order $u$ contribution to
$\hat\omega$, Eq.~(\ref{eq:omega0}), reads
\begin{equation}
\hat\omega_0(x,y,\theta_+) =  {w}(x,y) \, \sin^2 \theta_+ \ . \label{eq:omega0prime}
\end{equation} 
For later use we also denote by
$x^{\scriptscriptstyle>}_y$  and $x^{\scriptscriptstyle<}_y$ 
respectively the {\sl largest} and {\sl smallest}  values of 
$x$ satisfying the equation
\begin{equation}
{w}(x^{\scriptscriptstyle>}_y,y)= {w}(x^{\scriptscriptstyle<}_y,y)=\bar{\hat{\omega}}_0 , \label{eq:xupperlower}
\end{equation} 
where $\bar{\hat{\omega}}_0$ is a given value of $\hat\omega_0 \in [-\frac14, \frac12]$. These two values of $x$ are easily 
determined to be 
\begin{equation}
x^{\scriptscriptstyle\gtrless}_y = 
\frac{1 - 8 \bar{\hat{\omega}}_0 \pm 8 \sqrt{y^2 \, \left(2 (1 - 4 \bar{\hat{\omega}}_0) \bar{\hat{\omega}}_0 + y^2\right)}}{
      1 + 8 y^2} . \label{eq:xupperlowersol}
\end{equation}
Note also that they are reached if and only if 
$\sin^2 \theta_+=1$. 

Using Eq.~(\ref{eq:omega0prime}) to eliminate $\sin^2 \theta_+$ 
from Eq.~(\ref{eq:zeta0}) one obtains straightforwardly a 
relation between $\hat\omega_0$ and the $0^{th}$ order 
$u$ contribution to
$\hat\zeta$,
\begin{equation}
\hat\zeta_0(x, y, \hat\omega_0) = 1+ c_1 \, \hat\omega_0 + c_2 \, \hat\omega_0^2  \label{eq:zeta0prime}
\end{equation}
with
\begin{eqnarray}
c_1 &=& - \frac{(1 - x)}{2 \, {w}(x,y)}, \\
c_2 &=& - \frac{(1 - x) \, (1 + 3 x)}{8 {w}(x,y)^2} \ .
\end{eqnarray}
$x$, $y$ and $\hat\omega_0$ can be varied independently of each
other only locally, but they have global correlations due
to Eq.~(\ref{eq:omega0prime}): From $0 \leq \sin^2 
\theta_+  \leq 1$, one must require
\begin{eqnarray}
&{w}(x,y) \leq \hat\omega_0 \leq 0& \label{eq:wboundneg}\\
&\text{or} &\nonumber \\
&0 \leq \hat\omega_0 \leq {w}(x,y)& \label{eq:wboundpos}
\end{eqnarray}

Apart from the special cases 
$\{x=1, \hat\omega_0=0\}$ and $\{x=-1, \hat\omega_0=(1/4) \sin^2 \theta_+ \}$ where $y$ varies
freely in $[-1, +1]$,  the above constraints dictate in general that the allowed ranges for 
$y$ depend on $x(\neq -1, +1)$ and $\hat\omega_0$ as follows:
\begin{eqnarray}
&&\text{if} \, \hat\omega_0 \geq \ 0, \,\text{then} \, 
\max\left\{-1, -\frac{1}{2 \, \sqrt{2}}\sqrt{\frac{1 - x}{1 + x}} + 
    \frac{2 \, \sqrt{2} \hat\omega_0}{\sqrt{1 - x^2}}\right\} \leq y \leq 1\ , \label{eq:consist1}\\
&&\text{if} \, \hat\omega_0 \leq \ 0, \,\text{then} \, -1 \leq y  \leq 
 -\frac{1}{2 \, \sqrt{2}}\sqrt{\frac{1 - x}{1 + x}} + 
    \frac{2 \, \sqrt{2} \hat\omega_0}{\sqrt{1 - x^2}} \leq 0 \ .
\label{eq:consist2}
\end{eqnarray}
We now show the following key property:
\begin{eqnarray} 
\text{\sl $\hat\zeta_0(x, y, \hat\omega_0)$, taken as a function
of $y$, is increasing for $\hat\omega_0 \geq 0$ and decreasing
for $\hat\omega_0 \leq 0$.} \nonumber \\
\label{eq:keyprop}
\end{eqnarray}
 The derivative of $\hat\zeta_0$ reads
\begin{eqnarray}
\frac{\partial \hat\zeta_0}{\partial y}&=& \kappa^2\, \hat\omega_0\, \left(
1+ \frac{\hat\omega_0}{2 {w}(x,y)} \,(1 + 3 x)\right)
\end{eqnarray}
where $\kappa^2$ is a positive definite $x$- and $y$-dependent 
prefactor. Using Eq.~(\ref{eq:omega0prime}), one finds that 
the last factor to the right is also 
positive, since $\left(
1+ \frac{(1 + 3 x)}{2} \sin^2 \theta_+ \right) \geq \cos^2 \theta_+$ for $x \in [-1, +1]$. 

\subsubsection{Upper boundary}
It follows from (\ref{eq:keyprop}) that the maximum
of $\hat\zeta_0$ for fixed $x$ and $\hat\omega_0$ is given
by $\hat\zeta_0(x, +1, \hat\omega_0)$ (resp. $\hat\zeta_0(x, -1, \hat\omega_0)$) when $\hat\omega_0 \geq 0$ (resp. $\hat\omega_0 \leq 0$). This suggests the study of these two functions
in the corresponding
negative and positive ranges of $\hat\omega_0$, 
which we will treat as families of functions of 
$\hat\omega_0$ parameterized by $x$:
\begin{numcases}{\hat\zeta_0^{(x)}(\hat\omega_0) = }
\hat\zeta_0(x, +1, \hat\omega_0) \, , \text{for} \, 0 \leq \hat\omega_0 \leq {w}(x,+1)
\, \text{and} \, x \in [-1, +1] \ , \label{eq:brancH1} \\
\hat\zeta_0(x, -1, \hat\omega_0) \, , \text{for} \, {w}(x,-1) \leq \hat\omega_0 \leq 0 \, \text{and} \, x \in [-\frac79, +1] \ , \label{eq:brancH2}\\
1 \, , \text{for} \, 
 \hat\omega_0 = 0 \, \text{and} \,  x \in [-1,-\frac79] \ . \label{eq:branch0}
\end{numcases}
In writing the above we took into account the consistency conditions Eqs.~(\ref{eq:consist1}, \ref{eq:consist2}) and noted
that $\hat\omega_0 < 0$ cannot be satisfied when
$x \in [-1, -\frac79]$.
Obviously the upper boundary
in the $(\hat\omega_0,\hat\zeta_0)$ plane, that is the function
$\hat{\zeta_0}^{max}(\hat\omega_0)$ giving the maximal allowed
value of $\hat\zeta_0$ for a given $\hat\omega_0$, is obtained
by determining the upper envelope of the family of functions
$\hat\zeta_0^{(x)}(\hat\omega_0)$ defined in Eqs.~(\ref{eq:brancH1}, \ref{eq:brancH2}). We will show below that 
this envelope
is given by
\begin{numcases}{\hat{\zeta_0}^{max}(\hat\omega_0)= } 
\hat\zeta_0^{(x^{\scriptscriptstyle>}_{+1})}\left(\hat\omega_0\right)_{|\hat\omega_0={w}(x^{\scriptscriptstyle>}_{+1}, +1)}, \, \text{for}  \, \hat\omega_0
\geq 0, \label{eq:zetamaX1} \\
\hat\zeta_0^{(x^{\scriptscriptstyle>}_{-1})}\left(\hat\omega_0\right)_{|\hat\omega_0={w}(x^{\scriptscriptstyle>}_{- 1}, - 1)}, \, \text{for}  \, \hat\omega_0 \leq 0, \label{eq:zetamaX2}
\end{numcases}
where $x^{\scriptscriptstyle>}_{\pm 1}$ have been defined in Eqs.~(\ref{eq:xupperlower}).
 
In other terms, the upper boundary is
traced when $\hat\omega_0$ sits at the non-vanishing end-points
of its allowed domains given in Eqs.~(\ref{eq:brancH1}, 
\ref{eq:brancH2}), thus corresponding to $\sin^2 \theta_+=1$ as noted after Eq.~(\ref{eq:xupperlowersol}),  and for the largest value of $x$ that allows 
to reach each end-point. This result is a consequence of certain
 properties that can be easily shown by direct analytical (as well as numerical) inspection of the relevant functions and their
 first derivative, summarized hereafter without proof: 
\begin{itemize}
\item[i)]
${w}(x, -1)$ is $\leq 0$ if and only if $x \in [-\frac79, 1]$, and
 \begin{itemize}
\item ${w}(x, -1)$ is {\sl strictly  decreasing} for  $x\in [-\frac79, \frac13]$, spanning the full negative $\hat\omega_0$ domain $[-\frac14, 0]$,
\item ${w}(x, -1)$ is {\sl strictly  increasing} for $x\in [\frac13, 1]$, spanning the full negative $\hat\omega_0$ domain 
$[-\frac14, 0]$ .
\end{itemize}
It follows that 
$x^{\scriptscriptstyle<}_{-1} \in [-\frac79, \frac13]$
and 
$x^{\scriptscriptstyle>}_{-1} \in [\frac13, 1]$, (cf. Eq.~(\ref{eq:xupperlower})).
\item[ii)] 
${w}(x, +1)$ is $\geq 0$ in the entire $x$ domain $[-1, +1]$, and
\begin{itemize}
\item ${w}(x, +1)$ is {\sl strictly increasing} for 
$x\in [-1, -\frac13]$, spanning partially the positive $\hat\omega_0$ domain $[\frac14, \frac12]$,
\item ${w}(x, +1)$ is {\sl strictly decreasing} for $x\in [-\frac13, 1]$, spanning the full positive $\hat\omega_0$ domain $[0, \frac12]$.
\end{itemize}
It follows that 
$x^{\scriptscriptstyle<}_{+1} \in [-1, -\frac13]$
and 
$x^{\scriptscriptstyle>}_{+1} \in [-\frac13, 1]$, (cf. Eq.~(\ref{eq:xupperlower})).
 \item[iii)] in the domains of $x^{\scriptscriptstyle>}_{- 1}$ and $x^{\scriptscriptstyle>}_{+1}$,
that is respectively for $x\in[\frac13, 1], 
\hat\omega_0 \in[-\frac14, 0]$, and $x\in[-\frac13, 1], 
\hat\omega_0\in[0, \frac12]$, 
$\hat\zeta_0^{(x)}(\hat\omega_0)$ is a 
{\sl strictly increasing} function of $x$. ($\frac{\partial}{\partial x}\hat\zeta_0^{(x)}(\hat\omega_0)$ vanishes only at
 the two isolated points $\{\hat\omega_0= 0, \forall x\}$ and $\{\hat\omega_0= \frac12, x=-\frac13 \}$, where $\hat\zeta_0$ takes its two extreme values $1$ and $\frac13$).
\item[iv)] in the domains of $x^{\scriptscriptstyle<}_{- 1}$ and $x^{\scriptscriptstyle<}_{+1}$,
that is respectively for $x\in[-\frac79, \frac13], 
\hat\omega_0 \in[-\frac14, 0]$, and $x\in[-1,-\frac13], 
\hat\omega_0\in[\frac14, \frac12]$, 
$\hat\zeta_0^{(x)}(\hat\omega_0)$ taken as a function of $x$
can be either {\sl strictly} increasing or {\sl strictly}  decreasing, but it 
changes its monotonicity {\sl at most} once depending on the value
of $\hat\omega_0$.
\item[v)] $\hat\zeta_0^{(x^{\scriptscriptstyle>}_{\pm 1})}(\hat\omega_0) - \hat\zeta_0^{(x^{\scriptscriptstyle<}_{\pm 1})}(\hat\omega_0) = \frac{16}{27}(1 - 2 \, \hat\omega_0)^{3/2} \, \sqrt{1 + 4 \, \hat\omega_0} \geq 0$, valid for all $\hat\omega_0 \in [-\frac14, \frac12]$. 
\end{itemize}
Consider the value of  $\hat\zeta_0 = \hat\zeta_0^{(x^{\scriptscriptstyle>}_{-1})}\left(\hat\omega_0\right)_{|\hat\omega_0={w}(x^{\scriptscriptstyle>}_{- 1}, - 1)}$ for a given $\hat\omega_0 \leq 0$,
cf. Eq.~(\ref{eq:zetamaX2}). We now show that varying $x$ in the
vicinity of $x^{\scriptscriptstyle>}_{-1}$ does not allow to find 
for the same $\hat\omega_0$
a larger value for $\hat\zeta_0$: 
To find another value of $\hat\zeta_0$ for the same $\hat\omega_0$
one should, according to Eq.~(\ref{eq:brancH2}), 
choose an $x$ such that 
\begin{equation}
{w}(x, -1) < \hat\omega_0 ={w}(x^{\scriptscriptstyle>}_{- 1}, - 1) ={w}(x^{\scriptscriptstyle<}_{- 1}, - 1).
\end{equation}
(Note that the last equality is simply due to the definition of 
$x^{\scriptscriptstyle>}_{-1}$ and $x^{\scriptscriptstyle<}_{-1}$.)
 If
$x$ is taken sufficiently close to $x^{\scriptscriptstyle>}_{- 1}$ so that
$x \in [\frac13, 1]$, then the above inequality is satisfied only if $x$ is {\sl strictly smaller} than $x^{\scriptscriptstyle>}_{- 1}$ since by property i) ${w}$ is a strictly increasing function in the considered domain. It then follows from
property iii) that the new value of $\hat\zeta_0$ is necessarily
{\sl strictly smaller} than the initial $\hat\zeta_0^{(x^{\scriptscriptstyle>}_{-1})}\left(\hat\omega_0\right)_{|\hat
\omega_0={w}(x^{\scriptscriptstyle>}_{- 1}, - 1)}$. Thus the latter
is indeed a local maximum. But $x$ can also be in the domain
$[-\frac79, \frac13]$. In this case property i) implies 
that $x$ should be {\sl strictly greater} than $x^{\scriptscriptstyle<}_{- 
1}$ since $w$ is a strictly decreasing function of $x$ 
in the considered domain. Then according to property iv):
\begin{itemize}
\item[--] 
Either $\hat\zeta_0^{(x)}\left(\hat\omega_0\right)$ did not
change its monotonicity for the given value of $\hat\omega_0$
and the considered range for $x$ within the $[-\frac79, \frac13]$ 
domain, which means it is still a strictly increasing function of $x$ (cf.
property iii) ). In this case one has
$\hat\zeta_0^{(x)}\left(\hat\omega_0\right) < \hat\zeta_0^{(\frac13)}\left(\hat\omega_0\right) < \hat\zeta_0^{(x^{\scriptscriptstyle>}_{-1})}\left(\hat\omega_0\right)$
with $\hat\omega_0={w}(x^{\scriptscriptstyle>}_{- 1}, - 1)$
.
\item[--] Or $\hat\zeta_0^{(x)}\left(\hat\omega_0\right)$ changed
once its monotonicity becoming a strictly decreasing function of 
$x$.
In this case one has $\hat\zeta_0^{(x)}\left(\hat\omega_0\right) < \hat\zeta_0^{(x^{\scriptscriptstyle<}_{-1})}\left(\hat\omega_0\right)$ because $x > x^{\scriptscriptstyle<}_{-1}$ as shown
above. But then property v) implies $\hat\zeta_0^{(x)}\left(\hat\omega_0\right) < \hat\zeta_0^{(x^{\scriptscriptstyle>}_{-1})}\left(\hat\omega_0\right)$, which holds for any $\hat\omega_0$
 including 
$\hat\omega_0={w}(x^{\scriptscriptstyle>}_{- 1}, - 1)$.
\end{itemize}
It follows that in all cases $\hat\zeta_0^{(x^{\scriptscriptstyle>}_{-1})}\left(\hat\omega_0\right)_{|\hat\omega_0={w}(x^{\scriptscriptstyle>}_{- 1}, - 1)}$ is indeed a global maximum. 

A similar proof holds for the branch
$\hat\omega_0 \geq 0$ noting though the reversed inequality 
in Eq.~(\ref{eq:brancH1}) as compared to Eq.~(\ref{eq:brancH2}),
and the reversed behavior of $w$ in property ii) as compared
to property i). More specifically,
one should look for an $x$ 
such that ${w}(x, +1) > \hat\omega_0 ={w}(x^{\scriptscriptstyle>}_{+ 1}, + 1)
={w}(x^{\scriptscriptstyle<}_{+ 1}, + 1)$, cf. Eqs.~(\ref{eq:zetamaX1}) and (\ref{eq:brancH1}), and consider separately
the cases $x \in [-\frac13, 1]$ and $x \in [-1, -\frac13]$.
In the first case the above inequality
implies, using properties ii) and iii), that $x < x^{\scriptscriptstyle>}_{+ 1}$ and 
$\hat\zeta_0^{(x^{\scriptscriptstyle>}_{+1})}\left(\hat\omega_0\right)_{|\hat\omega_0={w}(x^{\scriptscriptstyle>}_{+1}, +1)}$ is
a local maximum. In the second case property ii) and the above
inequality imply $x > x^{\scriptscriptstyle<}_{+ 1}$ and the result that $\hat\zeta_0^{(x^{\scriptscriptstyle>}_{+1})}\left(\hat\omega_0\right)_{|\hat\omega_0={w}(x^{\scriptscriptstyle>}_{+1}, +1)}$ is a maximum is again obtained upon use of properties iv) and v). 

We can now write explicitly $\hat{\zeta_0}^{max}(\hat\omega_0)$.
First, from Eqs.~(\ref{eq:zeta0prime}, \ref{eq:zetamaX2}, \ref{eq:zetamaX1}) and properties i) and ii) one finds the simple form
\begin{numcases}{\hat{\zeta_0}^{max}(\hat\omega_0)= } 
Z(x^{\scriptscriptstyle>}_{+1}), \, \text{for}  \, 0 \leq \hat\omega_0
\leq \frac12, \label{eq:zetamaX1prime} \\
Z(x^{\scriptscriptstyle>}_{-1}), \, \text{for}  \, -\frac14 \leq \hat\omega_0 
\leq 0, \label{eq:zetamaX2prime}
\end{numcases}
where
\begin{equation}
Z(x) = \frac18 \, (3 + x (2 + 3 x)) . \label{eq:Z}
\end{equation}
The explicit dependence on $\hat\omega_0$ is obtained by solving
$\hat\omega_0 ={w}(x^{\scriptscriptstyle>}_{+ 1}, + 1)$ for $x^{\scriptscriptstyle>}_{+1}$ and $\hat\omega_0 ={w}(x^{\scriptscriptstyle>}_{- 1}, - 1)$ for $x^{\scriptscriptstyle>}_{-1}$ and plugging the result back in
Eqs.~(\ref{eq:zetamaX1prime}) and (\ref{eq:zetamaX2prime}). 
In fact a further simplification occurs because the two solutions
are found to have exactly the same functional dependence on $\hat\omega_0$, cf. Eq.~(\ref{eq:xupperlowersol}), even though they correspond to different ranges of the latter:
\begin{equation}
x^{\scriptscriptstyle>}_{\pm 1}= \frac19 \left( 1 - 8 \, \hat\omega_0 + 8 \, \sqrt{(1- 2 \, \hat\omega_0)(1 + 4 \, \hat\omega_0)}\right) . 
\end{equation}
Equations  (\ref{eq:zetamaX1prime}, \ref{eq:zetamaX2prime}) can thus be merged into one single form for the full 
$\hat\omega_0$ range
$[-\frac14, \frac12]$, 
\begin{equation}
\hat{\zeta_0}^{max}(\hat\omega_0)=\frac13 + \frac{2}{27} \, \left(1 - 2 \, \hat\omega_0 + 
2 \, \sqrt{(1- 2 \, \hat\omega_0)(1 + 4 \, \hat\omega_0)}
\right)^2, \; \; \hat\omega_0 \in  [-\frac14, \frac12]
\label{eq:upperbound0}
\end{equation}
which reproduces the upper boundary given \cite{Hartling:2014zca} (note however that we deal with the inverse
function wrt to the function considered in reference \cite{Hartling:2014zca}), see also Fig.~\ref{fig:omega-zeta}.  

\subsubsection{Lower boundary}
We turn now to the determination of the lower boundary of the 
domain. In contrast with the previous case we cannot just
study  $\hat\zeta_0(x, +1, 
\hat\omega_0)$ and $\hat\zeta_0(x, -1, \hat\omega_0)$ as being the minima in the $y$ domain, 
respectively for  $\hat\omega_0 \leq 0$ and $\hat\omega_0 \geq 0$ 
as suggested by the property (\ref{eq:keyprop}). Indeed, it is 
obvious from Eqs.~(\ref{eq:omega0prime}, \ref{eq:wdef}), see also
Eq.~(\ref{eq:brancH1}), that 
$\hat\zeta_0(x, +1, \hat\omega_0)$ and more generally 
$\hat\zeta_0(x, y \geq 0, \hat\omega_0)$ are never compatible with
$\hat\omega_0 < 0$. Moreover,  $\hat\zeta_0(x, -1, \hat\omega_0)$
is compatible with $\hat\omega_0 \geq 0$ only in the reduced 
domain of $x \in [-1, -\frac79]$ as already discussed after
Eq.~(\ref{eq:branch0}). This means that there could exist $y > -1$
and $x$ outside this reduced domain for which values
of $\hat\zeta_0$ smaller than $\hat\zeta_0(x, -1, \hat\omega_0)$
could be reached. Thus for both domains, $\hat\omega_0 \geq 0$
and $\hat\omega_0 \leq 0$, $y$ should be varied away from
$y=+1$ or $-1$ to determine the lower boundary function $\hat{\zeta_0}^{min}(\hat\omega_0)$ that gives for each $\hat\omega_0$
the minimal allowed value for $\hat{\zeta_0}$. It is easy to see that for given $\hat\omega_0$
and $x$, the minimal value of $\hat{\zeta_0}$ is reached only when
$\hat\omega_0 = {w}(x,y)$. This is a consequence of combining 
property (\ref{eq:keyprop}) with Eqs.~(\ref{eq:wboundneg}, 
\ref{eq:wboundpos}) and the fact that ${w}(x,y)$ is an increasing 
function of $y$. E.g. for a given positive $\hat\omega_0$ that 
should satisfy Eq.~(\ref{eq:wboundpos}) for say $y=+1$, decreasing 
$y$ will monotonically decrease simultaneously 
$\hat{\zeta_0}$, cf. (\ref{eq:keyprop}), and ${w}(x,y)$. Since values of $y$ such that $\hat\omega_0 > {w}(x,y)$ are forbidden
by Eq.~(\ref{eq:wboundpos}), the minimum of $\hat{\zeta_0}$ is indeed reached when $\hat\omega_0 = {w}(x,y)$. A similar reasoning holds for negative $\hat\omega_0$ satisfying Eq.~(\ref{eq:wboundneg}) so that $\hat{\zeta_0}$ is reached when and only when $\hat\omega_0 = {w}(x,y)$. Thus in both cases the relevant
functions are obtained for $x= x^{\scriptscriptstyle>}_y$ or $x^{\scriptscriptstyle<}_y$. Denoting by $\hat\zeta_0^\pm(y, \hat\omega_0)$ the two functions $\hat\zeta_0(x=x^{\scriptscriptstyle\gtrless}_y, y, \hat\omega_0)$  and using Eqs.~(\ref{eq:Z},\ref{eq:xupperlower}), we find after some algebra, 
\begin{eqnarray}
\hat\zeta_0^\pm(y, \hat\omega_0)=\frac13+ \frac23 \left(\frac{1-6 \hat\omega_0+2 \left(y^2 \pm 3 
\sqrt{y^2(2 \hat\omega_0-8 \hat\omega_0^2+y^2)}\right)}{1+8 y^2}\right)^2, \label{eq:zeta0+-}\\
\text{with}\; y^2 \in[-2\hat\omega_0 + 8 \hat\omega_0^2,1] . \nonumber
\end{eqnarray}
We note that these functions do not depend on the sign of $y$.
Starting from Eq.~(\ref{eq:zeta0+-}) it is straightforward to 
determine the configurations where $\hat\zeta_0$ reaches its absolute minimum value $\frac13$. One finds,
\begin{eqnarray}
\hat\zeta_0^+=\frac13 \; \text{iff} \; \hat\omega_0=\frac16 + \frac{\sqrt{y^2}}{3} , \\
\hat\zeta_0^-=\frac13 \; \text{iff} \; \hat\omega_0=\frac16 - \frac{\sqrt{y^2}}{3} , 
\end{eqnarray}
where these values of $\hat\omega_0$ always lie within 
the validity domain of Eq.~(\ref{eq:zeta0+-}). Varying $y^2$ in
$[0, 1]$  we see that $\hat\zeta_0$ reaches
the value of $\frac13$ through either $\hat\zeta_0^+$ or $\hat\zeta_0^-$ for any value of $\hat\omega_0$
in $[-\frac16, +\frac12]$, while $\frac13$ is never reached
when $\hat\omega_0 \in [-\frac14,-\frac16[$ . 
Thus for the $[-\frac16, +\frac12]$ sub-domain, the lower boundary 
$\hat{\zeta_0}^{min}(\hat\omega_0)$ is simply given by
\begin{equation}
\hat{\zeta_0}^{min}(\hat\omega_0)=\frac13, \; \;
\hat\omega_0 \in  [-\frac16, +\frac12] .
\label{eq:lowerbound01}
\end{equation}
To treat the $[-\frac14,-\frac16[$
sub-domain we first note from Eq.~(\ref{eq:zeta0+-}) the  
obvious inequality,   
\begin{equation}
\hat\zeta_0^-(y, \hat\omega_0) < \hat\zeta_0^+(y, \hat\omega_0), \;
\text{for all} \;   \hat\omega_0 < 0.
\end{equation}
The lower boundary for the portion $\hat\omega_0 \in[-\frac14, -\frac16]$ is thus to be found  within the $\hat\zeta_0^-$ branch.
A straightforward analytical study shows that $\hat\zeta_0^-(y, \hat\omega_0)$ is a strictly {\sl decreasing} function of $y^2$
for any $\hat\omega_0 \in[-\frac14, -\frac16]$.\footnote{More specifically, we find that the derivative 
$\frac{\partial}{\partial y^2} \hat\zeta_0^-(y, \hat\omega_0)$ 
vanishes only when $y^2= (1/4) (1 - 6 \hat\omega_0)^2$, a value
$\geq 1$ for $\hat\omega_0 \in[-\frac14, -\frac16]$, that is outside the $y^2$ domain. Thus $\frac{\partial}{\partial y^2} \hat\zeta_0^-(y, \hat\omega_0)$ does not
change sign in the considered domain of $\hat\omega_0$. This sign
is determined by choosing any value of 
$y^2 \in[2 (-\hat\omega_0 + 4 \hat\omega_0^2),1]$; e.g. 
for $y^2=2 (-\hat\omega_0 + 4 \hat\omega_0^2)$ 
it is given by 
$\sgn \{-8 \hat\omega_0 (-1 + 8 \hat\omega_0)^3 (1 - 6 \hat\omega_0 + 8 \hat\omega_0^2) = -$ for $\hat\omega_0 \in[-\frac14, -\frac16]$.} It follows that the lower boundary $\hat{\zeta_0}^{min}(\hat\omega_0)$ is given by $\hat\zeta_0^-(y, \hat\omega_0)$ at 
$y^2=1$ (strictly speaking at $y=-1$ since  $\hat\omega_0 < 0$), 
\begin{equation}
\hat{\zeta_0}^{min}(\hat\omega_0)=\frac13 + \frac{2}{27} \, \left(1 - 2 \, \hat\omega_0 - 
2 \, \sqrt{1 + 2 \, \hat\omega_0 - 8 \, \hat\omega_0^2}
\right)^2,  \; \; \hat\omega_0 \in  [-\frac14, -\frac16]
\label{eq:lowerbound02}
\end{equation}

\subsubsection{Comments \label{appendix:comments}}
The functions given in Eqs.~(\ref{eq:upperbound0}, \ref{eq:lowerbound01}, \ref{eq:lowerbound02}) provide the full boundary
in the $(\hat\omega_0, \hat\zeta_0)$ domain. Given that 
$\chi^{++}$ and $\operatorname{Im} \chi^+$ are put to zero by a gauge choice, i.e. Eq.~(\ref{eq:Xchoice}), we have proven
{\sl under the working assumption} 
$\operatorname{Re} \chi^+ \equiv u= 0$
in Eq.~(\ref{eq:Xchoice}), that this boundary is obtained when 
$y=\pm 1$  and $sin^2 \theta_+=1$, that is for 
$\operatorname{Im} \chi^0=\xi^+=0$, cf. Eqs.~(\ref{eq:canxydef}, \ref{eq:defteta+}). This agrees with \cite{Hartling:2014zca} where the domain was determined by a numerical
scan. There is however more to the proofs we provided: $sin^2 \theta_+=1$ is not only sufficient but also necessary; indeed 
as one can see from the various steps of the proofs given above, 
all the inequalities and monotonicity are strict. 

It is important to stress
that there is {\sl a priori} no simple reason to believe that 
the domain $(\hat\omega_0, \hat\zeta_0)$ will be identical to 
the full domain of $(\hat\omega, \hat\zeta)$, i.e. when relaxing 
the working assumption $u=0$.
The necessity of $sin^2 \theta_+=1$ proved instrumental while completing the 
determination of the domain when $u \neq 0$, see Sec.\ref{sec:GM-BFB}.

\section{Resolved NAS conditions for Eqs.~(\ref{eq:omega-alphas}, \ref{eq:zeta-alphas}), \label{subsec:resolved}}
Here we give without proof the necessary and sufficient conditions on the {\sl $\alpha$-parameters} in order
for the trajectories $(\hat\omega(t),\hat\zeta(t))$ given by  Eqs.~(\ref{eq:omega-alphas}, \ref{eq:zeta-alphas}) 
to go through a given point~$(\hat\omega,\hat\zeta)$: 
\begin{align}
\displaystyle \left\{ \hat\zeta \geq \alpha_{AB}  \, \lor \, \hat\zeta \geq \frac32 - \alpha_A \right\}
 &\land \,  
\displaystyle \hat\zeta \geq \frac{ 2 \alpha_A + 2 \alpha_{AB}^2  -3}{2 \alpha_A + 4 \alpha_{AB} -5 }  \nonumber \\
&\land  \nonumber \\ 
\displaystyle \bigg\{ \hat\omega \times \alpha_{ABH}  \geq 0 \, \lor \, 
 \min\{0, &\frac14 \beta_{AH}\} \leq \left. \hat\omega \leq \max\{0, \frac14 \beta_{AH}\} \right\} \nonumber \\
 &\land  \label{eq:resolved} \\ 
 \displaystyle \frac18 \left(\beta_{AH} - \sqrt{4 \alpha_{ABH}^2 + \beta_{AH}^2} \right) \leq &\hat\omega \leq 
 \frac18 \left(\beta_{AH} + \sqrt{4 \alpha_{ABH}^2 + \beta_{AH}^2} \right) \nonumber \\
 &\land  \nonumber \\
(r_1 \hat\omega^2 +r_2 \hat\omega +r_3 \hat\zeta +r_4) \, \hat\omega^2 &+ (r_5+ r_6 (\hat\zeta+1) +r_7 \hat\omega )\,(\hat\zeta-1)=0, \nonumber
\end{align}
with
\begin{align}
&r_1 = 4 (\beta_A + 2 \beta_{AB} -2 )^2, \nonumber \\
&r_2 = 4 (1 - \beta_{AB}) (\beta_A + 2 \beta_{AB} - 2 ) \beta_{AH}, \nonumber \\
&r_3 = (\beta_A + 2 \beta_{AB} -2  ) (\beta_{AH}^2 - 4 \alpha_{ABH}^2), \nonumber \\
&r_4 =  2 \alpha_{ABH}^2 \left(8 - (\beta_A-4)(\beta_{AB} -3) \right)+ \left((\beta_{AB} -2)^2 - \beta_A -1 \right)\beta_{AH}^2, \nonumber \\
&r_5 = \frac18 \left(4 \alpha_{ABH}^4 (\beta_A - 4 ) - 2 \alpha_{ABH}^2 (3 + \beta_{AB}) \beta_{AH}^2 - 
     \beta_{AH}^4 \right), \nonumber \\
&r_6 = \frac{1}{16} (\beta_{AH}^2 + 4 \alpha_{ABH}^2)^2, \nonumber \\
&r_7 = \frac12 \beta_{AH} \left(4 \alpha_{ABH}^2 (\beta_A + \beta_{AB} - 1) + 
     (1 - \beta_{AB}) \beta_{AH}^2\right) \nonumber
\end{align}
where we defined
\begin{equation}
\beta_X \equiv 2 \alpha_X -1, \ X=A,AB, AH.
\end{equation}

The first three lines in Eq.~(\ref{eq:resolved}) are the NAS conditions that ensure the existence of at least one real-valued $t$ solution 
to Eq.~(\ref{eq:zeta-alphas}) and at least one real-valued positive $t$ solution to Eq.~(\ref{eq:omega-alphas}). The last condition in
Eq.~(\ref{eq:resolved}) guarantees a common $t$ solution to both equations (\ref{eq:omega-alphas}) and (\ref{eq:zeta-alphas}). Note that Eq.~(\ref{eq:resolved}) is always satisfied for $\hat\omega =0, \hat\zeta=1$ for all {\sl $\alpha$-parameters} in the
$\alpha$-potatoid, which can be seen in particular from Eq.~(\ref{eq:dom-alfa-alfab}).
This corresponds to the fact that the point $(\hat\omega =0, \hat\zeta=1)$
is always reached when $t \to \infty$, as evident from Eqs.~(\ref{eq:omega-alphas}, \ref{eq:zeta-alphas}).

The {\sl $\alpha$-parameters} sets that are excluded by the $\omega$-$\zeta$--chips, (see the discussion in Sec.~\ref{sec:peeling} and footnote \ref{foot:7}), correspond to those that satisfy Eq.~(\ref{eq:resolved}) when substituting therein
$\hat\zeta$ by $\hat\zeta_0^{max}(\hat\omega) +\epsilon $ or by $\hat\zeta_0^{min}(\hat\omega) -\epsilon $, with $\epsilon$ an arbitrarily small positive number (cf. Eqs.~(\ref{eq:chips}, \ref{eq:upperbound} -- \ref{eq:lowerbound2}) ).

 

\section{New NAS positivity conditions for quartic polynomials on $\mathbb{R}$ \label{appendix:generalquartic}}

In this section we consider the general conditions 
on the set of {\sl real} coefficients $a_{i=0,1,2,3,4}$ that are necessary and sufficient to ensure
\begin{equation}
P(\xi) > 0, \forall \xi \in (-\infty, + \infty)
\label{eq:Ppositive}
\end{equation}  
where $P(\xi)$ is a quartic polynomial:
\begin{equation}
 P(\xi) \equiv a_0 + a_1 \xi + a_2 \xi^2 + a_3 \xi^3 + a_4 \xi^4 .
\label{eq:P}
\end{equation}
Our derivation does not rely on the known form of the four roots of $P(\xi)=0$, and will actually allow to cast the conditions in a simpler and more compact form than the ones usually relied upon in the 
literature, \cite{Kannike:2016fmd,osti_7041561}.
To achieve this we take a different path than just writing down the 
well-known expressions of the four roots of $P(\xi)$.  

\noindent
We are interested in determining the exact $\{a_i\}$ space region for which $P(\xi)$ is positive valued
for any $\xi$ in $(-\infty, + \infty)$. Recalling a classic theorem on positive definiteness of even degree polynomials defined on $\mathbb{R}$ and having all their coefficients real-valued,   
if $P(\xi)$ satisfies Eq.~(\ref{eq:Ppositive}) then it can be written in the form
\begin{equation}
 P(\xi) \equiv Q(\xi)^2 + R(\xi)^2,  \; \forall \xi \in (-\infty, + \infty)
\label{eq:A2}
\end{equation}
with
\begin{equation}
 Q(\xi) = x_1 + y_1 \xi + z_1 \xi^2 \; {\rm and} \; R(\xi) = x_2 + y_2 \xi + z_2 \xi^2,
\label{eq:A3}
\end{equation}
where the $x_i, y_i$ and $z_i$ denote real numbers.\footnote{Note that taking
$Q$ and $R$ as in Eq.~(\ref{eq:A3}) is more general than actually
needed. Indeed, $P(\xi)$ will satisfy
Eq.~(\ref{eq:Ppositive}) if and only if its four roots are non-real complex-valued, that is $P(\xi)$ of the form
$P(\xi) = r (\xi - s)(\xi -\bar{s})(\xi - t)(\xi -\bar{t})=
r |(\xi - s)(\xi -t)|^2$, with $Im(s), Im(t) \neq 0$,
 $s$, $t$ and their complex conjugates $\bar s$, $\bar t$  being the four roots, 
 and $r$ a positive real number. Expanding this form as the squared modulus of a complex number, leads to Eq.~(\ref{eq:A3}) but with
 one of the two polynomials $Q$ and $R$ being only linear in $\xi$. The symmetric choice made in Eq.~(\ref{eq:A3}) lends itself however 
 to a more convenient geometric discussion. Its equivalence with the more specific case above, results from
the invariance of  Eq.~(\ref{eq:A2}) under any rigid rotation
of the three vectors ${\bf x}, {\bf y}$ and ${\bf z}$ defined in Eq.~(\ref{eq:A4}).
\label{foot:8}}. 

The exact $\{a_i\}$ space is then defined by the NAS conditions on the
$a_i$ coefficients such that there exist real numbers $x_i, y_i$ and $z_i$ satisfying eq.~(\ref{eq:A2}).
To determine these conditions we find useful to geometrize this statement.
Introducing the vectors,
\begin{equation}
 {\bf x} = ({x_1}, {x_2}), \;\;
{\bf y} = (y_1,y_2), \; \;
{\bf z} = ({z_1}, {z_2}) ,
\label{eq:A4} 
\end{equation}
\noindent
the identification of the coefficients of each $\xi$ monomial in Eq~.(\ref{eq:A2}) leads to
\begin{eqnarray}
\|{\bf x}\|^2 &=& a_0  , \label{eq:mod2x}\\
\|{\bf z}\|^2 &=& a_4  , \label{eq:mod2z} \\
2 {\bf x}.{\bf y} &=& {a_1} , \label{eq:xdoty}\\
2 {\bf y}.{\bf z} &=& {a_3} , \label{eq:ydotz} \\
\|{\bf y}\|^2 &=&  a_2 - 2 {\bf x}.{\bf z} ,  \label{eq:mod2y}
 \end{eqnarray}
\noindent
so that the problem is equivalent to determining three vectors knowing some of their moduli and scalar
products and relations among them. The NAS conditions on the $a_i$ will thus be determined by requiring
consistent moduli of and angles between the three vectors ${\bf x, y, z}$. 
Equations (\ref{eq:mod2x}, \ref{eq:mod2z}) imply trivially the
NAS conditions for the existence of the moduli
of ${\bf x}$ and ${\bf z}$, namely $a_0 \geq 0 \; \land \; a_4 \geq 0$ . 
However, the strict inequality Eq.~(\ref{eq:Ppositive}) forbids
$a_0=0$ and $a_4=0$ (in the first case $P(\xi=0)=0$ and in the second $P$ is cubic and possesses at least one real root). 
The conditions should thus read
\begin{equation}
 a_0 > 0 \;\; \land \;\; a_4 > 0 . \label{eq:condition1}
\end{equation}
\noindent
Rewriting Eq.~(\ref{eq:mod2y}) as
\begin{equation}
 \|{\bf y}\|^2 =  a_2 - 2 \sqrt{a_0 a_4} \cos \widehat{ ({\bf x}, {\bf z })}
\end{equation}
\noindent 
and using the boundedness of the cosine one finds the {\sl necessary} condition for the existence of the modulus
of ${\bf y}$: 
\begin{equation}
 a_2 + 2 \sqrt{a_0 a_4}  \ge 0 . \label{eq:condition2}
\end{equation}
\noindent
It should be stressed that while this condition is necessary to ensure the 
existence of at least one choice of the angle 
$\widehat{ ({\bf x}, {\bf z })}$, not knowing the sign of $a_2$,  for which the modulus of ${\bf y}$
exists, Eqs.~(\ref{eq:condition1}, \ref{eq:condition2}) are in general not sufficient
to guarantee the existence of the vectors themselves (apart from the special case $a_1=a_3=0$); one has still to check for the
consistency of the three scalar products: Eqs.~(\ref{eq:xdoty},
\ref{eq:mod2x}, \ref{eq:mod2y}) lead to  
\begin{eqnarray}
 a_1 &=& 2 \sqrt{a_0} \sqrt{a_2 - 2 \sqrt{a_0 a_4} \cos \widehat{ ({\bf x}, {\bf z })}   } 
\cos \widehat{ ({\bf y}, {\bf x })} ,  \label{eq:A6.1}
\end{eqnarray}
and Eqs.~(\ref{eq:ydotz},
\ref{eq:mod2z}, \ref{eq:mod2y}) to
\begin{eqnarray}
a_3 &=& 2 \sqrt{a_4} \sqrt{a_2 - 2 \sqrt{a_0 a_4} \cos \widehat{ ({\bf x}, {\bf z })}} 
\cos \widehat{ ({\bf y}, {\bf z })} . \label{eq:A6.2}
\end{eqnarray}
Again, using $-1 \leq \cos \leq 1$,  one retrieves two {\sl necessary 
conditions} from these two equations that can be summarized as
\begin{equation}
\displaystyle a_2 + 2 \sqrt{a_0 a_4} \geq \max\{\frac{a_1^2}{4 a_0}, \frac{a_3^2}{4 a_4}\} \label{eq:cond2strong} \ .
\end{equation}
These conditions are stronger than condition (\ref{eq:condition2}). There is however a further constraint
that correlates Eqs.~(\ref{eq:A6.1}, \ref{eq:A6.2}), namely
$\widehat{ ({\bf y}, {\bf z })} = \widehat{ ({\bf y}, {\bf x })} + 
\widehat{ ({\bf x}, {\bf z })}$. This transforms 
Eqs.~(\ref{eq:A6.1}, \ref{eq:A6.2}) into 
\begin{eqnarray}
 \frac{a_1}{\sqrt{a_0}} \eta - \frac{a_3}{\sqrt{a_4}} &=&  2 \epsilon_{{\bf y x}}
\sqrt{1 - \eta^2} ( a_2 - \frac{a_1^2}{4 a_0} - 2 \eta \sqrt{a_0 a_4})^{\frac{1}{2}} , \label{eq:A7.1}\\
\frac{a_3}{\sqrt{a_4}} \eta -\frac{a_1}{\sqrt{a_0}} &=&  2 \epsilon_{{\bf y z}}
\sqrt{1 - \eta^2} ( a_2 - \frac{a_3^2}{4 a_4} - 2 \eta \sqrt{a_0 a_4})^{\frac{1}{2}} , \label{eq:A7.2}
\end{eqnarray}
where $\eta \equiv \cos \widehat{ ({\bf x}, {\bf z })}$, and $\epsilon_{{\bf y x}}$ (resp. $\epsilon_{{\bf y z}}$) indicates 
the {\sl relative} sign between $\sin \widehat{ ({\bf 
y}, {\bf z })}$ and $\sin \widehat{ ({\bf 
z}, {\bf x })}$ (resp. between $\sin \widehat{ ({\bf 
y}, {\bf x })}$ and $\sin \widehat{ ({\bf 
x}, {\bf z })}$). 
Note also that Eqs.~(\ref{eq:A7.1}, \ref{eq:A7.2}) are obtained 
from one another under the exchange
$a_1 \leftrightarrow a_3$ and  
$a_0 \leftrightarrow a_4$. 
The invariance of these conditions as well as any other
positivity condition such as e.g. Eq.~(\ref{eq:cond2strong}), under $(a_1 \leftrightarrow a_3,  
a_0 \leftrightarrow a_4)$, corresponds to the invariance
of the positivity condition under
the duality transformation $\xi \rightarrow \xi^{-1}$:
$$P(\xi) > 0, \forall \xi \in (-\infty, + \infty) \; \; 
\Leftrightarrow \; \xi^4 P(\xi^{-1}) > 0, \forall \xi \in ( -
\infty, + \infty ).$$ When the necessary conditions (\ref{eq:condition1}, \ref
{eq:cond2strong}) are verified one still has to check for the existence of at least one $\eta$  satisfying Eqs.~(\ref{eq:A7.1}, \ref{eq:A7.2}). Moreover, $\eta$ has to satisfy
\begin{equation}
\eta\in[-1, \min\{1, \eta^*\}], \label{eq:etadomain}
\end{equation}
 where
\begin{equation}
\eta^* \equiv \frac{1}{2 \sqrt{a_0 a_4}}
(a_2  - \max\{\frac{a_1^2}{4 a_0}, \frac{a_3^2}{4 a_4}\})
\label{eq:etastar}
\end{equation}
is the critical value above which at least one of the square roots
in Eqs.~(\ref{eq:A7.1}, \ref{eq:A7.2}) turns complex and thus
becomes invalid.\footnote{Note that a necessary condition for the 
existence of $\eta$ is obviously $\eta^* \geq -1$, leading
back  to Eq.~(\ref{eq:cond2strong}).}
To study further the conditions for the existence of $\eta$ 
we square both sides of 
Eq.~(\ref{eq:A7.1}). This leads to a cubic equation in $\eta$:
\begin{eqnarray}
I(\eta) = \widehat{I}, \label{eq:A8.1}
\end{eqnarray}
where we define for later use
\begin{eqnarray}
I(\eta) &\equiv& \Big(2 \sqrt{a_0 a_4}  \left(2 \sqrt{a_0 a_4}\eta-a_2 \right)(\eta+1)+a_1 a_3 \Big) (\eta-1), \label{eq:Idef}\\
\widehat{I} &\equiv& \frac{(\sqrt{a_0} a_3 - a_1 \sqrt{a_4})^2}{2 \sqrt{a_0 a_4}} \label{eq:hatIdef} .
\end{eqnarray}

It is important to note that Eq.~(\ref{eq:A8.1}) would equally 
result from squaring Eq.~(\ref{eq:A7.2}) due to the invariance
under the permutation $(a_1 \leftrightarrow a_3,  
a_0 \leftrightarrow a_4)$.
It follows that (\ref{eq:A8.1}) encodes by itself the 
information contained in (\ref{eq:A7.1}) as well as that contained 
in (\ref{eq:A7.2}), except for the one that is lost by squaring, namely the signs
$\epsilon_{{\bf y x}}, \epsilon_{{\bf y z}}$. This loss of 
information is however not problematic, as the signs  
can be retrieved by plugging back in Eqs.~(\ref{eq:A7.1}, \ref
{eq:A7.2}) whatever valid solutions for $\eta$ are found 
by solving (\ref{eq:A8.1}). Moreover, the constraint that only 
the solutions satisfying Eqs.~(\ref{eq:etadomain}, \ref{eq:etastar}) are valid, is implicitly embedded in Eq.~(\ref{eq:A8.1}): Whenever a solution is found satisfying $\eta \in [-1, +1]$, it automatically satisfies (\ref{eq:etadomain}, \ref{eq:etastar}). The reason is that squaring both sides of Eq.~(\ref{eq:A7.1})
enforces the positivity of the term under the square-root.
We thus conclude that the sought-after NAS conditions are those which guarantee
the existence of (at least one) real-valued $\eta$ satisfying
simultaneously (\ref{eq:A8.1}) and $\eta \in [-1, +1]$, together
with Eq.(\ref{eq:condition1}).
The function $I(\eta)$ being a cubic polynomial
in $\eta$, one can in principle solve 
$\displaystyle I(\eta)=\widehat{I}$ which has at least one, and up to three, real-valued solutions. 
One could of course proceed numerically, but this is not
our aim.  On the other hand, extracting an information from the 
analytical expressions of the three roots of this cubic equation 
is not particularly tractable. In fact $I(\eta)$ has some 
interesting properties listed below, that are straightforward to establish and 
will allow us to determine analytically the NAS conditions without solving the equation. A straightforward calculation shows that
one {\sl always} has: 
\begin{itemize}
\item[(a)] $I(\eta=-1) = -2 a_1 a_3 \leq \widehat{I}$,
\item[(b)] $I(\eta=+1) = 0 \leq \widehat{I}$,
\item[(c)] $\displaystyle I(\eta=\eta^*) \leq \widehat{I}$.
\end{itemize}
(Property (c) is valid for the two configurations of the Max in 
Eq.~(\ref{eq:etastar}) .)
\noindent
Being a cubic polynomial, $I(\eta)$ possesses at most two stationary points $\eta_\pm$
given by 
\begin{equation}
\eta_\pm = \frac{a_2 \pm \sqrt{\Delta_0}}{6 \sqrt{a_0 a_4}}, 
\end{equation}
where we define
\begin{equation}
\Delta_0 = a_2^2 + 12 a_0 a_4 - 3 a_1 a_3 \label{eq:Delta0}.
\end{equation}
Moreover, the coefficient of $\eta^3$ in $I(\eta)$ Eq.~(\ref{eq:Idef}) being always positive, cf. Eq.~(\ref{eq:condition1}), 
one also has that
\begin{itemize}
\item[(d)] when $\Delta_0 > 0$, i.e. $\eta_\pm$ exist and are distinct turning points, then $\eta_{-} < \eta_{+}$ and 
$I(\eta)$ increases monotonically in $(-\infty, \eta_{-}[ \;  \cup \;  ]\eta_{+}, +\infty)$ and  
decreases monotonically in $]\eta_{-}, \eta_{+}[$; $\eta_{-}, \eta_{+}$ correspond to local maximum, minimum, respectively,
\item[(e)] if it does not possess turning points ($\Delta_0 \leq 0$), $I(\eta)$ increases monotonically everywhere.
\end{itemize}

We can now write down the full NAS conditions. As clear from Eq.~(\ref{eq:A8.1}), they correspond to ensuring  
all possible configurations in the $(\eta, I)$ plane for which $I(\eta)$ crosses (at least once) the positive horizontal line 
$\displaystyle I = \widehat{I}$ within the $\eta$ domain given by 
Eq.~(\ref{eq:etadomain}). We will refer to these configurations as existence configurations (EC). To proceed, we begin by identifying 
a set of four {\sl necessary} conditions for the EC, then show that they form together with Eq.~(\ref{eq:cond2strong}) a set of {\sl sufficient} conditions as well.

We note first that, when it exists, $\eta_{+}$ is always the position of the local minimum of $I(\eta)$. Properties (b) and (d) then imply 
that this minimum is necessarily negative whenever $\eta_{+} \leq 1$. But since $\widehat{I}$ is positive definite it follows that
when $\eta_{+}$ lies in the relevant domain $[-1,+1]$ it never plays a role in the realization of the EC. We thus concentrate 
hereafter on $\eta_{-}$ and $\eta^*$.

Properties (b) and (e) imply that the EC are never realized if  $\Delta_0 \leq 0$, since in this case $I$ would reach $\widehat{I}$ 
only for $\eta > 1$, that is outside its allowed domain, cf. Eq.~(\ref{eq:etadomain}),
(except possibly for the non-generic case where  $a_3 \sqrt{a_0}  = a_1 \sqrt{a_4}$);
\begin{equation}
 \text{\sl a necessary condition is thus}
 \ \Delta_0 > 0. 
 \label{eq:NAS1}
 \end{equation}
 It follows that $\eta_{\pm}$ exist and are turning points. Similarly, properties (b) and (d) imply
 that the EC cannot be realized if $\eta_{-} > 1$ since again $I$ cannot reach $\widehat{I}$ within the allowed $\eta$ domain
 Eq.~(\ref{eq:etadomain});  
 \begin{equation}
 \text{\sl a necessary condition is thus} \ \eta_{-} \leq 1.
 \label{eq:NAS2}
 \end{equation}  Furthermore, the EC cannot be realized if $\eta_{-} > \eta^*$, since, according to property (d), $I(\eta)$ would be in this case monotonically increasing at $\eta^*$, and for it to reach  $\widehat{I}$ one would still have to increase $\eta$ above $\eta^*$ as implied by property (c), which is outside its allowed domain, cf. Eq.~(\ref{eq:etadomain}); 
 \begin{equation}
 \text{\sl a necessary condition is thus} \ \eta_{-} \leq \eta^*.
 \label{eq:NAS3}
 \end{equation} Since among the two turning points 
 $\eta_\pm$, only $\eta_{-}$ plays a role and is a local minimum, obviously if $I(\eta_{-}) < \widehat{I}$ then EC are never realized in the relevant 
 $\eta$ domain. $I(\eta)$ still reaches $\widehat{I}$ but outside this domain as a consequence of property (b);
 \begin{equation}
 \text{\sl a necessary condition is thus} \ 
 I(\eta_{-}) \geq \widehat{I} .
 \label{eq:NAS4}
 \end{equation}
 
 It is now easy to see that the latter condition, in conjunction
 with the necessary conditions Eq.~(\ref{eq:condition1}) and 
 (\ref{eq:NAS1} --\ref{eq:NAS3}),  would form also a set of {\sl sufficient} conditions
  if and only if $\eta_{-} \geq -1$. 
 Indeed, if $\eta_{-} < -1$ then to ensure that Eq.~(\ref{eq:A8.1}) can be fulfilled for an $\eta$
 in the allowed domain would also require
 $I(\eta=-1) \geq \widehat{I}$ which is generically in contradiction with property (a). Fortunately, however, $\eta_{-} < -1$ is anyway
 forbidden by the necessary condition Eq.~(\ref{eq:cond2strong}).
[This can be proven by showing, upon use of  Eq.~(\ref{eq:cond2strong}) which implies in particular
$a_2 + 6 \sqrt{a_0 a_4} \geq 0$, that $\eta_{-} < -1$ would lead to
$\displaystyle \left(a_2 + 2 \sqrt{a_0 a_4} \right)^2 < \frac{a_1^2 a_3^2}{16 a_0 a_4}$ that contradicts Eq.~(\ref{eq:cond2strong}).]
Thus $\eta_{-}$ always satisfies $\eta_{-} \geq -1$.

We therefore conclude that adding the necessary condition Eq.~(\ref{eq:cond2strong}) to Eq.~(\ref{eq:condition1}) and (\ref{eq:NAS1} --\ref{eq:NAS4}), one obtains 
a set of necessary and sufficient conditions. There is however more to it. One can show that (\ref{eq:NAS3}) actually implies 
Eq.~(\ref{eq:cond2strong}). The latter can hence be discarded without loss of generality.\footnote{The proof consists in showing that (\ref{eq:NAS3}), more explicitly Eq.~(\ref{eq:condb}),  together with Eq.~(\ref{eq:conda}), leads to Eq.~(\ref{eq:cond2strong}). We just sketch here the main steps:
If $2 a_2  -  \frac34 \max\{\frac{a_1^2}{a_0}, \frac{a_3^2}{a_4}\} >0$ then obviously $a_2  -  \frac14 \max\{\frac{a_1^2}{a_0}, \frac{a_3^2}{a_4}\} >0$ and  Eq.~(\ref{eq:cond2strong}) is satisfied.
If $2 a_2  -  \frac34 \max\{\frac{a_1^2}{a_0}, \frac{a_3^2}{a_4}\} <0$, then one can nonambiuously square the inequality in Eq.~(\ref{eq:condb}) and study it as a quadratic polynomial in $a_2$. One then finds that it is satisfied only in a closed domain of $a_2$
for which Eq.~(\ref{eq:cond2strong}) is always satisfied whatever
the configuration of the $\max$.} Putting everything together, the NAS conditions read finally:
 \begin{numcases}{P(\xi) > 0,  \forall \xi {\scriptstyle \in  (-\infty, + \infty)} \Leftrightarrow} 
a_0 >0   \;\; \land \;\; a_4 >0 \;\; \land \;\; \Delta_0 > 0 \label{eq:conda} \\ 
\;\;\;\;\;\;\;\;\;\;\;\; \land  \; \;  \nonumber \\
             \sqrt{\Delta_0} + 2 a_2  -  \frac34 \max\{\frac{a_1^2}{a_0}, \frac{a_3^2}{a_4}\} >0 \label{eq:condb}\\
                   \;\;\;\;\;\;\;\;\;\;\;\;    \land  \; \; \nonumber \\
                   \displaystyle 
\sqrt{\Delta_0} -a_2 + 6 \sqrt{a_0 a_4} > 0 \label{eq:condc}\\
                   \;\;\;\;\;\;\;\;\;\;\;\;    \land  \; \; \nonumber \\
    2 \Delta_0^{\frac{3}{2}} -\Delta_1 >0 ,
        \label{eq:condd}
\end{numcases}
where we defined
\begin{equation}
\Delta_1= 2 a_2^3 +27 (a_0 a_3^2 + a_4 a_1^2)-72 a_0 a_2 a_4  - 9 a_1 a_2 a_3 \label{eq:Delta1} .
\end{equation}
Note that we have switched all the inequalities over to 
{\sl strict}.
The non generic equality cases can lead to different conditions.
However, as argued at the beginning of Section \ref{sec:p-c-BFB},
 only strict positivity will be relevant. 
 We have performed a numerical check of the above NAS conditions
 by scanning randomly over $a_0, a_4 \in [0, 100]$ and
 $a_1, a_2,a_3 \in [-100, 100]$ for $10^5$ points, then solved numerically $P(\xi)=0$ for each point and checked that whenever
 Eqs.~(\ref{eq:conda} -- \ref{eq:condd}) are satisfied, $P(\xi)$
 has no real roots, and whenever one of the conditions is violated
 $P(\xi)$ has at least one real root. We also performed another
 non-trivial check based on the obvious fact that 
 a translation of $P(\xi)$ to $P(\xi+\xi_0)$ for any $\xi_0 \in \mathbb{R}^*$ should not affect
the positivity. It follows that the NAS conditions obtained
after the translation, where the modified coefficients $\tilde a_{0,1,2,3}$ depend explicitly on $\xi_0$ while $\tilde a_4=a_4$, should be equivalent to the initial ones. Incidentally we find that $\xi_0$ cancels out in the modified
$\Delta_0$ and $\Delta_1$, which means that these two quantities can be re-expressed as functions of {\sl differences} of the four roots
of $P(\xi)$, and lead to the same conditions as before. In contrast, $\tilde a_0$ and the modified Eqs.~(\ref{eq:condb}, \ref{eq:condc}) still depend on $\xi_0$. That $\tilde a_0 > 0$ is valid when the initial NAS conditions
Eqs.~(\ref{eq:conda} -- \ref{eq:condd}) are satisfied follows immediately from the fact that $\tilde a_0=P(\xi_0)$. It remains to be checked that the involved dependence on $\xi_0$ in the modified Eqs.~(\ref{eq:condb}, \ref{eq:condc}) does not lead to
further NAS conditions. We verified that this is indeed the case 
through a numerical scan over $5 \times 10^3$ points in the
$a_i$ space satisfying Eqs.~(\ref{eq:conda} -- \ref{eq:condd}) 
followed by a scan over $2 \times 10^3$ values of $\xi_0$ for
each of these points; the modified Eqs.~(\ref{eq:condb}, \ref{eq:condc}) were found to be automatically satisfied 
for all values of $\xi_0$.

In order to appreciate the simplification arrived at with Eqs.~(\ref{eq:conda}, \ref{eq:condd}), one can compare with common knowledge
\cite{osti_7041561,wiki}:
$\Delta_0$ and $\Delta_1$ being defined as in \cite{wiki}, we note that the discriminant of $P(\xi)$
can be factorized as follows, $\Delta=(2 \Delta_0^{\frac{3}{2}} -\Delta_1)(2 \Delta_0^{\frac{3}{2}} +\Delta_1)/27$.
Equation (\ref{eq:condd}) requires the positivity of the first factor.
It should then be clear that instead of relying on the signs of $\Delta$, $D$ and $P$ in the notations of \cite{wiki}, where the first
two are complicated expressions, with an 'and/or' structure as summarized in \cite{wiki}, we only need the signs of $\Delta_0$ and just one 
of the two factors of $\Delta$ and two other simple relations involving $\Delta_0$ with exclusively an 'and' structure. Moreover, the 'and' 
structure leads to unambiguous determination of necessary conditions. Another benefit of our approach is that it leads almost immediatly
to the conditions established in the following section.

\section{New NAS positivity conditions for quartic polynomials on $\mathbb{R}^+$\label{appendix:generalquarticplus}}

In this section we consider the NAS conditions on the parameters
of the quartic
polynomial Eq.~(\ref{eq:P}), that ensure its positivity 
for all non-negative $\xi$,

\begin{equation}
P(\xi) > 0, \forall \xi \in [0, + \infty).
\label{eq:PpositiveNew} 
\end{equation}  
Here, the form given by Eq.~(\ref{eq:A2}), although sufficient, 
is no more necessary. It should
be replaced by the necessary and sufficient form \cite{Polya:1976},
\cite{Powers:2000,Benoist:2017}:
\begin{equation}
 P(\xi) \equiv Q(\xi)^2 + R(\xi)^2 + \left( A(\xi)^2 + B(\xi)^2\right) \xi ,
 \label{eq:A2new}
\end{equation}
where, since $P(\xi)$ is a quartic polynomial, $Q$ and $R$ keep
 the same form as in Eq.(\ref{eq:A3}), and
\begin{equation}
 A(\xi) = u_1 + v_1 \xi  \;, \; B(\xi) = u_2 + v_2 \xi,
\label{eq:A3new}
\end{equation}
with $u_i, v_i$ denoting real numbers. Equating the coefficients of identical monomials
in $\xi$ on both sides of Eq.(\ref{eq:A2new}), 
one finds that Eqs.~(\ref{eq:mod2x},\ref{eq:mod2z}) remain unchanged while Eqs.~(\ref{eq:xdoty} -- \ref{eq:mod2y}) are
slightly modified: 
\begin{eqnarray}
\|{\bf x}\|^2 &=& a_0  , \label{eq:mod2xnew}\\
\|{\bf z}\|^2 &=& a_4  , \label{eq:mod2znew} \\
2 {\bf x}.{\bf y} &=& {a_1} - \|{\bf u}\|^2, \label{eq:xdotynew}\\
2 {\bf y}.{\bf z} &=& {a_3} -\|{\bf v}\|^2, \label{eq:ydotznew} \\
\|{\bf y}\|^2 &=&  a_2 - 2 {\bf x}.{\bf z} - 2 {\bf u}.{\bf v},  \label{eq:mod2ynew}
 \end{eqnarray}
 where we introduced the vectors
 \begin{equation}
 {\bf u} = ({u_1}, {u_2}), \;\;
{\bf v} = (v_1,v_2) .
\label{eq:A4new} 
\end{equation}
The study carried out in Appendix \ref{appendix:generalquartic}
can thus be taken over unchanged to the present case with the following
replacements:
\begin{eqnarray}
&&a_1 \to a_1 - u^2 \label{eq:repa1}\\
&&a_3 \to a_3 - v^2 \label{eq:repa3}\\
&&a_2 \to a_2  - 2 u v c \label{eq:repa2}
\end{eqnarray}
where $u\equiv \|{\bf u}\|$,  $v\equiv \|{\bf v}\|$ and
$-1 \leq c\equiv \cos \widehat{ ({\bf u}, {\bf v })} \leq 1$ 
can be chosen arbitrarily in their domains.
We thus reach the general solution to our problem:
 
\begin{equation}
\label{statmnt:Rplus}
\begin{aligned}
&\text{\sl The NAS conditions on $a_{i=0,1,2,3,4}$
for Eq.~(\ref{eq:PpositiveNew}) are obtained from Eqs.~(\ref{eq:conda} -
\ref{eq:condd})}& \\ 
&\text{\sl in which the replacements Eqs.~(\ref{eq:repa1} -- \ref{eq:repa2}) should lead to
satisfied inequalities} &  \\
&\text{\sl for at least one choice of $u\geq0, v\geq0$ and $-1 \leq c \leq 1$.}&
\end{aligned}
\end{equation}

This shows in what sense Eq.~(\ref{eq:PpositiveNew}) is less 
constraining than Eq.~(\ref{eq:Ppositive}). Indeed, consider the
domain $\mathcal{S}$ of all points in the  $(a_0, a_1,a_2,a_3,a_4)$ space that satisfy 
conditions (\ref{eq:conda} -- \ref{eq:condd}), thus 
Eq.~(\ref{eq:Ppositive}). Obviously $\mathcal{S}$ will satisfy 
also Eq.~(\ref{eq:PpositiveNew}) since the latter is contained in 
Eq.~(\ref{eq:Ppositive}). But now, any point 
$(a_0, a_1',a_2',a_3',a_4)$ lying outside of $\mathcal{S}$ and 
thus not satisfying 
Eq.~(\ref{eq:Ppositive}), will satisfy Eq.~(\ref{eq:PpositiveNew}) 
if it can be related to a point in $\mathcal{S}$ through the 
relations $a_1'> a_1$ and  $a_3' > a_3$, and $a_2' = a_2 + 2 c \sqrt
{a_1'- a_1} \sqrt{a_3'- a_3}$ with arbitrary $c \in [-1, +1]$.
This is so because using Eqs.~(\ref{eq:repa1} -- \ref{eq:repa2})
will bring the point back into the $\mathcal{S}$ domain. The additional
set of points $(a_0, a_1',a_2',a_3',a_4)$ together with $\mathcal{S}$ lead obviously to a
domain for which Eq.~(\ref{eq:PpositiveNew}) is satisfied larger than that for which Eq.~(\ref{eq:Ppositive})
is.

\bibliographystyle{utphys} 
\bibliography{references-triplet}

\providecommand{\href}[2]{#2}\begingroup\raggedright\begin{thebibliography}{10}

\bibitem{Aad:2012tfa}
{\bfseries ATLAS Collaboration} Collaboration, G.~Aad {\em et~al.},
  ``{Observation of a new particle in the search for the Standard Model Higgs
  boson with the ATLAS detector at the LHC},''
  \href{http://dx.doi.org/10.1016/j.physletb.2012.08.020}{{\em Phys.Lett.}
  {\bfseries B716} (2012) 1--29},
\href{http://arxiv.org/abs/1207.7214}{{\ttfamily arXiv:1207.7214 [hep-ex]}}.

\bibitem{Chatrchyan:2012ufa}
{\bfseries CMS Collaboration} Collaboration, S.~Chatrchyan {\em et~al.},
  ``{Observation of a new boson at a mass of 125 GeV with the CMS experiment at
  the LHC},'' \href{http://dx.doi.org/10.1016/j.physletb.2012.08.021}{{\em
  Phys.Lett.} {\bfseries B716} (2012) 30--61},
\href{http://arxiv.org/abs/1207.7235}{{\ttfamily arXiv:1207.7235 [hep-ex]}}.

\bibitem{Pich:2019pzg}
A.~Pich, ``{Flavour Anomalies},''
  \href{http://dx.doi.org/10.22323/1.350.0078}{{\em PoS} {\bfseries LHCP2019}
  (2019) 078}, \href{http://arxiv.org/abs/1911.06211}{{\ttfamily
  arXiv:1911.06211 [hep-ph]}}.

\bibitem{ExpHiggsLHC}
J.~Olsen, ``{Experimental overview of Higgs physics, HIGGS2020, 26-30 Oct.
  2020}.''
  \url{https://indico.cern.ch/event/900384/contributions/3795997/attachments/2129919/3586941/Experimental_Overview.pdf}.

\bibitem{Konetschny:1977bn}
W.~Konetschny and W.~Kummer, ``{Nonconservation of Total Lepton Number with
  Scalar Bosons},''
\href{http://dx.doi.org/10.1016/0370-2693(77)90407-5}{{\em Phys. Lett.}
  {\bfseries B70} (1977) 433}.

\bibitem{Cheng:1980qt}
T.~P. Cheng and L.-F. Li, ``{Neutrino Masses, Mixings and Oscillations in SU(2)
  x U(1) Models of Electroweak Interactions},''
\href{http://dx.doi.org/10.1103/PhysRevD.22.2860}{{\em Phys. Rev.} {\bfseries
  D22} (1980) 2860}.

\bibitem{Lazarides:1980nt}
G.~Lazarides, Q.~Shafi, and C.~Wetterich, ``{Proton Lifetime and Fermion Masses
  in an SO(10) Model},''
\href{http://dx.doi.org/10.1016/0550-3213(81)90354-0}{{\em Nucl. Phys.}
  {\bfseries B181} (1981) 287}.

\bibitem{Schechter:1980gr}
J.~Schechter and J.~W.~F. Valle, ``{Neutrino Masses in SU(2) x U(1)
  Theories},''
\href{http://dx.doi.org/10.1103/PhysRevD.22.2227}{{\em Phys. Rev.} {\bfseries
  D22} (1980) 2227}.

\bibitem{Mohapatra:1979ia}
R.~N. Mohapatra and G.~Senjanovic, ``{Neutrino mass and spontaneous parity
  nonconservation},''
\href{http://dx.doi.org/10.1103/PhysRevLett.44.912}{{\em Phys. Rev. Lett.}
  {\bfseries 44} (1980) 912}.

\bibitem{Mohapatra:1980yp}
R.~N. Mohapatra and G.~Senjanovic, ``{Neutrino Masses and Mixings in Gauge
  Models with Spontaneous Parity Violation},''
\href{http://dx.doi.org/10.1103/PhysRevD.23.165}{{\em Phys. Rev.} {\bfseries
  D23} (1981) 165}.

\bibitem{Georgi:1985nv}
H.~Georgi and M.~Machacek, ``{Doubly Charged Higgs Bosons},''
\href{http://dx.doi.org/10.1016/0550-3213(85)90325-6}{{\em Nucl. Phys.}
  {\bfseries B262} (1985) 463}.

\bibitem{Chanowitz:1985ug}
M.~S. Chanowitz and M.~Golden, ``{Higgs Boson Triplets With M ($W$) = M ($Z$)
  $\cos \theta \omega$},''
\href{http://dx.doi.org/10.1016/0370-2693(85)90700-2}{{\em Phys. Lett.}
  {\bfseries B165} (1985) 105}.

\bibitem{Gluza:2020qrt}
J.~Gluza, M.~Kordiaczynska, and T.~Srivastava, ``{Discriminating HTM and MLRSM
  models in colliders studies of a doubly charged Higgs boson pair production
  and its subsequent leptonic decays},''
  \href{http://arxiv.org/abs/2006.04610}{{\ttfamily arXiv:2006.04610
  [hep-ph]}}.

\bibitem{Padhan:2019jlc}
R.~Padhan, D.~Das, M.~Mitra, and A.~Kumar~Nayak, ``{Probing doubly and singly
  charged Higgs bosons at the $pp$ collider HE-LHC},''
  \href{http://dx.doi.org/10.1103/PhysRevD.101.075050}{{\em Phys. Rev. D}
  {\bfseries 101} no.~7, (2020) 075050},
  \href{http://arxiv.org/abs/1909.10495}{{\ttfamily arXiv:1909.10495
  [hep-ph]}}.

\bibitem{Primulando:2019evb}
R.~Primulando, J.~Julio, and P.~Uttayarat, ``{Scalar phenomenology in type-II
  seesaw model},'' \href{http://dx.doi.org/10.1007/JHEP08(2019)024}{{\em JHEP}
  {\bfseries 08} (2019) 024}, \href{http://arxiv.org/abs/1903.02493}{{\ttfamily
  arXiv:1903.02493 [hep-ph]}}.

\bibitem{Fuks:2019clu}
B.~Fuks, M.~Nemev\v{s}ek, and R.~Ruiz, ``{Doubly Charged Higgs Boson Production
  at Hadron Colliders},''
  \href{http://dx.doi.org/10.1103/PhysRevD.101.075022}{{\em Phys. Rev. D}
  {\bfseries 101} no.~7, (2020) 075022},
  \href{http://arxiv.org/abs/1912.08975}{{\ttfamily arXiv:1912.08975
  [hep-ph]}}.

\bibitem{Ghosh:2017pxl}
D.~K. Ghosh, N.~Ghosh, I.~Saha, and A.~Shaw, ``{Revisiting the high-scale
  validity of the type II seesaw model with novel LHC signature},''
  \href{http://dx.doi.org/10.1103/PhysRevD.97.115022}{{\em Phys. Rev. D}
  {\bfseries 97} no.~11, (2018) 115022},
  \href{http://arxiv.org/abs/1711.06062}{{\ttfamily arXiv:1711.06062
  [hep-ph]}}.

\bibitem{Ouazghour:2018mld}
B.~A. Ouazghour, A.~Arhrib, R.~Benbrik, M.~Chabab, and L.~Rahili, ``{Theory and
  phenomenology of a two-Higgs-doublet type-II seesaw model at the LHC run
  2},'' \href{http://dx.doi.org/10.1103/PhysRevD.100.035031}{{\em Phys. Rev. D}
  {\bfseries 100} no.~3, (2019) 035031},
  \href{http://arxiv.org/abs/1812.07719}{{\ttfamily arXiv:1812.07719
  [hep-ph]}}.

\bibitem{Ait-Ouazghour:2020slc}
B.~Ait-Ouazghour and M.~Chabab, ``{The Higgs Potential in 2HDM extended with a
  Real Triplet Scalar: A roadmap},''
  \href{http://arxiv.org/abs/2006.12233}{{\ttfamily arXiv:2006.12233
  [hep-ph]}}.

\bibitem{Dev:2013ff}
P.~Bhupal~Dev, D.~K. Ghosh, N.~Okada, and I.~Saha, ``{125 GeV Higgs Boson and
  the Type-II Seesaw Model},''
  \href{http://dx.doi.org/10.1007/JHEP03(2013)150}{{\em JHEP} {\bfseries 1303}
  (2013) 150},
\href{http://arxiv.org/abs/1301.3453}{{\ttfamily arXiv:1301.3453 [hep-ph]}}.

\bibitem{Dev:2017ouk}
P.~S.~B. Dev, C.~M. Vila, and W.~Rodejohann, ``{Naturalness in testable type II
  seesaw scenarios},''
  \href{http://dx.doi.org/10.1016/j.nuclphysb.2017.06.007}{{\em Nucl. Phys. B}
  {\bfseries 921} (2017) 436--453},
  \href{http://arxiv.org/abs/1703.00828}{{\ttfamily arXiv:1703.00828
  [hep-ph]}}.

\bibitem{Dev:2018kpa}
P.~Bhupal~Dev and Y.~Zhang, ``{Displaced vertex signatures of doubly charged
  scalars in the type-II seesaw and its left-right extensions},''
  \href{http://dx.doi.org/10.1007/JHEP10(2018)199}{{\em JHEP} {\bfseries 10}
  (2018) 199}, \href{http://arxiv.org/abs/1808.00943}{{\ttfamily
  arXiv:1808.00943 [hep-ph]}}.

\bibitem{Dev:2019hev}
P.~B. Dev, S.~Khan, M.~Mitra, and S.~K. Rai, ``{Doubly-charged Higgs boson at a
  future electron-proton collider},''
  \href{http://dx.doi.org/10.1103/PhysRevD.99.115015}{{\em Phys. Rev. D}
  {\bfseries 99} no.~11, (2019) 115015},
  \href{http://arxiv.org/abs/1903.01431}{{\ttfamily arXiv:1903.01431
  [hep-ph]}}.

\bibitem{Frank:2020mqh}
M.~Frank, B.~Fuks, K.~Huitu, S.~Mondal, S.~K. Rai, and H.~Waltari,
  ``{Left-right supersymmetric option at a high-energy upgrade of the LHC},''
  \href{http://dx.doi.org/10.1103/PhysRevD.101.115014}{{\em Phys. Rev. D}
  {\bfseries 101} no.~11, (2020) 115014},
  \href{http://arxiv.org/abs/2003.08443}{{\ttfamily arXiv:2003.08443
  [hep-ph]}}.

\bibitem{Huitu:2020qxm}
K.~Huitu, ``{A minimal supersymmetric left-right model, dark matter and signals
  at the LHC},'' \href{http://dx.doi.org/10.1140/epjst/e2020-000039-9}{{\em
  Eur. Phys. J. ST} {\bfseries 229} no.~21, (2020) 3187--3203}.

\bibitem{Aaboud:2017qph}
{\bfseries ATLAS} Collaboration, M.~Aaboud {\em et~al.}, ``{Search for doubly
  charged Higgs boson production in multi-lepton final states with the ATLAS
  detector using proton\textendash{}proton collisions at $\sqrt{s}=13\,\text
  {TeV}$},'' \href{http://dx.doi.org/10.1140/epjc/s10052-018-5661-z}{{\em Eur.
  Phys. J. C} {\bfseries 78} no.~3, (2018) 199},
  \href{http://arxiv.org/abs/1710.09748}{{\ttfamily arXiv:1710.09748
  [hep-ex]}}.

\bibitem{ATLAS:2020ius}
{\bfseries ATLAS} Collaboration, ``{Search for doubly- and singly-charged Higgs
  bosons decaying into vector bosons in multi-leptons final states with the
  ATLAS detector using proton-proton collisions at $\sqrt{s} = 13$ TeV}.''
  {ATLAS-CONF-2020-056}, 11, 2020.

\bibitem{Chatrchyan:2012ya}
{\bfseries CMS Collaboration} Collaboration, S.~Chatrchyan {\em et~al.}, ``{A
  search for a doubly-charged Higgs boson in $pp$ collisions at $\sqrt{s}=7$
  TeV},'' \href{http://dx.doi.org/10.1140/epjc/s10052-012-2189-5}{{\em
  Eur.Phys.J.} {\bfseries C72} (2012) 2189},
\href{http://arxiv.org/abs/1207.2666}{{\ttfamily arXiv:1207.2666 [hep-ex]}}.

\bibitem{CMS:2017pet}
{\bfseries CMS} Collaboration, ``{A search for doubly-charged Higgs boson
  production in three and four lepton final states at
  $\sqrt{s}=13~\mathrm{TeV}$}.'' {CMS-PAS-HIG-16-036}, 1, 2017.

\bibitem{ElKaffas:2006nt}
A.~W. El~Kaffas, W.~Khater, O.~M. Ogreid, and P.~Osland, ``{Consistency of the
  Two Higgs Doublet Model and CP violation in top production at the LHC},''
  \href{http://dx.doi.org/10.1016/j.nuclphysb.2007.03.041}{{\em Nucl. Phys.}
  {\bfseries B775} (2007) 45--77},
\href{http://arxiv.org/abs/hep-ph/0605142}{{\ttfamily arXiv:hep-ph/0605142}}.

\bibitem{Arhrib:2011uy}
A.~Arhrib, R.~Benbrik, M.~Chabab, G.~Moultaka, M.~Peyranere, J.~Ramadan, and
  L.~Rahili, ``{The Higgs Potential in the Type II Seesaw Model},''
  \href{http://dx.doi.org/10.1103/PhysRevD.84.095005}{{\em Phys.Rev.}
  {\bfseries D84} (2011) 095005},
  \href{http://arxiv.org/abs/1105.1925}{{\ttfamily arXiv:1105.1925 [hep-ph]}}.

\bibitem{Bonilla:2015eha}
C.~Bonilla, R.~M. Fonseca, and J.~W.~F. Valle, ``{Consistency of the triplet
  seesaw model revisited},''
  \href{http://dx.doi.org/10.1103/PhysRevD.92.075028}{{\em Phys. Rev.}
  {\bfseries D92} no.~7, (2015) 075028},
\href{http://arxiv.org/abs/1508.02323}{{\ttfamily arXiv:1508.02323 [hep-ph]}}.

\bibitem{Hartling:2014zca}
K.~Hartling, K.~Kumar, and H.~E. Logan, ``{The decoupling limit in the
  Georgi-Machacek model},''
  \href{http://dx.doi.org/10.1103/PhysRevD.90.015007}{{\em Phys. Rev.}
  {\bfseries D90} no.~1, (2014) 015007},
\href{http://arxiv.org/abs/1404.2640}{{\ttfamily arXiv:1404.2640 [hep-ph]}}.

\bibitem{Blasi:2017xmc}
S.~Blasi, S.~De~Curtis, and K.~Yagyu, ``{Effects of custodial symmetry breaking
  in the Georgi-Machacek model at high energies},''
  \href{http://dx.doi.org/10.1103/PhysRevD.96.015001}{{\em Phys. Rev.}
  {\bfseries D96} no.~1, (2017) 015001},
\href{http://arxiv.org/abs/1704.08512}{{\ttfamily arXiv:1704.08512 [hep-ph]}}.

\bibitem{Krauss:2017xpj}
M.~E. Krauss and F.~Staub, ``{Perturbativity Constraints in BSM Models},''
  \href{http://dx.doi.org/10.1140/epjc/s10052-018-5676-5}{{\em Eur. Phys. J.}
  {\bfseries C78} no.~3, (2018) 185},
\href{http://arxiv.org/abs/1709.03501}{{\ttfamily arXiv:1709.03501 [hep-ph]}}.

\bibitem{Maniatis:2006fs}
M.~Maniatis, A.~von Manteuffel, O.~Nachtmann, and F.~Nagel, ``{Stability and
  symmetry breaking in the general two-Higgs-doublet model},''
  \href{http://dx.doi.org/10.1140/epjc/s10052-006-0016-6}{{\em Eur. Phys. J. C}
  {\bfseries 48} (2006) 805--823},
  \href{http://arxiv.org/abs/hep-ph/0605184}{{\ttfamily arXiv:hep-ph/0605184}}.

\bibitem{Maniatis:2014oza}
M.~Maniatis and O.~Nachtmann, ``{Stability and symmetry breaking in the general
  three-Higgs-doublet model},''
  \href{http://dx.doi.org/10.1007/JHEP10(2015)149}{{\em JHEP} {\bfseries 02}
  (2015) 058}, \href{http://arxiv.org/abs/1408.6833}{{\ttfamily arXiv:1408.6833
  [hep-ph]}}. [Erratum: JHEP 10, 149 (2015)].

\bibitem{Maniatis:2015gma}
M.~Maniatis and O.~Nachtmann, ``{Stability and symmetry breaking in the general
  $n$-Higgs-doublet model},''
  \href{http://dx.doi.org/10.1103/PhysRevD.92.075017}{{\em Phys. Rev. D}
  {\bfseries 92} no.~7, (2015) 075017},
  \href{http://arxiv.org/abs/1504.01736}{{\ttfamily arXiv:1504.01736
  [hep-ph]}}.

\bibitem{Ivanov:2006yq}
I.~Ivanov, ``{Minkowski space structure of the Higgs potential in 2HDM},''
  \href{http://dx.doi.org/10.1103/PhysRevD.75.035001}{{\em Phys. Rev. D}
  {\bfseries 75} (2007) 035001},
  \href{http://arxiv.org/abs/hep-ph/0609018}{{\ttfamily arXiv:hep-ph/0609018}}.
  [Erratum: Phys.Rev.D 76, 039902 (2007)].

\bibitem{Kannike:2012pe}
K.~Kannike, ``{Vacuum Stability Conditions From Copositivity Criteria},''
  \href{http://dx.doi.org/10.1140/epjc/s10052-012-2093-z}{{\em Eur. Phys. J.}
  {\bfseries C72} (2012) 2093},
\href{http://arxiv.org/abs/1205.3781}{{\ttfamily arXiv:1205.3781 [hep-ph]}}.

\bibitem{Chakrabortty:2013mha}
J.~Chakrabortty, P.~Konar, and T.~Mondal, ``{Copositive Criteria and
  Boundedness of the Scalar Potential},''
  \href{http://dx.doi.org/10.1103/PhysRevD.89.095008}{{\em Phys. Rev. D}
  {\bfseries 89} no.~9, (2014) 095008},
  \href{http://arxiv.org/abs/1311.5666}{{\ttfamily arXiv:1311.5666 [hep-ph]}}.

\bibitem{Kannike:2016fmd}
K.~Kannike, ``{Vacuum Stability of a General Scalar Potential of a Few
  Fields},'' \href{http://dx.doi.org/10.1140/epjc/s10052-016-4160-3}{{\em Eur.
  Phys. J.} {\bfseries C76} no.~6, (2016) 324},
  \href{http://arxiv.org/abs/1603.02680}{{\ttfamily arXiv:1603.02680
  [hep-ph]}}.
[Erratum: Eur.Phys.J.C 78, 355 (2018)].

\bibitem{Ivanov:2018jmz}
I.~P. Ivanov, M.~K\"opke, and M.~M\"uhlleitner, ``{Algorithmic
  Boundedness-From-Below Conditions for Generic Scalar Potentials},''
  \href{http://dx.doi.org/10.1140/epjc/s10052-018-5893-y}{{\em Eur. Phys. J. C}
  {\bfseries 78} no.~5, (2018) 413},
  \href{http://arxiv.org/abs/1802.07976}{{\ttfamily arXiv:1802.07976
  [hep-ph]}}.

\bibitem{BFBpackage}
M.~K\"opke, ``{BFB, a Mathematica package to check boundedness of general Higgs
  potentials}.'' Available at \url{https://git.io/vFQvi}, 2017.

\bibitem{Babu:2016gpg}
K.~Babu, I.~Gogoladze, and S.~Khan, ``{Radiative Electroweak Symmetry Breaking
  in Standard Model Extensions},''
  \href{http://dx.doi.org/10.1103/PhysRevD.95.095013}{{\em Phys. Rev. D}
  {\bfseries 95} no.~9, (2017) 095013},
  \href{http://arxiv.org/abs/1612.05185}{{\ttfamily arXiv:1612.05185
  [hep-ph]}}.

\bibitem{Deshpande:1977rw}
N.~G. Deshpande and E.~Ma, ``{Pattern of Symmetry Breaking with Two Higgs
  Doublets},'' \href{http://dx.doi.org/10.1103/PhysRevD.18.2574}{{\em Phys.
  Rev. D} {\bfseries 18} (1978) 2574}.

\bibitem{Chauhan:2019fji}
G.~Chauhan, ``{Vacuum Stability and Symmetry Breaking in Left-Right Symmetric
  Model},'' \href{http://dx.doi.org/10.1007/JHEP12(2019)137}{{\em JHEP}
  {\bfseries 12} (2019) 137}, \href{http://arxiv.org/abs/1907.07153}{{\ttfamily
  arXiv:1907.07153 [hep-ph]}}.

\bibitem{Gunion:1990dt}
J.~F. Gunion, R.~Vega, and J.~Wudka, ``{Naturalness problems for rho = 1 and
  other large one loop effects for a standard model Higgs sector containing
  triplet fields},''
\href{http://dx.doi.org/10.1103/PhysRevD.43.2322}{{\em Phys. Rev.} {\bfseries
  D43} (1991) 2322--2336}.

\bibitem{Aoki:2007ah}
M.~Aoki and S.~Kanemura, ``{Unitarity bounds in the Higgs model including
  triplet fields with custodial symmetry},''
  \href{http://dx.doi.org/10.1103/PhysRevD.77.095009}{{\em Phys. Rev. D}
  {\bfseries 77} no.~9, (2008) 095009},
  \href{http://arxiv.org/abs/0712.4053}{{\ttfamily arXiv:0712.4053 [hep-ph]}}.
  [Erratum: Phys.Rev.D 89, 059902 (2014)].

\bibitem{Powers:2000}
V.~Powers and B.~Reznick, ``{Polynomials That Are Positive on an Interval},''
  {\em Transactions of the American Mathematical Society} {\bfseries Vol. 352,
  No. 10} (2000) 4677--4692.

\bibitem{Benoist:2017}
O.~Benoist, ``{Writing Positive Polynomials as Sums of (Few) Squares},'' {\em
  European Mathematical Society NewsLetter} {\bfseries Issue September 2017}
  (2017) 8--13.

\bibitem{Polya:1976}
G.~P\'olya and G.~Szeg\'o, {\em Problems and Theorems in Analysis II}.
\newblock Classics in Mathematics. Springer-Verlag, New York, 1976.

\bibitem{Mathematica}
W.~R. Inc., ``Mathematica, {V}ersion 11.3.''. Champaign, IL, 2018.

\bibitem{Staub:2017ktc}
F.~Staub, ``{Reopen parameter regions in Two-Higgs Doublet Models},''
  \href{http://dx.doi.org/10.1016/j.physletb.2017.11.065}{{\em Phys. Lett. B}
  {\bfseries 776} (2018) 407--411},
  \href{http://arxiv.org/abs/1705.03677}{{\ttfamily arXiv:1705.03677
  [hep-ph]}}.

\bibitem{Krauss:2018orw}
M.~E. Krauss and F.~Staub, ``{Unitarity constraints in triplet extensions
  beyond the large s limit},''
  \href{http://dx.doi.org/10.1103/PhysRevD.98.015041}{{\em Phys. Rev. D}
  {\bfseries 98} no.~1, (2018) 015041},
  \href{http://arxiv.org/abs/1805.07309}{{\ttfamily arXiv:1805.07309
  [hep-ph]}}.

\bibitem{Bando:1992np}
M.~Bando, T.~Kugo, N.~Maekawa, and H.~Nakano, ``{Improving the effective
  potential},'' \href{http://dx.doi.org/10.1016/0370-2693(93)90725-W}{{\em
  Phys. Lett. B} {\bfseries 301} (1993) 83--89},
  \href{http://arxiv.org/abs/hep-ph/9210228}{{\ttfamily arXiv:hep-ph/9210228}}.

\bibitem{Ford:1992mv}
C.~Ford, D.~Jones, P.~Stephenson, and M.~Einhorn, ``{The Effective potential
  and the renormalization group},''
  \href{http://dx.doi.org/10.1016/0550-3213(93)90206-5}{{\em Nucl. Phys. B}
  {\bfseries 395} (1993) 17--34},
  \href{http://arxiv.org/abs/hep-lat/9210033}{{\ttfamily
  arXiv:hep-lat/9210033}}.

\bibitem{Chataignier:2018aud}
L.~Chataignier, T.~Prokopec, M.~G. Schmidt, and B.~Swiezewska, ``{Single-scale
  Renormalisation Group Improvement of Multi-scale Effective Potentials},''
  \href{http://dx.doi.org/10.1007/JHEP03(2018)014}{{\em JHEP} {\bfseries 03}
  (2018) 014}, \href{http://arxiv.org/abs/1801.05258}{{\ttfamily
  arXiv:1801.05258 [hep-ph]}}.

\bibitem{Chakrabortty:2013zja}
J.~Chakrabortty, P.~Konar, and T.~Mondal, ``{Constraining a class of
  B\ensuremath{-}L extended models from vacuum stability and perturbativity},''
  \href{http://dx.doi.org/10.1103/PhysRevD.89.056014}{{\em Phys. Rev. D}
  {\bfseries 89} no.~5, (2014) 056014},
  \href{http://arxiv.org/abs/1308.1291}{{\ttfamily arXiv:1308.1291 [hep-ph]}}.

\bibitem{Chakrabortty:2016wkl}
J.~Chakrabortty, J.~Gluza, T.~Jelinski, and T.~Srivastava, ``{Theoretical
  constraints on masses of heavy particles in Left-Right Symmetric Models},''
  \href{http://dx.doi.org/10.1016/j.physletb.2016.05.092}{{\em Phys. Lett. B}
  {\bfseries 759} (2016) 361--368},
  \href{http://arxiv.org/abs/1604.06987}{{\ttfamily arXiv:1604.06987
  [hep-ph]}}.

\bibitem{osti_7041561}
G.~Ulrich and L.~T. Watson, ``Positivity conditions for quartic polynomials,''
  \href{http://dx.doi.org/10.1137/0915035}{{\em SIAM Journal on Scientific and
  Statistical Computing (Society for Industrial and Applied Mathematics);
  (United States)} }.

\bibitem{wiki}
See for instance \url{https://en.wikipedia.org/wiki/Quartic_function}.

\end{thebibliography}\endgroup

\end{document}